\documentclass[a4paper,fleqn]{cas-dc}
\usepackage[authoryear]{natbib}
\usepackage{amssymb}
\usepackage{caption}
\usepackage{multirow}
\usepackage{longtable}
\usepackage[format=plain,font=normalsize,labelfont=bf]{caption}

\begin{document}
\let\WriteBookmarks\relax
\def\floatpagefraction{1}
\def\textpagefraction{.001}
\shorttitle{Extragalactic Science with the ASTRI Mini-Array}
\shortauthors{F. G. Saturni et al.}

\title[mode = title]{Extragalactic Observatory Science with the ASTRI Mini-Array at the {\itshape Observatorio del Teide}}

\author[1,2]{F. G. Saturni}[type=editor,orcid=0000-0002-1946-7706]
\cormark[1]
\ead{francesco.saturni@inaf.it}
\cortext[cor1]{Corresponding author}
\author[3,4,5,6]{C. H. E. Arcaro}[orcid=0000-0002-1998-9707]
\author[7]{B. Balmaverde}[orcid=0000-0002-0690-0638]
\author[8,9]{J. {Becerra Gonz{\'a}lez}}[orcid=0000-0002-6729-9022]
\author[10]{A. Caccianiga}[orcid=0000-0002-2339-8264]
\author[11]{M. Capalbi}[orcid=0000-0002-9558-2394]
\author[1]{A. Lamastra}[orcid=0000-0003-2403-913X]
\author[1,2]{S. Lombardi}[orcid=0000-0002-6336-865X]
\author[1,2]{F. Lucarelli}[orcid=0000-0002-6311-764X]
\author[12]{R. {Alves Batista}}[orcid=0000-0003-2656-064X]
\author[1,2]{L. A. Antonelli}[orcid= 0000-0002-5037-9034]
\author[13]{E. M. {de Gouveia Dal Pino}}[orcid=0000-0001-8058-4752]
\author[10]{R. {Della Ceca}}[orcid=0000-0001-7551-2252]
\author[1,2,14]{J. G. Green}[orcid=0000-0002-1130-6692]
\author[11]{A. Pagliaro}[orcid=0000-0002-6841-1362]
\author[15]{C. Righi}[orcid=0000-0002-1218-9555]
\author[15]{F. Tavecchio}[orcid=0000-0003-0256-0995]
\author[15]{S. Vercellone}[orcid=0000-0003-1163-1396]
\author[10]{A. Wolter}[orcid=0000-0001-5840-9835]
\author[16]{E. Amato}[orcid=0000-0002-9881-8112]
\author[1,2]{C. Bigongiari}[orcid=0000-0003-3293-8522]
\author[4]{M. B{\"o}ttcher}[orcid=0000-0002-8434-5692]
\author[17]{G. Brunetti}[orcid=0000-0003-4195-8613]
\author[18]{P. Bruno}[orcid=0000-0003-3919-9611]
\author[19]{A. Bulgarelli}[orcid=0000-0001-6347-0649]
\author[20]{M. Cardillo}[orcid=0000-0001-8877-3996]
\author[19]{V. Conforti}[orcid=0000-0002-0007-3520]
\author[18]{A. Costa}[orcid=0000-0003-0344-8911]
\author[11]{G. Cusumano}[orcid=0000-0002-8151-1990]
\author[19]{V. Fioretti}[orcid=0000-0002-6082-5384]
\author[21]{S. Germani}[orcid=0000-0002-2233-6811]
\author[22]{A. Ghedina}[orcid=0000-0003-4702-5152]
\author[19]{F. Gianotti}[orcid=0000-0003-4666-119X]
\author[18]{V. Giordano}[orcid=0000-0001-8865-5930]
\author[23]{A. Giuliani}[orcid=0000-0002-4315-1699]
\author[18]{F. Incardona}[orcid=0000-0002-2568-0917]
\author[11]{A. {La Barbera}}[orcid=0000-0002-5880-8913]
\author[18]{G. Leto}[orcid=0000-0002-0040-5011]
\author[24,25]{F. Longo}[orcid=0000-0003-2501-2270]
\author[16]{G. Morlino}[orcid=0000-0002-5014-4817]
\author[26]{B. Olmi}[orcid=0000-0001-6022-8216]
\author[19]{N. Parmiggiani}[orcid=0000-0002-4535-5329]
\author[15]{P. Romano}[orcid=0000-0003-0258-7469]
\author[18]{G. Romeo}[orcid=0000-0003-3239-6057]
\author[1]{A. Stamerra}[orcid=0000-0002-9430-5264]
\author[15]{G. Tagliaferri}[orcid=0000-0003-0121-0723]
\author[1]{V. Testa}[orcid=0000-0003-1033-1340]
\author[10,21]{G. Tosti}[orcid=0000-0002-0839-4126]
\author[23]{P. A. Caraveo}[orcid=0000-0003-2478-8018]
\author[15]{G. Pareschi}[orcid=0000-0003-3967-403X]

\address[1]{INAF -- Osservatorio Astronomico di Roma, Via Frascati 33, I-00078 Monte Porzio Catone (RM), Italy}
\address[2]{ASI -- Space Science Data Center, Via del Politecnico snc, I-00133 Roma, Italy}
\address[3]{INAF -- Osservatorio Astronomico di Padova, V.lo Osservatorio 5, I-35122 Padova, Italy}
\address[4]{North-West University, Centre for Space Research, SA-2520 Potchefstroom, South Africa}
\address[5]{Universit{\`a} di Padova, Dip. di Fisica, Via F. Marzolo 8, I-35121 Padova, Italy}
\address[6]{INFN -- Sezione di Padova, Via F. Marzolo 8, I-35121 Padova, Italy}
\address[7]{INAF -- Osservatorio Astrofisico di Torino, Via Osservatorio 20, I-10025 Pino Torinese (TO), Italy}
\address[8]{Instituto de Astrof{\'i}sica de Canarias, C/ V{\'i}a L{\'a}ctea s/n, E-38205 La Laguna (Tenerife), Spain}
\address[9]{Universidad de La Laguna, Dep.to de Astrof{\'i}sica, Av.da Astrof{\'i}sico F. S{\'a}nchez s/n, E-38206 La Laguna (Tenerife), Spain}
\address[10]{INAF -- Osservatorio Astronomico di Brera, Via Brera 28, I-20121 Milano, Italy}
\address[11]{INAF -- Istituto di Astrofisica Spaziale e Fisica Cosmica di Palermo, Via U. La Malfa 153, I-90146 Palermo, Italy}
\address[12]{UAM -- CSIC, Instituto de F{\'i}sica Te{\'o}rica, C/ N. Cabrera 13-15, E-28049 Madrid, Spain}
\address[13]{Univ. de S{\~a}o Paulo, Inst. de Astronomia, Geof{\'i}sica e Ci{\^e}ncias Atmosf{\'e}ricas, Cid. Universitaria, R. do Mat{\~a}o 1226, BR-05508-090 S{\~a}o Paulo (SP), Brazil}
\address[14]{Max-Planck-Institut F{\"u}r Physik, F{\"o}hringer Ring 6, D-80805 M{\"u}nchen, Germany}
\address[15]{INAF -- Osservatorio Astronomico di Brera, Via E. Bianchi 46, I-23807 Merate (LC), Italy}
\address[16]{INAF -- Osservatorio Astrofisico di Arcetri, L.go E. Fermi 5, I-50125 Firenze, Italy}
\address[17]{INAF -- Istituto di Radioastronomia, Via P. Gobetti 101, I-40129 Bologna, Italy}
\address[18]{INAF -- Osservatorio Astrofisico di Catania, Via S. Sofia 78, I-95123 Catania, Italy}
\address[19]{INAF -- Osservatorio di Astrofisica e Scienza dello Spazio di Bologna, Via P. Gobetti 93/3, I-40129 Bologna, Italy}
\address[20]{INAF -- Istituto di Astrofisica e Planetologia Spaziali di Roma, Via del Fosso del Cavaliere 100, I-00133 Roma, Italy}
\address[21]{Universit{\`a} di Perugia, Dip. di Fisica e Geologia, Via G. Pascoli snc, I-06123 Perugia, Italy}
\address[22]{INAF -- Fundaci{\'o}n Galileo Galilei, Rbla. J. A. Fern{\'a}ndez P{\'e}rez 7, ES-38712 San Antonio de Bre{\~n}a (TF), Spain}
\address[23]{INAF -- Istituto di Astrofisica Spaziale e Fisica Cosmica di Milano, Via A. Corti 12, I-20133 Milano, Italy}
\address[24]{Universit{\`a} degli Studi di Trieste, Dip. di Fisica, Via A. Valerio 2, I-34127 Trieste, Italy}
\address[25]{INFN -- Sezione di Trieste, Via A. Valerio 2, I-34127 Trieste, Italy}
\address[26]{INAF -- Osservatorio Astronomico di Palermo, P.zza del Parlamento 1, I-90134 Palermo, Italy}

\begin{abstract}
The ASTRI Mini-Array is a next-generation system of nine imaging atmospheric Cherenkov telescopes that is going to be built at the {\itshape Observatorio del Teide} site. After a first phase, in which the instrument will be operated as an experiment prioritizing a schedule of primary science cases, an observatory phase is foreseen in which other significant targets will be pointed. We focus on the observational feasibility of extragalactic sources and on astrophysical processes that best complement and expand the ASTRI Mini-Array core science, presenting the most relevant examples that are at reach of detection over long-term time scales and whose observation can provide breakthrough achievements in the very-high energy extragalactic science. Such examples cover a wide range of $\gamma$-ray emitters, including the study of AGN low states in the multi-TeV energy range, the possible detection of Seyfert galaxies with long exposures and the searches of dark matter lines above 10 TeV. Simulations of the presented objects show that the instrument performance will be competitive at multi-TeV energies with respect to current arrays of Cherenkov telescopes.
\end{abstract}


\maketitle

\tableofcontents

\section{Introduction}\label{sec:intro}
Observations from Earth with arrays of imaging air Cher\-enkov telesc\-opes \citep[IACTs; e.g.,][]{aha92} play a par\-amount role in the future development of the $\gamma$-ray astronomy. In this context, the ``Astronomia con Specchi a Tecnologia Replicante Italiana'' (ASTRI) Mini-Array, a system composed of 9 ASTRI Small-Sized Telescopes (SSTs) originally proposed as a precursor for the Southern site of the Cherenkov Telescope Array \citep[CTA;][]{par16}, is now under construction at the {\itshape Observatorio del Teide} (Tenerife, Canary Islands).

The ASTRI project is an international collaboration led by the Italian National Institute for Astrophysics (INAF), that involves the Instituto de Astrof{\'i}sica de Canarias (IAC, Spain) as strategic partner and scientific partnerships from other Italian institutes, Brazil and South Africa. It points towards the realization of an IACT array of dual-mirror SSTs with Schwarzschild-Couder optical configuration. Such telescopes are characterized by a large field of view (FoV) of $\sim$10$^\circ$ with a spatial and energy resolution of $\lesssim$0.1$^\circ$ and $\sim$10\%, respectively, for energies $\gtrsim$1 TeV, and are equipped with Cherenkov cameras based on silicon photo-multiplier (Si\-PM) detectors and an innovative readout electronics. The first ASTRI prototype (ASTRI-{\itshape Horn D'Artu\-ro}) has been operating at the Serra La Nave observing station on Mt. Etna (Catania, Italy) since 2014. The full functionality of its optical design and camera for Cherenkov observations has been recently demonstrated through the detection of the $\gamma$-ray emission from the Crab Nebula at TeV energies \citep{lom20}.

We highlight that this is the fourth article of a series of papers devoted to the comprehensive description of the ASTRI Mini-Array project from a technological, managerial and scientific point of view. The full technical description of the ASTRI Mini-Array and the expected performance of the system are reported in Scuderi et al. (2022, {\bfseries Paper I} hereafter) and Vercellone et al. (2022, {\bfseries Paper II}), respectively. Since the ASTRI Mini-Array will start to operate as an experiment, it will prioritize observations of core-science cases, which are outlined in {\bfseries Paper II}. Observations of additional sources will be either carried out simultaneously to the core science ones, exploiting the large instrumental FoV, or performed in a subsequent observatory phase. D'A{\`i} et al. (2022, {\bfseries Paper III}) focuses on the potential science outcome from observations of Galactic targets; in this document ({\bfseries Paper IV}), we aim to highlight the scientific prospects of the ASTRI Mini-Array for the observations of extragalactic sources during the observatory phase of the instrument.

In view of the analysis and scientific exploitation of the ASTRI Mini-Array data, the ASTRI Comprehensive Data Challenge project (ACDC) started in 2018 with the goal of producing a representative data set of the ASTRI Mini-Array capabilities, based on a state-of-the-art model of the $\gamma$-ray sky and a realistic observing plan. Details can be found in \citet{pin20}. Although the simulations presented in \citet{pin20} were performed within the framework of the ASTRI Mini-Array located at the CTA Southern site (thus taking into account astrophysical objects that may be unfavorably observable from the Northern hemisphere), they nevertheless provided a useful benchmark to infer the capabilities of the instrument in observing high-energy processes in astrophysical sources.

The paper is organized as follows: we provide an overv\-iew on the extragalactic science at TeV energies in Sect. \ref{sec:targets}; we discuss the possibility to perform serendipitous observations of some extragalactic sources simultaneously to the core-science targets (see {\bfseries Paper II}) in Sect. \ref{sec:simobs}; then, we briefly describe the analysis and simulation setup adopted for each of the proposed targets and present the corresponding results in Sect. \ref{sec:results} and \ref{sec:dmctools}, also comparing the results obtained with ASTRI Mini-Array simulated observations to the existing literature and outlining potential observing strategies to improve the future scientific exploitation of the instrument; finally, we summarize our most important results in Sect. \ref{sec:conc}.

Throughout the article, we evaluate the scientific pros\-pects of observation of potential ASTRI Mini-Array targets by performing $\gamma$-ray observing simulations using the most updated versions of the public software {\ttfamily ctools} \citep{kno16} and {\ttfamily GammaPy} \citep{gammapy:2017}, coupled with a suitable set of ASTRI Mini-Array instrument response functions (IRFs). We refer to {\bfseries Paper II} for a detailed description about the IRF production and validation process. In such simulations, we make use of the most recent model for extragalactic background light (EBL) by \citet{fra17} unless otherwise stat\-ed (in which cases, the adopted EBL model is indicated). We note that the adopt\-ed IRFs were produced for a fixed zenith angle (ZA) of 20$^\circ$. While these IRFs are appropriate for sources that are observable at low ZAs ($\sim$30$^\circ$), they may not be entirely adequate for sources whose culmination is at significantly higher ZAs. In order to avoid significant bias in our analysis, we therefore limit our panoramic view of the ASTRI Mini-Array extragalactic targets to objects that can be observed under low-to-intermediate ZAs ($\lesssim$45$^\circ$). This choice ensures that the energy threshold -- a particularly important quantity for extragalactic VHE studies, due to EBL absorption -- is at most a factor of $\sim$2 greater than that of the adopted IRFs\footnote{Waiting for the production IRFs at intermediate (40$^\circ$) and large ZAs (60$^\circ$), we estimate, by means of an {\itshape ad-hoc} Monte-Carlo simulation of $\gamma$-rays at various ZAs up to 60$^\circ$, that the energy threshold of the ASTRI Mini-Array can be approximated by the empirical formula $E_{\rm thr}({\rm ZA}) = E_{\rm thr}(0^\circ) \times \cos{({\rm ZA})}^{-2.5}$. Therefore, the energy threshold at ZA $= 45^\circ$ is a factor of $\sim$2 greater than that at ZA $= 20^\circ$.}, while the other performance quantities should not be much affected (thus making the impact on our obtained results quite limited). Finally, we adopt a $\Lambda$-CDM cosmology with $H_0 = 70$ km s$^{-1}$ Mpc$^{-1}$, $\Omega_{\rm M} = 0.3$ and $\Omega_\Lambda = 0.7$.

\section{Overview of the extragalactic science at TeV energies}\label{sec:targets}
The main scientific prospects of extragalactic astronomy with the ASTRI Mini-Array mainly rely on deep observations of active galactic nuclei \citep[AGN; e.g.,][]{lyn69} and galaxy clusters at energies $\gtrsim$1 TeV, and on cosmology and fundamental phys\-ics studies. Since the search for Lorentz invariance violation (LIV) effects, the TeV observations and constraints on the EBL and the test on the existence of axion-like particles have been already presented in {\bfseries Paper II} {\bfseries (and refs. therein)}, here we focus on the search of $\gamma$-ray signals produced by dark matter annihilation or decay into Standard Model (SM) pairs \citep[e.g.,][]{ber05} from halos around extragalactic astrophysical sources, such as the dwarf spheroidal galaxies \citep[dSphs; e.g.,][]{str08}. In the following, we therefore provide an overview on such fields of extragalactic science at very high energies (VHE), and also briefly outline additional science cases that are worth considering for future observations.

\subsection{Emission of $\gamma$-rays from active galactic nuclei}
AGN are one of the primary $\gamma$-rays emitters located outside the Milky Way. In these objects, the gravitational energy released by matter falling on the central supermassive black hole (SMBH) through accretion processes \citep{sal64,zel65} is released in the form of radiation and/or kinetic energy powering gas outflows. An exhaustive review on $\gamma$-ray observations of AGN is given e.g. in \citet{mad16}; here, we mainly focus on the capabilities of the ASTRI Mini-Array to detect:
\begin{itemize}
    \item the signature emission from the brightest and closest blazars Mkn 421 and Mkn 501 \citep[e.g.,][]{mar72};
    \item the signal from additional (extreme) high-synchrotron peaked blaz\-ars \citep[HSPs and EHSPs; e.g.,][]{pad95}, besides the sources mentioned above;
    \item the $\gamma$-ray emission from Seyfert galaxies \citep{FermiColl19}.
\end{itemize}

\subsubsection{TeV emission from blazars}
Blazars are extragalactic sources characterized by emission of radiation covering the whole electromagnetic spectrum and usually showing flux variability, often characterized by exceptional amplitude and, in some cases, by extremely short timescales down to few minutes \citep[e.g.,][]{aha07}. Their spectral energy distribution (SED) is dominated by non-thermal radiation attributed to a relativistic jet of plasma pointing close to our line of sight \citep[see e.g.][]{urr95}. In their SEDs we indeed identify a low-energy component associated to synchrotron radiation by relativistic electrons, and a component peaking at higher energy that, although widely interpreted as inverse Compton (IC) emission, could also be associated to hadronic processes involving high-energy protons or ions \citep[see e.g.][for  reviews of the blazar emission mechanisms and energetics]{cel08,bot13, 2020arXiv200306587M}. Protons (or nuclei) accelerated at very high energy could indeed emit through various processes, such as synchrotron radiation \citep{aha00,mue03}, photo-meson reactions \citep{man93,mue03} or pion production through collisions with low-energy protons \citep[e.g.,][]{kel06}. Blazars are also considered possible sources of ultra-high-energy cosmic rays (UHECRs), and have been recently associated with PeV neutrinos detected by IceCube \citep{pad16,TG15,res17,ice18}.

Recent VHE blazar studies within a multi-wavelength context have found evidence for a more complex blazar jet structure than assumed in classical one-zone models. The properties of VHE emitting blazars suggest that a spine-she\-ath structure characterizes their jets \citep{GTC05}. Moreover, in order to reproduce the observed emission, structured jets with multiple emission regions are required \citep[see e.g.] []{2011ApJ...730L...8A, 2014A&A...567A.135A, 2019A&A...623A.175M}. In particular, the extremely high flux variability of few minutes duration observed in some blazars at TeV energies -- e.g., Mkn 501 \citep{Albert2007} -- suggests the existence of extremely fast and compact acceleration/emission regions which could be plausibly explained by fast magnetic reconnection involving misaligned current sheets inside the jet  \citep{giannios_etal_09, 2019arXiv190308982D} that can be naturally excited by turbulence driven by kink instabilities in the underlying helical fields \citep{2016ApJ...824...48S,kad21,med21}. In such scenario, the observed VHE spectrum is supposed to be a superposition of more than one emitting region. However, no solid detection of the expected spectral features has been possible so far using the current generation of Cherenkov telescopes. Next-generation arrays with better sensitivity and energy resolution might help in the search of such characteristic signatures. 

Blazars have been empirically divided into two main clas\-ses on the basis of their optical spectral properties: flat-spec\-trum radio quasars (FSRQs), and BL Lac objects (BL Lacs). The former show strong and broad emission lines, whereas the latter are characterized by featureless optical spectra. A further classification proposed for the blazars is based on the position of the synchrotron peak in their SED that define low, intermediate or high-synchrotron peaked (HSP) blazars, when the peak falls below 10$^{14}$ Hz, between 10$^{14}$ and 10$^{15}$ Hz and above 10$^{15}$ Hz, respectively. Since HSPs typically have featureless optical spectra, i.e. they belong to the class of BL Lacs, the HSPs are also often called high-energy peak\-ed BL Lacs \citep[HBLs;][]{pad95}.

Within the class of HSPs/HBLs there is an important minority, called ``extreme HSPs'' (EHSPs), where the synchrotron emission peaks in the 0.1--10 keV X-ray band \citep{Costamante2001,Biteau2020}. Since the synchrotron and IC humps are usually correlated, in the class of HSPs also the $\gamma$-ray hump peaks at very high energies, typically above 100 GeV. \citet{cos18} recently recognized an even more extreme class of HSPs in which the $\gamma$-ray hump peaks above $\sim$1 TeV (the hard-TeV BL Lacs). As discussed by \citet[][see also the review in \citealt{Biteau2020}]{cos18}, the high value of the Compton peak is potentially challenging the standard one-zone leptonic model. The next generation of Cherenkov telescopes, like the ASTRI Mini-Array, will be fundamental to better understand the physics of these enigmatic sources.

To quantify the actual capabilities of ASTRI Mini-Array to detect and study in detail VHE spectral features in normal and extreme populations of blazars, we investigate simulated observations of the two BL Lac objects Mkn 421 and Mkn 501 . Such sources are the closest known VHE blazars (and HSPs/HBLs), and possibly the best sources to search for peculiar spectral features due to several reasons: (i) they are not strongly affected by the EBL absorption up to high energies; (ii) they are bright VHE sources, allowing a good signal-to-noise detection which is crucial to search for features, and (iii) they are likely only slightly affected by internal absorption within the source. Indeed, Mkn 501 is the only blazar for which a $\sim$4$\sigma$ hint of a narrow spectral feature has been detected so far \citep{2011MNRAS.410.2556D}.

Along with these targets, we also present a selection of (E)HBLs potentially detectable with the ASTRI Mini-Array. As concrete examples of how an (E)HBL should appear at the ASTRI Mini-Array we discuss the simulation of two so\-urces, i.e. the prototypical EHBL 1ES 0229$+$200 \citep{aharonian_etal_07} and the HSP RGB J1117$+$202 detected by {\itshape Fermi}-LAT \citep{abd10}.

\subsubsection{$\gamma$-ray emission from Seyfert galaxies}\label{1068_sec}
The SED of Seyfert galaxies is dominated by thermal emission in the optical-to-UV waveband produced by the accretion disk ar\-ound the SMBH (Seyfert 1). In addition, a corona of hot plasma forms above the accretion disc and can IC-scatter accretion disc photons up to X-ray energies. A large fraction of the optical-UV and X-ray radiation  may be  obscured by interstellar gas and dust close to the accretion disk and/or in the host galaxy (Seyfert 2). The absorbed radiation is reprocessed at some other waveband, most likely in the infrared.

Seyfert galaxies emit also non-thermal radiation in the $\gamma$-ray band, as indicated by the detection of the nearby Seyfert galaxies NGC 1068, NGC 4945, and the Circinus galaxy with the {\itshape Fermi}-LAT $\gamma$-ray space telescope \citep{ack12,hay13}. These galaxies exhibit characteristics of starburst activity, AGN-driven winds, and weak misaligned jets \citep{gal96,elm98,len09,kri11,Garcia14,mel15,zsc16,hen18}. Given the existence of several possible emission mechanisms operating at high-energy the origin of the $\gamma$-ray emission is still undetermined.

A potential mechanism for this emission could be the acceleration of relativistic particles by magnetic reconnection \citep{dgdp_lazarian_05} in the nuclear region of these sources, at the turbulent magnetized corona around the black hole \citep[e.g.,][]{2015ApJ...802..113K, 2015ApJ...799L..20S, 2016MNRAS.455..838K, 2019ApJ...879....6R}. Another possibility relies on the evidence of starburst activity in these systems. The standard paradigm for the origin of the $\gamma$-ray emission in star-forming galaxies is non-thermal emission from relativistic particles accelerated in the shocks produced by supernova explosions \citep[e.g.][]{per08,cea09,abd10,ack12}. Finally, a further possibility supported by recent UHECR observations \citep{augerSBG} is that the $\gamma$-ray emission could result from particles accelerated via other mechanisms. Regardless of how particles are accelerated in these galaxies, the $\gamma$-ray emission is predominantly hadronic, and it is produced by the decay of neutral pions created by collisions between relativistic proton and ambient protons. The detection of the nearby starburst galaxies M 82 and NGC 253 by VERITAS \citep{2009Natur.462..770V} and H.E.S.S. \citep{2009Sci...326.1080A} indicates that VHE photons can be produced in the nuclei of these galaxies.

Similarly to the shocks produced by supernovae explosions, the shocks produced by the interaction of AGN-driven winds with the surrounding interstellar matter are expected to accelerate protons and electrons to relativistic energies, with an efficiency that may exceed that of supernova remnants \citep{fau12,nim15,lam16}. In this scenario, the hadronic emission from pion decays following proton-proton interactions is dominant  above about 100 MeV. At lower energies leptonic processes like IC scattering and non-thermal brems\-strahlung can significantly contribute to the $\gamma$-ray emission. Relativistic particles can also be accelerated in misaligned jets; in the leptonic AGN jet scenario, the $\gamma$-ray emission is produced by IC emission, where the high-energy electrons that are accelerated in the jet up-scatter photons produced either through synchrotron emission from those same electrons  or external seed photons \citep{len10}.

Any hadronic interactions which produce $\gamma$-rays through neutral pion decay will also produce neutrinos through charg\-ed pion decay. Thus understanding the nature of the $\gamma$-ray emission in Seyfert/starburst galaxies has important implications for the neutrino signal expected from this astrophysical objects. A search for astrophysical point-like neutrino sources using 10 years of IceCube data finds an excess of neutrino events over expectations from the isotropic background 0.35 degrees away from the Seyf\-ert galaxy NGC 1068, with a 2.9$\sigma$ statistical significance \citep{ice10}. While the estimated IceCube neutrino flux appears higher than that predicted by  starburst and AGN wind models built on measured $\gamma$-ray flux, the large uncertainty from IceCube spectral measurement and the possible $\gamma$-ray absorption with\-in the source prevented a straightforward connection. Thus, studying possible $\gamma$-ray and neutrino production mechanisms in NGC 1068 is a timely task that may provide a key clue for unveiling the origin of the cosmic diffuse neutrino background flux \citep[see e.g.][]{MAGIC19}. The VHE emission from individual Seyfert galaxies is expected to be low and has yet to be directly observed. To quantify the  capabilities of ASTRI Mini-Array to detect the VHE emission in NGC 1068, in this paper we present dedicated simulation of observation of the $\gamma$-ray spectrum predicted by the model that envisages $\gamma$-ray emission in the energy band covered by the instrument.

A significant fraction of starburst galaxies may coexist with AGN as indicated by observational evidence and theoretical arguments \cite[see e.g.][and references therein]{Alexander12}. The performance of the ASTRI Mini-Array in detecting the VHE emission from starburst galaxies detected by {\itshape Fermi}-LAT in the Northern Hemisphere is analyzed in {\bfseries Paper II}. This analysis indicated that the most promising target for observations with the ASTRI Mini-Ar\-ray is the starburst galaxy M82, for which we performed dedicated simulations of the VHE spectrum. Here we investigate the potential discovery of the VHE emission from star\-burst- and AGN-driven outflows with the ASTRI Mini-Array by considering  a few more examples of starburst/Seyf\-ert galaxies which co\-uld benefit of long exposure times th\-anks to simultaneous observations with the extragalactic so\-urces presented in this paper and the ASTRI Mini-Array core sci\-ence targets. To this aim we used:
\begin{itemize}
    \item the list of starburst galaxies detected by {\itshape Fermi}-LAT \citep{aje20};
    \item a selection of starburst galaxies with {\itshape Fermi}-LAT in search of $\gamma$-ray emission \citep{ack12} that are extracted from a survey of the dense molecular gas tracer HCN \citep{gao04};
    \item a sample of local ($d_\odot < 130$ Mpc) starburst galaxies selected in the radio and infrared bands presented in \citet[][see therein for the sample selection criteria]{Lunardini19}.
\end{itemize}

\subsection{VHE emission from the intra-cluster medium}
During the hierarchical process of cluster formation, sh\-ocks and turbulence are generated in the intra-cluster medi\-um (ICM) and are expected to accelerate both electrons and protons at relativistic energies, leading to a non-thermal population of cosmic rays (CRs) that are confined within the cluster magnetic fields \citep[see e.g.][for a review]{bru14}. The presence of these components is proved by radio observations that detect cluster-scale radio emission from galaxy clusters in the form of radio halos and relics \citep[see e.g.][for a review]{van19}. An unavoidable consequence is the generation of high energy emission due to the decay of $\pi^0$ generated by CR proton-proton collisions in the ICM and IC scattering of CMB photons from primary and secondary electrons.

A $\gamma$-ray emission has recently been detected by {\itshape Fermi}-LAT in the vicinity of the Coma cluster \citep{xi18,ada21,bag21}, although its origin is unclear and contribution from discrete sources may be important. Overall, current observations of near\-by clusters with {\itshape Fermi}-LAT constrain the CR energy budget in these systems to less than few percent of that of the thermal ICM \citep{fer14,zan14,ada21} and shed light on the origin of the diffuse radio emission suggesting that radio emitting electrons are reaccelerated in the ICM, presumably by turbulence \citep{bru12,bru17,ada21}. Models based on turbulent reacceleration of primary and secondary particles in the ICM \citep{bru05,bru11,pin17} predict levels of $\gamma$-ray flux for a Coma-like cluster in the range $E^2 d\Phi/dE \sim 10^{-14} \div 10^{-13}$ erg s$^{-1}$ cm$^{-2}$ at $5 \div 10$ TeV, i.e. about one order of magnitude fainter than the sensitivity achievable by the ASTRI Mini-Array in 50 h of observation.

An additional mechanism to produce VHE photons in galaxy clusters is IC scatter from electron–posit\-ron pairs that are generated by photo-pair and photo-pion production due to the interaction between ultra high energy CR and photons of the CMB. If CR protons are accelerated at EeV energies and confined in the ICM \citep[see e.g.][and refs. therein]{bru14}, the high-energy pairs that are produced should radiate IC emission peaking in the TeV energy band \citep{ino05,van11}; the combination of cosmological and Monte Carlo CR simulations indicates that clusters can contribute substantially to the diffuse $\gamma$-ray flux beyond 100 GeV as observed by {\itshape Fermi}-LAT, HAWC, and CASA-MIA upper limits, depending on the power-law index and the maximum energy of the injected CR spectrum. The contribution amounts to up 100\% of the flux for a spectral index of $\sim$2 and a maximum energy around $10^{17}$ eV \citep{hus22}. Future observations with the ASTRI Mini-Array, in conjunction with other Cherenkov facilities (MAGIC, H.E.S.S., VERITAS, CTA), may allow us to obtain interesting constraints on these processes. Overall, the search of VHE emission from galaxy clusters will be the subject of forthcoming dedicated studies.

\subsection{Indirect dark matter searches with observations of extragalactic astrophysical sources}\label{indirectDM}

\begin{table*}[width=17cm,align=\centering]
\centering
\caption{Basic properties of the most relevant DM-dominated extragalactic sources (dSphs at $d_\odot \lesssim 100$ kpc and galaxy clusters) that can be observed from the {\itshape Observatorio del Teide} site. In the dSph classification, ``cls'' stands for ``classical'' and ``uft'' for ``ultra-faint''.}
\label{tab:dmnorth}
\resizebox{\textwidth}{!}{
\begin{tabular}{ccccccc}
\hline
\hline
\multicolumn{7}{l}{ }\\
Target & Class & R.A. J2000 &  dec. J2000 & Min. ZA & $d_\odot$ & Notes \\
(IAU Name) & & (deg) & (deg) & (deg) & (kpc) & \\
\multicolumn{7}{l}{ }\\
\hline
\multicolumn{7}{l}{ }\\
\multicolumn{7}{l}{{\itshape dSphs}}\\
Bo{\"o}tes I & dSph (uft) & 210.03 & $+$14.50 & 13.80 & $66 \pm 2$ & \\
Bo{\"o}tes II & dSph (uft) & 209.50 & $+$12.85 & 15.45 & $42 \pm 1$ & \\
Bo{\"o}tes III & dSph (uft) & 209.30 & $+$26.80 & 1.50 & $47 \pm 2$ & \\
Coma Berenices & dSph (uft) & 186.75 & $+$23.90 & 4.40 & $44 \pm 4$ & 1.6 deg distance from ON 246\\
Draco I & dSph (cls) & 260.05 & $+$57.92 & 29.62 & $76 \pm 6$ & \\
Draco II & dSph (uft) & 238.20 & $+$64.57 & 36.27 & $20 \pm 3$ & \\
Laevens 3 & dSph (uft) & 316.73 & $+$14.98 & 13.32 & $67 \pm 3$ & \\
Segue 1 & dSph (uft) & 151.77 & $+$16.08 & 12.22 & $23 \pm 2$ & \\
Segue 2 & dSph (uft) & 34.82 & $+$20.18 & 8.12 & $35 \pm 2$ & 3.2 deg distance from 1ES 0229$+$200\\
Sextans & dSph (cls) & 153.26 & $-$1.61 & 29.91 & $86 \pm 4$ & \\
Triangulum II & dSph (uft) & 33.32 & $+$36.18 & 7.88 & $30 \pm 2$ & \\
Ursa Major II & dSph (uft) & 132.88 & $+$63.13 & 34.83 & $32 \pm 4$ & \\
Ursa Minor & dSph (cls) & 227.29 & $+$67.22 & 38.92 & $76 \pm 3$ & \\
Willman 1 & dSph (uft) & 162.34 & $+$51.05 & 22.75 & $38 \pm 7$ & 1.7 deg distance from GB6 J1053$+$4930\\
\multicolumn{7}{l}{ }\\
\multicolumn{7}{l}{{\itshape Clusters}}\\
Abell 520 & Gal. Cluster & 73.58 & $+$2.95 & 25.35 & $9.746 \times 10^5$ & \\
Coma Berenices & Gal. Cluster & 194.95 & $+$27.98 & 0.32 & $1.007 \times 10^5$ & \\
Perseus & Gal. Cluster & 49.95 & $+$41.51 & 13.21 & $0.777 \times 10^5$ & \\
Virgo & Gal. Cluster & 187.70 & $+$12.34 & 15.96 & $0.154 \times 10^5$ & \\
NGC 5813 & Gal. Cluster & 225.30 & $+$1.70 & 26.60 & $0.281 \times 10^5$ & 1.3 deg distance from NGC 5846\\
NGC 5846 & Gal. Cluster & 226.62 & $+$1.61 & 26.69 & $0.246 \times 10^5$ & 1.3 deg distance from NGC 5813\\
\multicolumn{7}{l}{ }\\
\hline
\end{tabular}
}
\end{table*}

Dark matter \citep[DM;][]{zwi33} is the major component of the matter amount in the Universe \citep[$\Omega_{\rm DM} \sim 0.24$;][]{pla16}. Its existence is so far only inferred through indirect evidence of gravitational interaction with baryonic matter, such as the dynamical stability of galaxy clusters, the flattening of spiral-galaxy rotation curves at large distances from the central bulges \citep{rub80}, the different kinematic behavior of gas reservoirs and gravitational wells in events of cluster collisions \citep{clo04}, the formation of large-scale structures and the observed distribution of fluctuations in the Cosmic Microwave Background (CMB).

In past years, efforts aimed at observing the DM components under the form of baryonic matter concentrated in astrophysical objects with no or negligible luminosity at all wavelengths (the so-called Massive Compact Halo Objects, MaCHOs) have proven unfruitful \citep[e.g.,][]{tis07}. Therefore, the current frontier of the DM searches is represented by the identification of candidate elementary particles outside the Standard Model (SM). In particular, since particle DM is compatible with a collisionless fluid of cold Weakly Interacting Massive Particles (WIMPs), there could be the possibility of detecting $\gamma$-ray signals emitted from DM annihilation or decay into SM pairs \citep{ber97}:
\begin{equation}\label{eqn:dmfluxann}
\frac{d\Phi_{\rm ann}}{dE_\gamma} = B^{\rm (ann)}_{\rm F} \frac{\langle \sigma_{\rm ann} v\rangle}{8 \pi m_\chi^2} \sum_i {\rm BR}_i \frac{dN_\gamma^{(i)}}{dE_\gamma} \times J\left(
\Delta\Omega
\right)
\end{equation}
\begin{equation}\label{eqn:dmfluxdec}
    \frac{d\Phi_{\rm dec}}{dE_\gamma} = \frac{B^{\rm (dec)}_{\rm F}}{4 \pi m_\chi} \sum_i\Gamma_i\frac{dN_\gamma^{(i)}}{dE_\gamma} \times D\left(
\Delta\Omega
\right)
\end{equation}
where $m_\chi$ is the DM particle mass, $\langle \sigma_{\rm ann} v\rangle$ its velocity-aver\-aged cross section in annihilation processes, $dN_\gamma^{(i)}/dE_\gamma$ the specific number of final-state VHE photons produced in each SM interaction channel with branching ratio BR$_i$ and/or lifetime $\tau_i = 1/\Gamma_i$, and $B_{\rm F}$ a generalized (de)boost factor that summarizes all the processes that may enhance or quench the $\gamma$-ray emission -- for e.g. monochromatic lines we have $B_{\rm F} = \alpha^2$ because of loop-induced suppression, with $\alpha$ the fine structure constant. Measurements on the CMB power spectrum predict that $\langle \sigma_{\rm ann} v\rangle \lesssim 3 \times 10^{-26}$ cm$^3$ s$^{-1}$ for $100$ GeV $\lesssim m_\chi \lesssim$ 100 TeV, the order of magnitude of those proper of SM electroweak interactions. The sensitivity to such cross-section values is at reach of the next-generation $\gamma$-ray Cherenkov telescopes \citep[e.g.,][]{pie14}, making DM-dominated astrophysical sour\-ces compelling targets for observations with these instruments. 

The evaluation of the potential goodness of a source is quantified by the so-called astrophysical factors $J$ (for DM annihilation) and $D$ \citep[for DM decay;][]{eva04}, i.e. the integral quantities of functions of the DM density profile $\rho_{\rm DM}$ along the line of sight to the targets and the projected angular dimensions $\Delta\Omega$ of the DM halos:
\begin{eqnarray}\label{eqn:jdfact}
J\left(
\Delta\Omega
\right) = \int_{\Delta\Omega} d\Omega \int_{\rm l.o.s.} \rho^2_{\rm DM}\left(
\ell, \Omega
\right) d\ell \\
D\left(
\Delta\Omega
\right) = \int_{\Delta\Omega} d\Omega \int_{\rm l.o.s.} \rho_{\rm DM}\left(
\ell, \Omega
\right) d\ell
\end{eqnarray}
Due to the unobservability of DM with direct astronomical techniques, its spatial distribution around galaxies must be inferred by the study of the kinematic properties of the baryonic matter moving in the DM potential wells \citep[see e.g.][]{bon15a,bon15c}.

A discussion of the prospects of DM searches in the Mil\-ky Way center and halo is made in {\bfseries Paper III}. Concerning the extragalactic science, the most DM-dominated sources are:
\begin{itemize}
    \item the dwarf spheroidal galaxies \citep[dSphs; e.g.,][]{str08,mcc12}, whose relative proximity ($d_\odot \lesssim 250$ kpc) and lack of background emission\footnote{See instead \citet{cro22} for the description of a case of non-background free dSph.} \citep[e.g.,][]{mat98} configure them among the best astrophysical targets to indirectly search for $\gamma$-ray signals from DM annihilation or decay;
    \item the nearest clusters of galaxies, which represent the largest gravitationally bound structures in the Universe ($M \sim 10^{15}$ M$_\odot$) formed up to $\sim$80\% by DM \cite[e.g.,][]{jel09,pin09}.
\end{itemize}
In Tab. \ref{tab:dmnorth}, we report the basic properties of the dSphs within a distance $d_\odot$ of 100 kpc and the clusters of galaxies that are visible from the {\itshape Observatorio del Teide} site under a maximum ZA of 45$^\circ$. The threshold distance of 100 kpc for dSphs has been chosen since the astrophysical factors $J$ and $D$ of these halos scale with $d_\odot^{-2}$ and $d_\odot^{-1}$ respectively \citep[e.g.,][]{pac19}, making their expected DM $\gamma$-ray signal very faint at larger distances. In the Northern sky, the dSphs with the highest values of $J$ and $D$ are the ``classical'' Ursa Minor \citep[UMi;][]{wil55}, the ``ultra-faint'' Coma Ber\-enices \citep[CBe;][]{tru35} and Ursa Major II \citep[UMa II;][]{ser60}. In Sect. \ref{sec:dmctools}, we analyze the prospects of such targets for DM detection with the ASTRI Mini-Array.

\section{Serendipitous observations of ancillary sources and optimized strategies for dedicated pointings of extragalactic targets}\label{sec:simobs}
The scientific prospects at VHE with the ASTRI Mini-Array presented in this paper deal with the observation of $\gamma$-ray emissions from several classes of extragalactic sources over time scales spanning from $\sim$10 to $\gtrsim$100~h. Such time scales are typical for $\gamma$-ray observations at multi-TeV regime and allow us to successfully detect peculiar spectral features (emission lines, bumps, hard cut-offs) or to strength\-en the constraints on the expected emission parameters. If, on the one hand, these goals are often at reach mostly when the studied extragalactic sources are in high-activity states or only with dedicated long-term observing campaigns, on the other hand the large FoV of the ASTRI Mini-Array can be fully exploited to perform simultaneous observations of sour\-ces located within an angular distance up to $\sim$5$^\circ$ from a given primary target. In fact, the reduction in sensitivity by a factor of $\lesssim$2 for off-axis observations at $\sim$5$^\circ$ with respect to on-axis exposures (see {\bfseries Paper II}) is still suitable for scientific purposes.

It is therefore clear that a consistent fraction of observation time of extragalactic sources may be obtained almost ``for free'' in the case those ancillary targets are contained in the same fiducial FoV of a given primary target. This, in turn, can be seen as an effective increase of the total duty cycle of the system, since no large amount of dedicated time for ancillary targets would be needed. In the following, we thus explore in detail some of the opportunities offered by the large FoV of the ASTRI Mini-Array, both for the observations of the core-science targets described in {\bfseries Paper II}, that will be performed during the first 2-3 years of the project, and for dedicated pointings to be proposed in the observatory phase of the instrument lifetime.

\begin{figure*}[pos=tbp,width=17cm,align=\centering]
\centering
\includegraphics[scale=0.7]{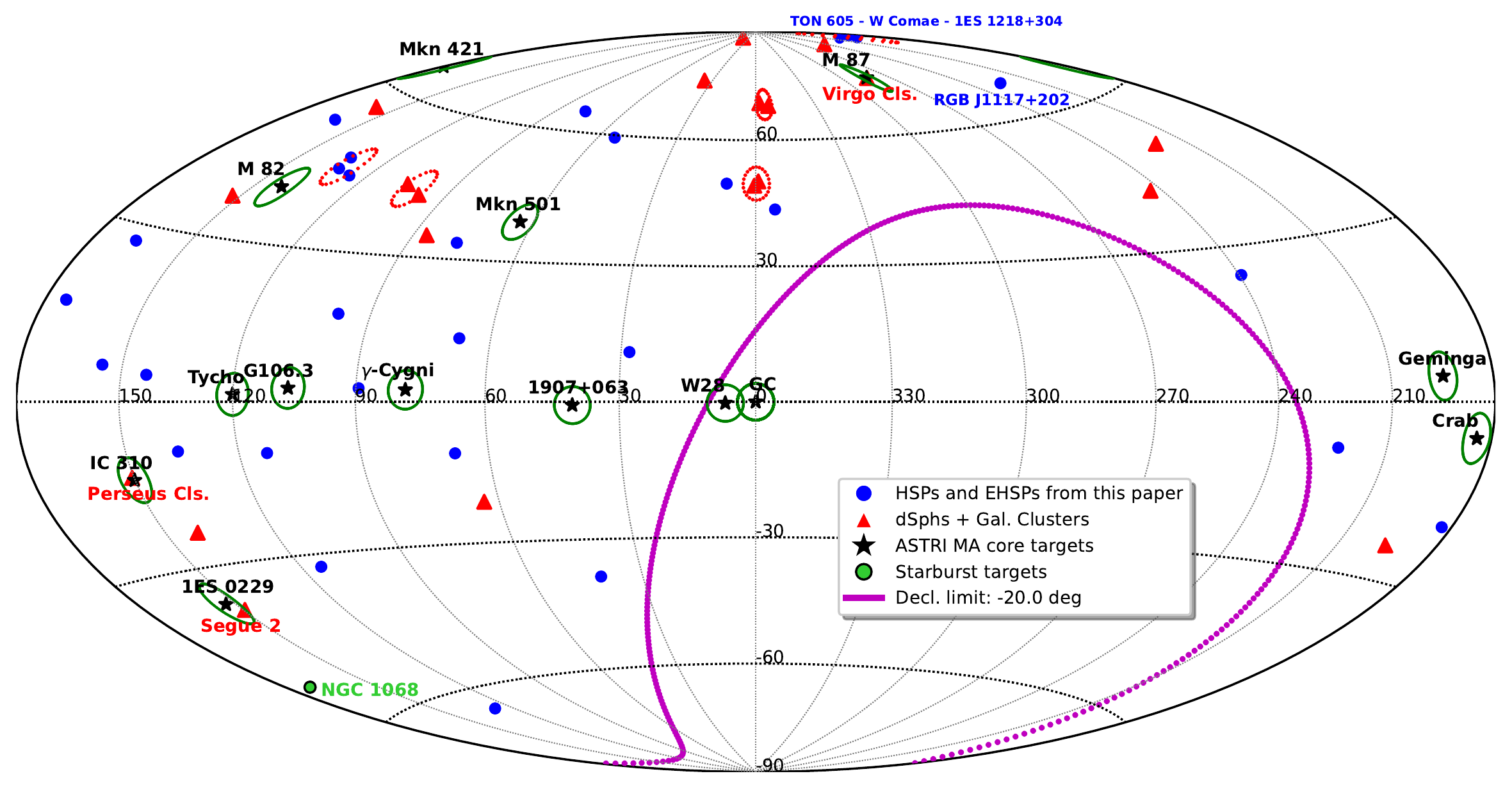}
\centering\caption{Sky distribution, in Galactic coordinates and Hammer-Aitoff projection, of the ASTRI Mini-Array extra-galactic targets presented in this paper: the samples of HSP/HBL blazars extracted from the 5BZCAT Catalogue and EHSP blazars extracted from the 3HSP Catalogue ({\itshape blue dots}), and presented in Tab.~\ref{tab:hsp-ma} and Tab.~\ref{tab:ehsp-ma} respectively; the sample of dSphs and galaxy clusters for DM searches ({\itshape red diamonds}), listed in Tab. \ref{tab:dsph}; the starburst galaxy NGC~1068 ({\itshape black green dots}; see Sect.~\ref{1068_sec}). The position in Galactic coordinates of the ASTRI core-science targets described in {\bfseries Paper II} ({\itshape black stars}) is also shown. A 4$^\circ$ radius circle is drawn around each core-science target ({\itshape green solid circles}) to evidence the fiducial ASTRI Mini-Array FoV around each object, and highlight possible extragalactic targets to be observed simultaneously. Clustered extragalactic targets (in sky projection) possibly to be observed in a single joint observation are also highlighted ({\itshape red dotted circles}). The assumed limit on source declination for the objects visible by the ASTRI Mini-Array from the {\itshape Observatorio del Teide} under a maximum ZA of 45$^\circ$ ({\itshape purple line}) is indicated.}
\label{fig:aitoff_alltargets}
\end{figure*}

First, we consider the possibility of optimizing the observation strategy in order to include more than one target in a single observation of a core-science object. That will be feasible only in those specific cases where two or more interesting targets are clustered in relatively small regions of the sky. The second possibility that we want to explore is the use of pointed observations to serendipitously detect relatively faint sources, not included in the lists of selected candidates presented in this paper, and that are undergoing particularly strong flares. To this end, we will present a large list of additional sources (mainly blazars and starburst/Seyfert galaxies) that fall within the FoV of core-science targets described in {\bfseries Paper II}.

Most of these sources are probably too faint to be detected in normal conditions but the presence of an ASTRI Mini-Array pointing will allow a ``free'' monitoring that could provide, in case of strong flares or unexpected conditions, a detection. Even in case of a non-detection, the observations will nonetheless provide a potentially useful upper-limit on the VHE emission from this class of objects. Finally, we will evaluate whether a similar pointing strategy might be adopted during the observatory phase of the ASTRI Mini-Array, to simultaneously observe sky-projected clusters of some of the extragalactic candidates proposed in this paper. In this way, we will be able to further optimize the dedicated exposure time with respect to the amount needed to observe such targets individually without losing any sources of interest.

Most of the ASTRI observations will be performed in the so-called {\it wobble} observation mode \citep{fom94}. In this observational method, the primary target is typically displaced by $\sim$0.5$^\circ$ from the FoV center. However, thanks to the large FoV of the ASTRI Mini-Array and the rather flat performance up to a few degrees off-axis, a wobble angle of $\sim$1$^\circ$ may be safely adopted for regular observations. Under this assumption, the cross-search radius between the ASTRI core-science targets and the ancillary extragalactic targets can hence reach a realistic value of 4$^\circ$: this ensures that each candidate target lies within $\sim$5$^\circ$ of the FoV center, regardless of the actual position of the main target observed in wobble mode. Such an observational scheme can be immediately applied during the first phase of the ASTRI Mini-Array operations to the regions around the core-science targets.

Fig. \ref{fig:aitoff_alltargets} shows the sky distribution, in Galactic coordinates and Hammer-Aitoff projection, of all the extragalactic ASTRI Mini-Array targets described in this paper (blazars of the HSP/HBL and EHSP sub-class, the sample of dSph and galaxy clusters listed in Tab.~\ref{tab:dsph}, and the Seyfert galaxy NGC 1068) along with the sky position of all the core-science targets described in {\bfseries Paper II}. To mark a possible ancillary extragalactic target within the same FoV, a 4$^\circ$ radius circles has been drawn around each of the core-science targets (solid green circles). A visual inspection already allows us to identify three pairs of main core-science target and a dSph/cluster within the same FoV: Segue 2/1ES 0229$+$200, Perseus Cluster/IC 310, and Virgo Cluster/M 87 (see Fig. \ref{fig:aitoff_alltargets}).

A more detailed analytical cross-correl\-ation study between the core-science targets and the dSph/clu\-ster catalogue reveals that the three targets reported above are within 3.2, 0.6 and 0.1 deg, respectively, from the core-sci\-ence targets. Given the small angular separation between these three couples of objects, a single wobble observation of one of this core-science target will already allow to gather useful exposure on each of these three DM candidates proposed here. As the possible simultaneous observability of HSP/HBL blazars along with the main ASTRI Mi\-ni-Arr\-ay targets is concerned, no relevant objects are found within the same fiducial FoV. Thus, dedicated pointings of some of the most interesting sources among the ones listed in this paper (see Tab.~\ref{tab:hsp-ma} and Tab.~\ref{tab:ehsp-ma}) shall necessarily be performed. Nevertheless, we notice that during the observation of one of the proposed HBL target (BZB J1217+3007, also known as TON 605 or 1ES 1215$+$303)), the large ASTRI FoV will allow to simultaneously observe and monitor other two close-by known TeV blazars, W Comae and 1ES 1218+304 (this last one classified as EHSP), which are, respectively, about 2$^\circ$ and 0.8$^\circ$ from the suggested target (see Fig. \ref{fig:aitoff_alltargets} for the sky location of this triplet).

\onecolumn

\begin{scriptsize}
\begin{center}
\captionsetup{width=17cm}
\begin{longtable}{m{0.25\textwidth}ccccccc}
\caption{\small List of potentially interesting candidates to be observed simultaneously by the ASTRI Mini-Array during the observations of the main core-science targets described in {\bfseries Paper II}. The selection of (ancillary) candidate targets has been performed using the list of blazars available from the {\it Open Universe -- Blazars} reference list \citep{gio19}, which is based on the BZCAT v5.0 \citep{mas15}, the 3HSP \citep{cha19}, and the {\itshape Fermi}-LAT 4LAC \citep{fer20} catalogues, the targets for DM searches reported in Tab. \ref{tab:dmnorth}, and a selection of starburst galaxies that are visible from the Northern hemisphere from \citet{ack12}, \citet{Lunardini19} and \citet{aje20}. For each core-science target (named in the first column) the table reports: the name of the blazar/DM-dominated object/starburst galaxy found within 4$^\circ$ from the main target (see text for discussion about the interplay between the adopted value of angular separation and the wobble angle); its celestial coordinates (J2000); the angular separation in degrees from the main target; source redshift (when available); optical classification for each object; and, for blazars, the SED classification extracted from the 4LAC catalogue.}\\
\label{tab:anc-targ}\\
\hline
\hline
\multicolumn{8}{c}{ }\\
Core-Science Target & Blazar/DM/Starburst Gal. Name & R.A. J2000 &  dec. J2000 & Ang. Sep. & $z$ & Optical Class & SED Class \\
& (within 4$^\circ$ of the main target) & (deg) & (deg) & (deg) & & & \\
\multicolumn{8}{c}{ }\\
\hline
\endfirsthead
\multicolumn{8}{c}{ }\\
{\scriptsize \tablename\ \thetable\ -- \textit{Continued from previous page}} \\
\hline
\hline
\multicolumn{8}{c}{ }\\
Core-Science Target & Blazar/DM/Starburst Gal. Name & R.A. J2000 &  dec. J2000 & Ang. Sep. & $z$ & Optical Class & SED Class \\
& (within 4$^\circ$ of the main target) & (deg) & (deg) & (deg) & & & \\
\multicolumn{8}{c}{ }\\
\hline
\endhead
\hline \multicolumn{4}{r}{\scriptsize \textit{Continued on next page}} \\
\endfoot
\hline
\endlastfoot
\multicolumn{8}{c}{ }\\
\multirow{2}{3cm}{Tycho}&3FGL J0014.6+6119      &     3.70     &     $+$61.30     &       3.1    &      --      &      BCU     &     LSP     \\
&   3HSP J005758.3+632639.3   &     14.49    &     63.44    &       $+$3.7      &     0.180     &      BLL     &     HSP     \\
\multicolumn{8}{c}{ }\\
\hline
\multicolumn{8}{c}{ }\\
\multirow{1}{3cm}{eHWC 1907+063} &   3HSP J191803.6+033031.1   &    289.52    &     $+$3.51     &       3.8      &     0.230     &      BLL     &     HSP     \\
\multicolumn{8}{c}{ }\\
\hline
\multicolumn{8}{c}{ }\\
\multirow{3}{3cm}{$\gamma$ Cygni}&   3FGL J2000.1+4212         &    300.00    &     $+$42.23     &        4.0       &      --     &      BCU     &     LSP     \\
&   5BZU J2015+3710           &     303.87    &     $+$37.18    &       3.7      &      --     &    BZU/FSRQ  &     LSP     \\
&   3FGL J2018.5+3851         &    304.63    &     $+$38.86    &       1.9      &      --     &      BCU     &     ISP     \\
\multicolumn{8}{c}{ }\\
\hline
\multicolumn{8}{c}{ }\\
\multirow{3}{3cm}{Crab} &   5BZB J0521+2112           &     80.44    &     $+$21.21    &       3.1      &     0.108    &      BLL     &     HSP     \\
&   3FGL J0528.3+1815         &     82.12    &     $+$18.28    &        4.0       &      --      &      BCU     &      --     \\
&   5BZB J0540+2507           &     85.06     &     $+$25.13    &       3.4      &     0.623    &      BLL     &      --     \\
\multicolumn{8}{c}{ }\\
\hline
\multicolumn{8}{c}{ }\\
\multirow{3}{3cm}{Geminga}&   5BZB J0621+1747           &     95.45    &     $+$17.79    &       2.9      &      --     &      BLL     &      --    \\
&   3FGL J0631.2+2019         &     97.75    &     $+$20.35     &       2.7      &      --      &      BCU     &      --    \\
&   3HSP J064813.9+160656.5   &    102.06    &     $+$16.12    &       3.8      &     0.350     &      BLL     &     HSP    \\
\multicolumn{8}{c}{ }\\
\hline
\multicolumn{8}{c}{ }\\
\multirow{9}{3cm}{M 82} &   3HSP J091429.7+684508.7   &    138.62    &     $+$68.75    &       3.8      &     0.450     &      BLL     &     HSP     \\
&   5BZQ J0921+7136           &     140.35    &     $+$71.60    &       3.4      &     0.594    &      FSRQ    &      --     \\
&   3HSP J092113.0+684902.2   &    140.30    &     $+$68.82    &       3.2      &      --      &      BLL     &      --     \\
&   3FGL J0928.7+7300         &     142.25    &     $+$72.95    &       3.9      &      --      &      BCU     &      --     \\
&   4FGL J0931.9+6737         &    142.99    &     $+$67.62    &        3.0       &    0.023    &      RDG     &      --     \\
&   3HSP J095849.8+703959.4   &    149.71    &     $+$70.67    &        1.0       &     0.310     &      BLL     &     HSP     \\
&   3HSP J100313.9+705912.6   &    150.81    &     $+$70.99    &       1.4      &      --      &      BLL     &     HSP     \\
&   5BZQ J1003+6813           &    150.78    &     $+$68.22    &       1.6      &     0.770     &      FSRQ    &      --     \\
&   3HSP J102704.3+671619.0   &    156.77    &     $+$67.27    &       3.7      &     0.270     &      BLL     &     HSP     \\
\multicolumn{8}{c}{ }\\
\hline
\multicolumn{8}{c}{ }\\
\multirow{10}{*}{IC 310} &   3HSP J030544.1+403510.5   &     46.43    &     $+$40.59    &       2.2      &     0.240     &      BLL     &     HSP     \\
&   5BZB J0310+4056           &     47.53    &     $+$40.95    &       1.3      &     0.137    &      BLL     &      --     \\
&   5BZQ J0310+3814           &     47.71    &     $+$38.25    &       3.3      &     0.816    &      FSRQ    &     LSP     \\
&   5BZU J0313+4120           &     48.26    &     $+$41.33    &       0.7      &     0.136    &    BZU/RDG   &     LSP     \\ &   5BZG J0313+4115           &     48.49     &     $+$41.26    &       0.5      &     0.029    &      BLL     &      --     \\
&   4FGL J0315.5+4231         &     48.86    &     $+$42.55    &       1.2      &      --      &      BCU     &      --     \\
&   5BZU J0319+4130           &     49.95    &     $+$41.51    &       0.6      &     0.018    &    BZU/RDG   &     LSP     \\ &   4FGL J0333.8+4007         &     53.45    &     $+$40.11    &       3.5      &      --      &      BCU     &      --     \\
&   4FGL J0334.3+3920         &     53.58    &     $+$39.36    &       3.9      &     0.021    &      RDG     &     ISP     \\ &   Perseus                   &     49.95     &     $+$41.51     &       0.6      &       -      &   Gal. Cluster &             \\
\multicolumn{8}{c}{ }\\
\hline
\multicolumn{8}{c}{ }\\
\multirow{8}{*}{M 87} &   3HSP J122307.2+110038.1   &     185.78    &     $+$11.01    &       2.3      &      0.500     &      BLL     &     HSP     \\
&   3HSP J122340.1+124203.6   &    185.92    &     $+$12.70    &       1.8      &     0.340     &      BLL     &     HSP     \\
&   4FGL J1223.3+1213         &    185.95    &     $+$12.04     &       1.8      &      --      &      BLL     &     LSP     \\
&   3HSP J122820.5+155655.1   &    187.09    &     $+$15.95    &       3.6      &     0.232    &      BLL     &     HSP     \\
&   5BZB J1231+1421           &     187.85    &     $+$14.36    &        2.0       &     0.256    &      BLL     &     ISP     \\
&   3HSP J123353.4+145925.7   &    188.47    &     $+$14.99     &       2.7      &     0.520     &      BLL     &     HSP     \\
&   Virgo             &     187.70     &     $+$12.34     &      0.1      &       0.004      &   Gal. Cluster &      --     \\ &   NGC 4254                  &    184.71   &    $+$14.43    &      3.6      &   0.008   &   Starburst Gal.  &      --     \\
\multicolumn{8}{c}{ }\\
\hline
\multicolumn{8}{c}{ }\\
\multirow{9}{*}{Mkn 501} &   4FGL J1639.2+4129         &    249.82    &     $+$41.48    &       3.3      &     0.691    &      FSRQ    &     LSP     \\
&   5BZQ J1642+3948           &    250.75    &     $+$39.81     &       2.1      &     0.593    &      FSRQ    &     LSP     \\
&   5BZQ J1646+4059           &    251.74    &     $+$40.99    &       1.8      &     0.835    &      FSRQ    &      --     \\
&   3HSP J164702.6+385001.6   &    251.76    &     $+$38.83    &       1.6      &     0.135    &      BLL     &      --     \\
&   5BZQ J1648+4104           &    252.12    &     $+$41.07    &       1.7      &     0.852    &      FSRQ    &      --     \\
&   4FGL J1648.2+4232         &    252.13    &     $+$42.56    &        3.0       &      --      &      BCU     &      --     \\
&   5BZQ J1650+4140           &    252.52    &     $+$41.68    &        2.0       &     0.585    &      FSRQ    &             \\
&   5BZB J1651+4212           &    252.79    &     $+$42.22    &       2.5      &     0.269    &      BLL     &      --     \\
&   5BZB J1652+3632           &    253.20    &     $+$36.54    &       3.2      &     0.648    &      BLL     &      --     \\
\multirow{6}{*}{Mkn 501} &   5BZB J1652+4023           &    253.21    &     $+$40.39    &       0.7      &     0.240     &      BLL     &     HSP     \\
&   5BZB J1655+3723           &    253.97    &     $+$37.39    &       2.4      &      --      &      BLL     &      --     \\
&   5BZQ J1659+3735           &    254.88    &     $+$37.59    &       2.4      &     0.771    &      FSRQ    &      --     \\
&   5BZB J1701+3954a          &    255.29    &     $+$39.91    &       1.4      &      --      &      BLL     &      --     \\
&   5BZB J1701+3954b          &    255.35    &     $+$39.91     &       1.5      &     0.507    &      BLL     &      --     \\
&   3HSP J170132.2+381103.9   &    255.38    &     $+$38.18    &       2.2      &      0.600     &      BLL     &     HSP     \\
\multicolumn{8}{c}{ }\\
\hline
\multicolumn{8}{c}{ }\\
\multirow{15}{*}{Mkn 421} &   5BZB J1051+3943           &    162.86    &     $+$39.72    &        3.0       &     0.498    &      BLL     &     ISP     \\
&   5BZG J1100+4210           &    165.09    &     $+$42.18    &        4.0       &     0.323    &      BLL     &      --     \\
&   5BZB J1100+4019           &    165.09    &     $+$40.32    &       2.3      &     0.225    &      BLL     &      --     \\
&   3FGL J1101.5+4106         &    165.35    &     $+$41.06    &       2.9      &      --      &      BCU     &      --     \\
&   5BZB J1101+4108           &    165.35    &     $+$41.15    &        3.0       &     0.580     &      BLL     &     LSP     \\
&   4FGL J1101.5+3904         &    165.38    &     $+$39.08    &        1.0       &      --      &      BCU     &     LSP     \\
&   5BZB J1102+3801           &    165.60    &     $+$38.02    &       0.4      &     0.392    &      BLL     &             \\
&   5BZG J1105+3946           &    166.47    &     $+$39.78    &       1.6      &     0.099    &      BLL     &     LSP     \\
&   3HSP J110600.3+375445.6   &    166.50    &     $+$37.91    &       0.4      &     0.640     &      BLL     &     HSP     \\
&   5BZB J1109+3736           &     167.41    &     $+$37.60    &       1.2      &     0.398    &      BLL     &     ISP     \\
&   5BZB J1110+3539           &    167.74    &     $+$35.65    &       2.9      &      --      &      BLL     &     ISP     \\
&   5BZB J1111+3452           &    167.88    &     $+$34.87    &       3.6      &     0.212    &      BLL     &      --     \\
&   3HSP J111603.4+371036.1   &    169.01    &     $+$37.18    &       2.5      &     0.269    &      BLL     &     HSP     \\
&   3HSP J111644.6+402635.8   &    169.19    &     $+$40.44    &       3.3      &     0.202    &      BLL     &     HSP     \\
&   Arp 148                   &    165.98    &      $+$40.85    &      2.6       &    0.035   &  Starburst Gal.   &      --     \\
\multicolumn{8}{c}{ }\\
\hline
\multicolumn{8}{c}{ }\\
\multirow{8}{3cm}{1ES 0229+200} &   4FGL J0226.7+2312         &     36.63     &     $+$23.19     &       3.2      &      --      &      BCU     &     ISP     \\
&   4FGL J0227.8+2246         &     36.94    &     $+$22.81    &       2.8      &     0.428    &      BCU     &     LSP     \\
&   3HSP J023005.9+194921.0   &     37.53    &     $+$19.82    &       0.8      &     0.530     &      BLL     &     HSP     \\
&   4FGL J0237.3+2000         &     39.33    &     $+$20.01    &       1.1      &      --      &      BLL     &      --     \\
&   5BZB J0238+1636           &     39.66    &     $+$16.62    &       3.9      &     0.940     &      BLL     &     LSP     \\
&   5BZU J0242+1742           &     40.60    &     $+$17.72    &       3.4      &     0.551    &      BZU     &      --     \\
&   3HSP J024507.8+184308.1   &     41.28    &     $+$18.72    &       3.3      &     0.430     &      BLL     &     HSP     \\
&   Segue 2                   &     34.82     &     $+$20.18     &      3.2      &   *   &   dSph (uft) &      --     \\
\multicolumn{8}{c}{ }\\
\end{longtable}
\begin{flushleft}
{\normalsize $^*$Distance of $35 \pm 2$ kpc (see Tab. \ref{tab:dmnorth}).}
\end{flushleft}
\end{center}
\end{scriptsize}

\twocolumn

In addition to the cross-search between the core-science and the extragalactic targets presented above, we also perform a search for other potentially interesting extragalactic sources (in particular known blazars, blazar candidates and starburst galaxies) within the large ASTRI Mini-Array FoV around the core-science targets. Although such additional sources do not satisfy the criteria to be considered as main candidate targets for the observatory phase of the instrument, their observation may nevertheless provide interesting insights, such as flux upper limits (ULs) on different source classes, useful to constrain their predicted $\gamma$-ray emission.

To this end, we cross-match the list of core-science targets with the {\it Open Universe} master list of known and candidate blazars\footnote{The v2.0 of the {\it Open Universe} blazars list is available at the following web address: {\ttfamily http://openuniverse.asi.it/OU4Blazars/MasterListV2/}.} \citep{gio19,Chang_2020} which has been assembled by combining the 5BZCAT \citep{mas15}, the 3HSP \citep{cha19}, and the {\itshape Fermi}-LAT 4LAC catalogs \citep{fer20}, as well as with the sample of starburst galaxies presented in \citet{ack12}, \citet{Lunardini19}, and \citet[][see Sect. \ref{1068_sec} for more details]{aje20}. Tab. \ref{tab:anc-targ} reports, for each core-science target, the list of blazars and/or starburst galaxies found within 4$^\circ$ from it, along with the already identified cross-matches with the DM-dominated astrophysical targets. All blazars with $z \gtrsim 1$ have been removed from the selection. The table also reports, when available, the optical and SED classification of each object in the last two columns.

Among the known BL Lac blazars found in the cross-match search (indicated as BLL in the ``Optical Class'' column), around 30\% of the objects listed in Tab. \ref{tab:anc-targ} belong to the HSP BLL sub-class. These sources, due to their high redshifts or their expected weakness as hard TeV emitters, did not pass the stringent criteria used to select the sample of HSPs and EHSPs shown in Tables \ref{tab:hsp-ma} and \ref{tab:ehsp-ma} and, thus, have to be considered only as ancillary targets of the main ASTRI Mini-Array core-science observations. Besides the relevant number of blazars within 4$^\circ$ from each of the main targets, the cross-match search with the starburst galaxy samples returns also two more objects of this source class which might be exposed during the initial ASTRI Mini-Array experiment phase: Arp 148, 2$^\circ$.6 away from Mkn 421, and NGC 4254, 3$^\circ$.6 away from M 87.

As we have seen above, the majority of the extragalactic targets presented in the paper will be necessarily observed by means of dedicated pointings, presumably during the second phase of the experiment. Nevertheless, as we stated at the beginning of the section, we can again take advantage of the large ASTRI FoV to perform a joint observation of close targets. The red dotted circles in Fig. \ref{fig:aitoff_alltargets} show some of the possible joint pointings which could allow to optimize the observation of two or more targets at the same time, like, for example: the (projected) triplet of blazars composed of Mkn 180, 3HSP J113630.0$+$673704 and 3HSP J122514.2$+$721447 Nor\-thwards of M 82; the cluster pairing NGC 5813/NGC 5846 (separated by an angular distance of $\sim$1$^\circ$.3) at $\sim$50$^\circ$ of Galactic latitude; the two dSphs Bo{\"o}tes I and II (with an angular separation of $\sim$1$^\circ$.7), at a Galactic latitude of $\sim$70$^\circ$.

It is clear that the feasibility of the presented plan of simultaneous observations critically depends on a number of collateral issues that must be preliminary addressed. In particular, the scheduling plan of the ASTRI Mini-Array experiment phase should account for the possibility to have multiple sour\-ces of interest in the same FoV when allocating observing time for the core-science targets. In this respect, a global optimization of the pointing strategy around a given core-science target may be adopted, e.g. defining a pointing region that maximizes the number of possible target in the FoV, while keeping the sensitivity on the primary target very close to the on-axis one.

\section{Results of the simulated observations of TeV-emitting AGN}\label{sec:results}
We present here a comprehensive view of the scientific prosp\-ects that can be achieved with long-term ASTRI Mini-Array observations of the VHE extragalactic sky. Such pros\-pects are related to challenging science cases for which the ASTRI Mini-Array can obtain breakthrough data at $E_\gamma \gtrsim 1$ TeV with exposures that overcome the experiment phase of the instrument. For the case of $\gamma$-ray emitting AGN, we identify the science cases reported below:
\begin{itemize}
    \item the bright and nearby ($z \sim 0.03$) BL Lac objects Mkn 421 and Mkn 501;
    \item two catalogues of HSPs and EHSPs that can represent potential scientific cases for the ASTRI Mini-Array;
    \item two representative science cases from such catalogues, like RGB J1117+202 and 1ES 0229+200;
    \item the $\gamma$-ray emitting Seyfert 2 galaxy NGC 1068.
\end{itemize}

The catalogue and simulation studies of AGN observable at VHE with the ASTRI Mini-Array highlight that the instrument is able to perform detections of $\gamma$-ray signals with expected exposure times in the range from $\sim$10 h (blazars) to $\sim$200 h (Seyferts and starburst galaxies), depending on the object class and activity state. In particular, telescope pointings at nearby blazars in high state or extreme $\gamma$-ray emitters may allow to better characterize peculiar spectral features (e.g., $\gamma$-ray lines, spectral breaks and cut-offs) and systematically study populations of rare and unusual objects. To this end, the catalogues of blazars potentially at reach of the ASTRI Mini-Array presented in this paper offer a powerful tool to immediately identify the best candidates to be targeted for dedicated observations. In the following, we describe for each target the expected scientific results from ASTRI Mini-Array observations.

\subsection{Mkn 421 and Mkn 501}

\begin{figure}
	\centering
	\includegraphics[width=0.5\textwidth]{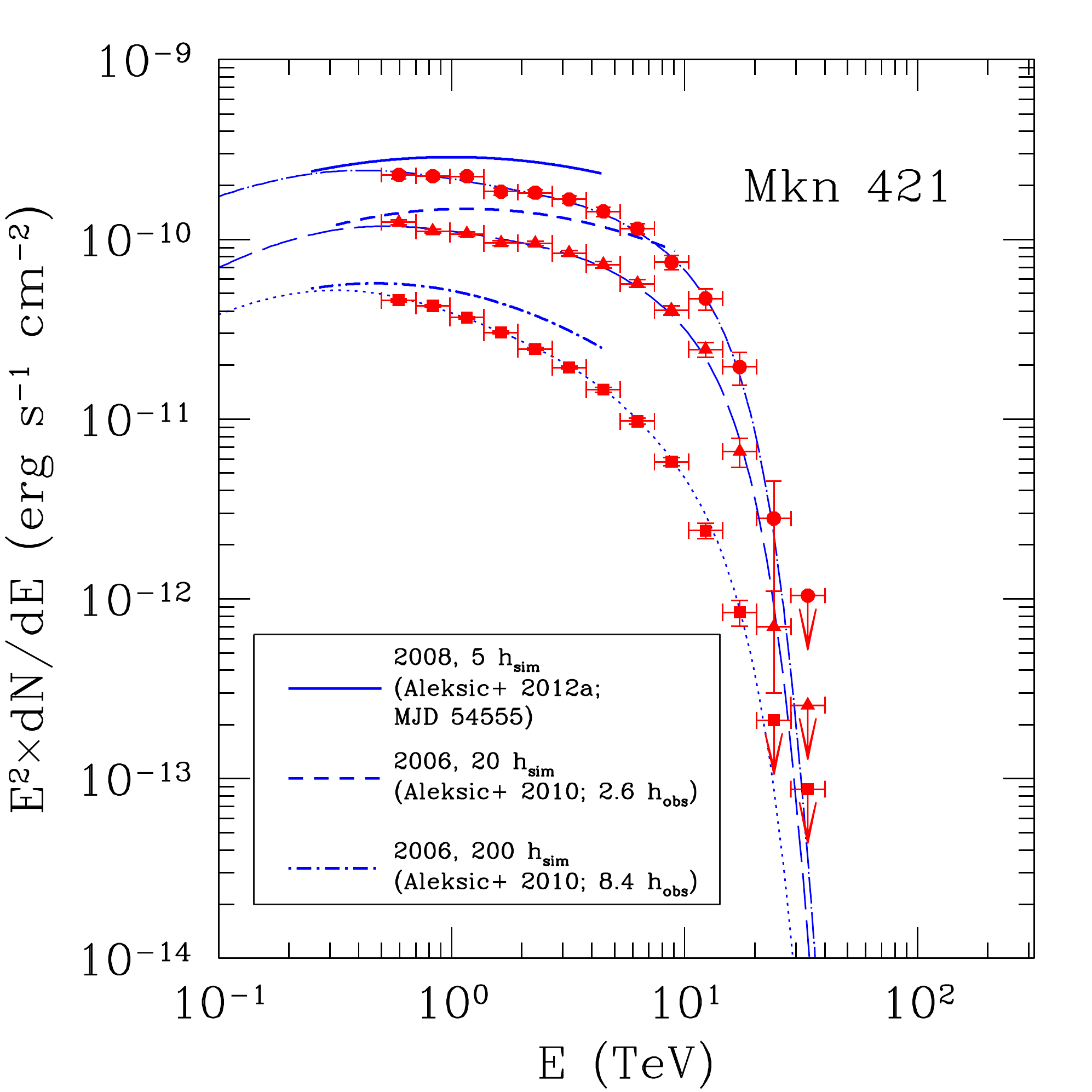} \\
	\caption{Simulations of different source states of Mkn 421 as observed by MAGIC in 2006 during a low (8.4 h of observations) and a high state \citep[2.6 h of observations;][]{Aleksic2010}, as well as in 2008 during a major flare \citep[MJD 54555;][]{Aleksic2012}. Depending on the source state, 5 h ({\itshape red dots}), 20 h ({\itshape red triangles}) and 200 h ({\itshape red squares}) were simulated respectively, considering the intrinsic source spectra ({\itshape blue solid, short-dashed and dot-short-dashed lines}) convolved with the EBL absorption ({\itshape blue dot-long-dashed, long-dashed and dotted lines}).}
	\label{FIG:Chap7_para1_Fig01}
\end{figure}

\begin{figure}
	\centering
	\includegraphics[width=0.5\textwidth]{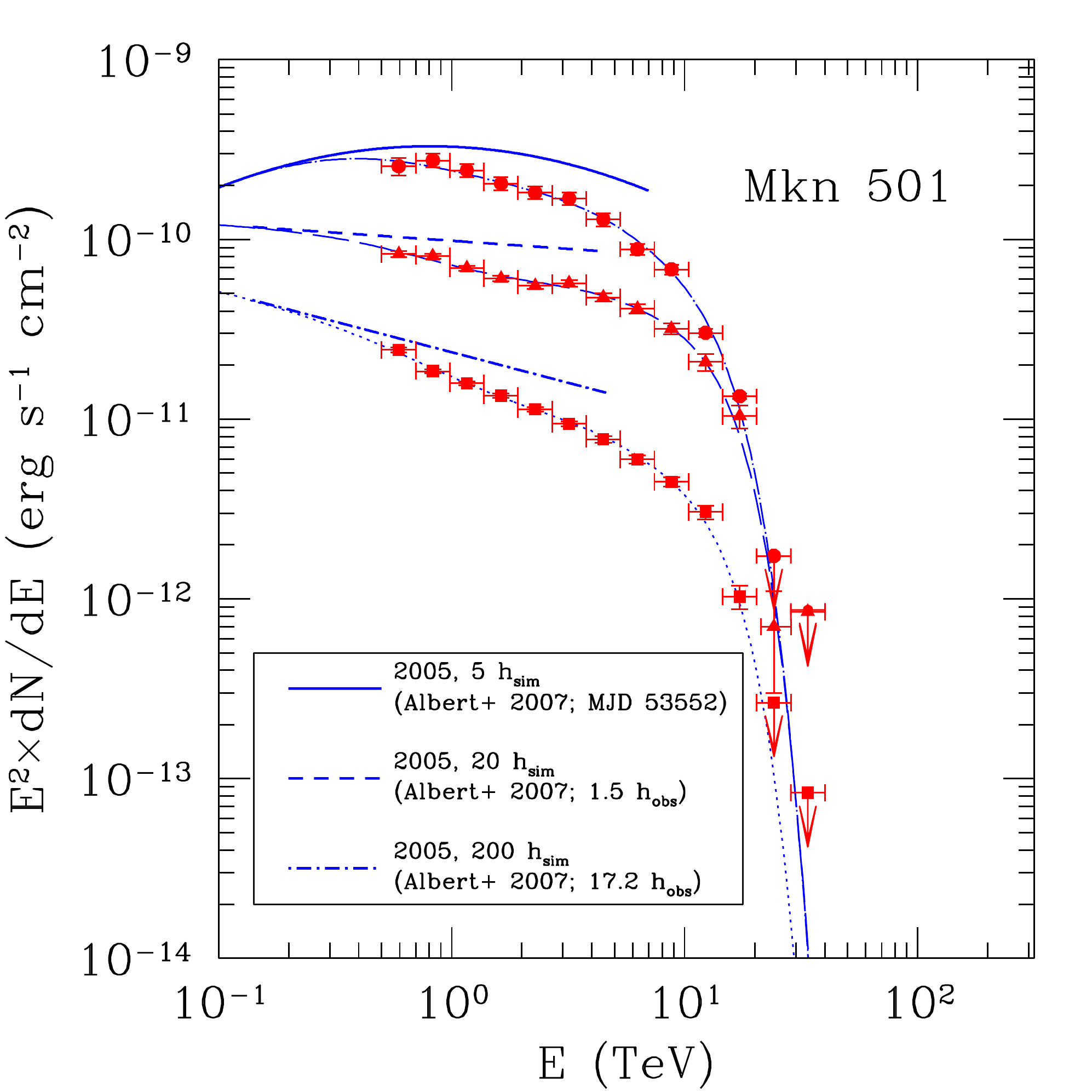} \\
	\caption{Simulations of different source states of Mkn 501 as observed by MAGIC in 2005 during a low (17.2 h of observations) and a high state (1.5 h of observations), as well as during a major flare in the same year \citep[MJD 53552;][]{Albert2007}. Depending on the source state, 5 h ({\itshape red dots}), 20 h ({\itshape red triangles}) and 200 h ({\itshape red squares}) were simulated respectively, considering the intrinsic source spectra ({\itshape blue solid, short-dashed and dot-short-dashed lines}) convolved with the EBL absorption ({\itshape blue dot-long-dashed, long-dashed and dotted lines}).}
	\label{FIG:Chap7_para1_Fig02}
\end{figure}

The blazar subgroup of BL Lac objects dominates the TeV sky as observed by the current generation of Cherenkov arrays. At these energies we can optimally probe the emission from the most energetic electrons emitting through IC scattering \citep[e.g.,][]{1998ApJ...509..608T} and, potentially, from the by-products of hadronic reactions \citep[possibly involved in the emission of high-energy neutrinos; e.g.,][]{2016APh....80..115P}. Although the spectral characterization of the VHE emission of blazars beyond tens of TeV is already at reach of IACTs during high-activity source states and flares -- see e.g. the cases of Mkn 501 detected by HEGRA \citep{aha99} and the Mkn 421 flare detected by MAGIC in 2013 \citep{mag20b} -- the study of the most energetic part of their $\gamma$-ray spectrum during quiescent states is currently hampered by the limited sensitivity of the present generation of arrays above 10 TeV.

The ASTRI Mini-Array will allow us to study in detail the emission from the most energetic particles, constraining the maximum energy attained by the acceleration process(es) and investigating the time-depend\-ent evolution. Complemented with multi-wavelength data, the spectrum record\-ed by the instrument can be used to derive tight constrains on the physical parameters of the emission region. In Tab. \ref{tab:mkn} we present the basic properties of the two closest ($z \sim 0.03$) BL Lac objects Mkn 421 and Mkn 501 for observations with the ASTRI Mini-Array.

\subsubsection{Spectral characterization of low and high flux states}
The goal of the proposed observations is to determine the spectrum of Mkn 421 and Mkn 501 from few TeV up to 30 TeV, above which the EBL absorption completely suppresses the observed emission. We propose ASTRI Mini-Array observations of Mkn 421 and Mkn 501, the closest BL Lacs, to probe: (i) the spectral slope, the maximum energy and the dynamics (through variability) of the most energetic particles; (ii) the optical depth, a key parameter for the potential multimessenger role \citep[e.g.,][]{2019MNRAS.488.4023T}; (iii) the existence of other spectral components, related to photo-meson and/or synchrotron losses of high-energy protons \citep[e.g.,][]{2017A&A...602A..25Z}. The observations above 10 TeV, where the absorption by EBL is rather important, can also be used to test several proposals of fundamental physics (see {\bfseries Paper II}).

\begin{table*}[width=17cm,align=\centering]
\centering
\caption{Basic properties of the BL Lac objects Mkn 421 and Mkn 501 for observations from the {\itshape Observatorio del Teide} site.}
\label{tab:mkn}
\resizebox{\textwidth}{!}{
\begin{tabular}{ccccccc}
\hline
\hline
\multicolumn{7}{c}{ }\\
Target & Class & R.A. J2000 &  dec. J2000 & Min. ZA & $z$ & Notes \\
(IAU Name) & & (deg) & (deg) & (deg) & & \\
\multicolumn{7}{c}{ }\\
\hline
\multicolumn{6}{c}{ }\\
Mkn 421 & Blazar & 166.11 & $+$38.21 & 9.91 & 0.030 & Better suited for ToO observations of high states\\
Mkn 501 & Blazar & 253.47 & $+$39.76 & 11.46 & 0.034 & Better suited for ToO observations of high states\\
\multicolumn{7}{c}{ }\\
\hline
\end{tabular}
}
\end{table*}

In Fig. \ref{FIG:Chap7_para1_Fig01} and \ref{FIG:Chap7_para1_Fig02} we report the VHE section of the SED of the proposed targets, Mkn 421 and Mkn 501. In both cases we report representative spectra for low, high and flare states; such spectra are simulated with observing times of 200 h, 20 h and 5 h. It is worth noting that, although not representative of a single observing run of the source, the 200-h simulation performed here must be interpreted as resulting from the combination of multiple data sets taken over years of ASTRI Mini-Array observations, during periods in which Mkn 421 and Mkn 501 remain at quiescent flux levels. This approach is commonly adopted in current IACTs in the framework of the multi-wavelength (MWL) and multi-messenger study of the blazar emission \citep[see e.g.][]{hec22}.

In particular, the better sensitivity above 10 TeV will allow the ASTRI Mini-Array to probe the potential emergence of additional spectral components (e.g., hadronic), possibly less variable than the leptonic IC component. In all cases the spectrum should be observable up to $\sim$30 TeV for both sources with moderate ($\gtrsim$20 h) exposures during intermediate states, in a similar way to the results achieved by HEG\-RA \citep{aha99}. On the other hand, for low states the detection of the steep spectrum above 10 TeV requires exposures larger than at least 100 h. For both sources, the best opportunities are offered by observations during (relatively frequent, especially for Mkn 421) high states. Historical records show that these states can last for several days, allowing to easily accumulate $\gtrsim$5 hours of data during a single event. Low states can be potentially relevant in view of the possible presence of slowly-varying hadronic components.

\subsubsection{Searches of very-high energy spectral features in Mkn 501}
Mkn 501 displayed a historical high flux and hard spectrum during a two-week flare detected with {\itshape Swift}-XRT in the X-rays and with the MAGIC telescopes in the VHE band from 2014 July 16 to 2014 July 31. On 2014 July 19 (MJD 56857.98), a narrow spectral feature centered around 3 TeV was detected at $\sim$4$\sigma$ confidence level, in coincidence with the day with the highest X-ray flux ($>$0.3~keV) measured during more than 14 years of operation of the {\itshape Swift} mission \citep{2020A&A...637A..86M}. If real, this VHE spectral feature can be interpreted within the context of three different theoretical scenarios: a two-zone emitting region model, a pile-up in the electron energy distribution or as the result of a pair cascade from electrons accelerated in a black hole magnetospheric vacuum gap.

\begin{figure}
    \centering
    \includegraphics[scale=0.18]{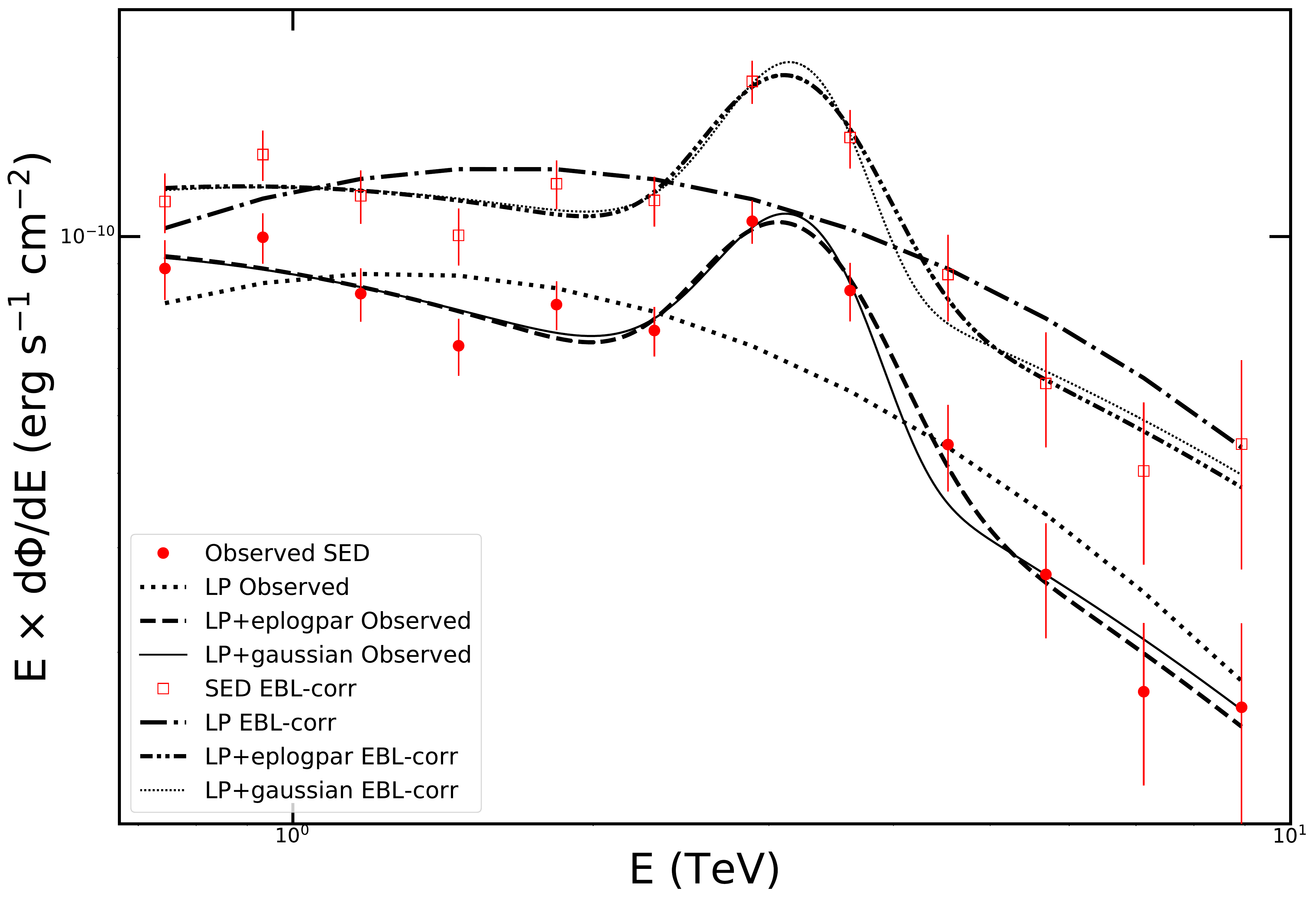}
    \caption{Simulation of Mkn 501 in the high state detected with the MAGIC telescopes on 2014 July 19 (MJD 56857.98), as it would be observed by the ASTRI Mini-Array. 1 h of simulated data taking shows the ability of the ASTRI Mini-Array to detect a spectral feature of the type hinted in the MAGIC data. Both observed and EBL-corrected spectra are shown. The different lines denote the fits performed to the spectral points.}
    \label{fig:mrk501_bump}
\end{figure}

Since the narrow spectral feature is centered at 3 TeV, within the observation window of the ASTRI Mini-Array, and being Mkn 501 one of the targets selected to be followed with the array, simulations of the ASTRI Mini-Array response have been carried out. The starting point of the simulation is the spectral shape observed by MAGIC after correcting for the EBL absorption using the model by \citet{2011MNRAS.410.2556D}. For the MAGIC spectral fit a log-parabola (LP) and an additional strongly curved LP or a Gaussian function were used. The best-fit parameters are given in table 4 of \citet{2020A&A...637A..86M}. For the ASTRI Mini-Array simulations, the spectral points observed with the MAGIC telescopes are used as input to avoid additional uncertainties from the spectral fit. From this input, we simulate the detectable events, that are in turn used to produce a SED by splitting them into independent energy slices that correspond to independent spectral points. In order to calculate the significance of a possible spectral feature, two types of fits are used and compared with a likelihood ratio test, for both the observed spectrum as well as for the intrinsic one after EBL correction \citep{2011MNRAS.410.2556D}. On one hand a broad-band LP is assumed as null hypothesis, and two distinct functions are used to test the hypothesis of an extra spectral component: (i) a LP plus a curved LP \citep[as described in equation 6 of][]{2020A&A...637A..86M} and (ii) a LP plus a Gaussian function. Following this procedure, a single realization of a 1-h observation of Mkn 501 with the ASTRI Mini-Array has been simulated.



The result is shown in Fig. \ref{fig:mrk501_bump}: while the observations with the MAGIC telescopes could only reveal the spectral feature at $\sim$4$\sigma$ confidence level, both assumptions of the narrow LP and of the Gaussian function are preferred with respect to a single broad LP fit with a significance of 5.8$\sigma$ and 5.8$\sigma$, respectively, for the observed spectrum, and 5.4$\sigma$ and 5.3$\sigma$ for the EBL-corrected (intrinsic) spectrum. In order to account for the statistical uncertainties of the spectral points observed by MAGIC, we have simulated a set of 200 realizations. At each iteration, every spectral point has been extracted from a normal distribution whose mean and standard deviations correspond to the individual observed spectral points and their corresponding uncertainties, respectively. The results are reported in Tab.~\ref{tab:mrk501bump}.

Therefore, the presence of a narrow feature in the spectrum of Mkn 501 assuming a similar behaviour as the one observed with the MAGIC telescopes, could be confirmed with the ASTRI Mini-Array in a 1-h exposure assuming the spectral points observed by MAGIC as shown in Fig.~\ref{fig:mrk501_bump}. However, when taking into account the statistical uncertainties, in order to have at least a $\sim$50\% probability of detection the feature 1.5 h of observation time would be required, increasing up to $\sim$80\% probability for 2 h of exposure. This set of simulations have been performed assuming the same characteristics of the MAGIC observations. However, future observations will be essential to search for spectral features with potentially different characteristics and/or in different $\gamma$-ray blazars. 


\begin{table}
\caption{Simulations of the Mkn 501 spectral feature hint. The results are reported as percentage of number of detections of the spectral feature with respect to a broad LP above 5$\sigma$ confidence level for 200 realizations, to account for the statistical uncertainties of the spectral points observed with the MAGIC telescopes.}
\label{tab:mrk501bump}
  \begin{tabular}{ccccc}
    \hline
    \hline
    \multicolumn{5}{l}{ }\\
    Obs. Time &
      \multicolumn{2}{c}{Observed} &
      \multicolumn{2}{c}{Intrinsic} \\
(h) & LP & Gaussian & LP & Gaussian \\
    \multicolumn{5}{l}{ }\\
    \hline
    \multicolumn{5}{l}{ }\\
    1.0 & 32\% & 27\% & 24\% & 19\% \\
    1.5 & 57\% & 53\% & 48\% & 44\% \\
    2.0 & 77\% & 78\% & 75\% & 70\% \\
    \multicolumn{5}{l}{ }\\
    \hline
  \end{tabular}
\end{table}

\subsection{Blazars beyond the local Universe}
In addition to the two (very local) HSPs Mkn 421 and Mkn 501, discussed in the previous section, more blazars will be likely within reach of detection by the ASTRI Mini-Array, even if the EBL absorption is expected to significantly reduce the observed VHE flux of more distant objects. As already discussed, HSPs represent the most promising class of extragalactic so\-urces to be detected. The actual probability of detecting them depends on their global brightness, the shape of the $\gamma$-ray spectrum and their redshift. About 50 HBLs have been currently detected at TeV energies by the former and current ground-based Cher\-enkov detectors\footnote{See the ASI-SSDC TeGeV web catalog ({\ttfamily https://www.ssdc.asi.it/tgevcat/}) or the TeVCat v2.0 ({\ttfamily http://tevcat2.uchicago.edu/}) for the complete list.}. The relatively small number of blazars currently detected at very high energies is mainly due to the limited sensitivity of the current Cher\-enkov instruments at the TeV energies and to the lack of systematic searches. Observations of HSP blazars by the ASTRI Mini-Array can be interesting for several reasons: they can be used as probes for the EBL distribution and fundamental physics studies (as shown in {\bfseries Paper II}), but they can also be observed to study particle acceleration processes up to the most extreme energies and to identify the origin sites of UHECRs and  cosmic neutrinos with energies beyond the PeV.

Given the expected energy threshold ($\sim$1 TeV) and sensitivity (e.g., a factor of $\gtrsim$2 better than the H.E.S.S. Cherenkov array at the highest energies above $\sim$10 TeV; see Sect. 8.2 of {\bfseries Paper II}), the ASTRI Mini-Array is a suitable instrument to perform observations and spectral characterization of HBL/HSP blazars and, in particular, of the so-called {\it extreme} blazars (EHSP), with the IC component peaking in the TeV band. Moreover, the large ASTRI Mini-Array FoV can be exploited to perform HSP/HBL and EHSP blazar surveys, possibly in joint observations with the other class of candidate and known TeV extragalactic sources (see Sect. \ref{sec:simobs}). Coupled with data from existing Cherenkov facilities (MAGIC, H.E.S.S., VERITAS, HAWC), the analysis of ASTRI Mini-Array exposures will be beneficial for the VHE astronomical community in order to characterize the TeV emission of these sources in a multi-instrument framework.

\begin{table*}[width=17cm,align=\centering]
\centering
\caption{List of HSPs potentially detectable with 50-h observations by the ASTRI Mini-Array on the basis of the method described by \citet{bal20}. The last column reports the value of the parameter $F$ which is proportional to the chance for the source to be detected by the ASTRI Mini-Array (see text for details).}
\label{tab:hsp-ma}
\resizebox{\textwidth}{!}{
\begin{tabular}{cccccccc}
\hline
\hline
\multicolumn{8}{l}{ }\\
Target & IAU name & R.A. J2000 & dec. J2000 & Min. ZA & $z$ & {\itshape Fermi} / TeVCat & $F$ \\
      &           & (deg)     & (deg) & (deg)  &  &  & \\
\multicolumn{8}{l}{ }\\
\hline
\multicolumn{8}{l}{ }\\
BZBJ0643+4214 &   B3 0639+423                    &                 100.86 &           $+$42.24 &                 13.94 &   0.089      &   N          /   N &         0.14 \\
BZBJ1104+3812 &   Mkn 421                        &                 166.11 &           $+$38.21 &                  9.91 &   0.030      &   Y          /   Y &         0.85 \\
BZBJ1117+2014 &   RGB J1117+202                  &                 169.28 &           $+$20.24 &                  8.06 &   0.138      &   Y          /   N &         0.14 \\
BZBJ1136+7009 &   Mkn 180                        &                 174.11 &           $+$70.16 &                 41.86 &   0.045      &   Y          /   Y &         0.48 \\
BZBJ1217+3007 &   TON605                         &                 184.47 &           $+$30.12 &                  1.82 &   0.130      &   Y          /   Y &         0.21 \\
BZBJ1428+4240 &   1ES 1426+428                   &                 217.14 &           $+$42.67 &                 14.37 &   0.129      &   Y          /   Y &         0.12 \\
BZBJ1653+3945 &   Mkn 501                        &                 253.47 &           $+$39.76 &                 11.46 &   0.033      &   Y          /   Y &         0.41 \\
BZBJ1728+5013 &   1ES 1727+650                   &                 262.08 &           $+$50.22 &                 21.92 &   0.055      &   Y          /   Y &         0.23 \\BZBJ1959+6508 &   1ES 1959+650                   &                 300.00 &           $+$65.15 &                 36.85 &   0.047      &   Y          /   Y &         0.46
\\
BZBJ2123$-$1036 &   RBS 1742                       &                 320.78 &          $-$10.61 &                 38.91 &   0.023      &   N          /   N &         0.19 \\
BZBJ2347+5142 &   1ES 2344+514                   &                 356.77 &           $+$51.70 &                 23.40 &   0.044      &   Y?     /   Y &         0.17 \\
\multicolumn{6}{l}{ }\\
\hline
\end{tabular}
}
\end{table*}

In order to select a list of HSPs potentially detectable with the ASTRI Mini-Array, we have followed two independent methods. In the first one we start from the catalog of blazars discovered so far \citep[the Roma-BZCAT Multifrequency Catalogue of Blazars;][]{mas09,mas15} to find all the HSPs (not necessarily ``extreme'') with declination $\delta \gtrsim -20^\circ$ -- which guarantees a good source visibility for Cherenkov observations under a maximum ZA of about $45^\circ$ -- that can have a VHE emission strong enough to be detected by the instrument. Since the BZCAT is not a complete catalog, we also present a second selection, this time based on the 3HSP catalog of EHSP candidates \citep{cha19}, specifically focused on the selection of more cases of extreme HSPs not yet detected at TeV energies.

\subsubsection{HSPs from the BZCAT}\label{sec:bzcat}
In order to select a reasonable list of good targets, we have first considered all the HSPs present in the BZCAT and followed the method discussed in \citet{bal20} to predict the VHE emission of each object. Unlike other methods, which are based on the extrapolation of $\gamma$-ray (typically {\itshape Fermi}-LAT) photometric points into the VHE regime, this technique is based just on low energy data, in the radio and in the X-ray band, respectively. This approach was motivated by the idea that some of the HSPs detectable by CTA could be faint at lower energies if they have a very hard $\gamma$-ray spectrum, and may not be yet detected by {\itshape Fermi}-LAT.

We now want to use the same approach for the ASTRI Mini-Array. The prediction of the VHE properties based on radio and X-ray data is possible because the synchrotron and IC humps observed in the SEDs of HSPs are mutually connected so that radio, X-ray and $\gamma$-ray properties are significantly correlated. For example, as discussed in detail in \citet{bal20}, from the intensity of radio emission it is possible to obtain a reasonable estimate of intensity of the $\gamma$-ray emission in the {\itshape Fermi}-LAT energy band, while from the X-ray-to-radio flux ratio parametrized with the two-point spectral index $\alpha_{\rm RX}$ we can derive the slope of the $\gamma$-ray spectrum and the position of the synchrotron and IC peaks. The predicted VHE emission can be then folded with the EBL absorption model \citep{fra17} in order to obtain a prediction of the number of photons actually observable at VHE. In order to take into account the large scatter on these statistical relations we produce a significant number (1000) of possible realizations of the VHE spectrum \citep[see][for more details]{bal20}. The fraction $F$ of VHE spectral realizations that lay above the 50-h ASTRI Mini-Array sensitivity curve then gives a reasonable estimate of the detection probability with this instrument.

We note that the large scatter insisting on the relations used in this method -- from $\sim$0.2 dex to $\sim$0.5 dex \citep{bal20} -- is not only due to the intrinsic variance of the properties of the population, but it also includes the strong variability of the sources. This means that a relatively low value of $F$, e.g. 0.1, does not necessarily mean that the object has a low (10\%) probability of being detected: if the source is variable and it is caught during a high state -- e.g., following a trigger from MWL monitoring -- then its chances of being detected could be significantly higher. At the same time, variability may lead to an overestimate of the value of F under some specific circumstances. This means that a high value of F will not necessarily guarantee the actual detection of the source.

\begin{figure}
\centering
\includegraphics[scale=0.4]{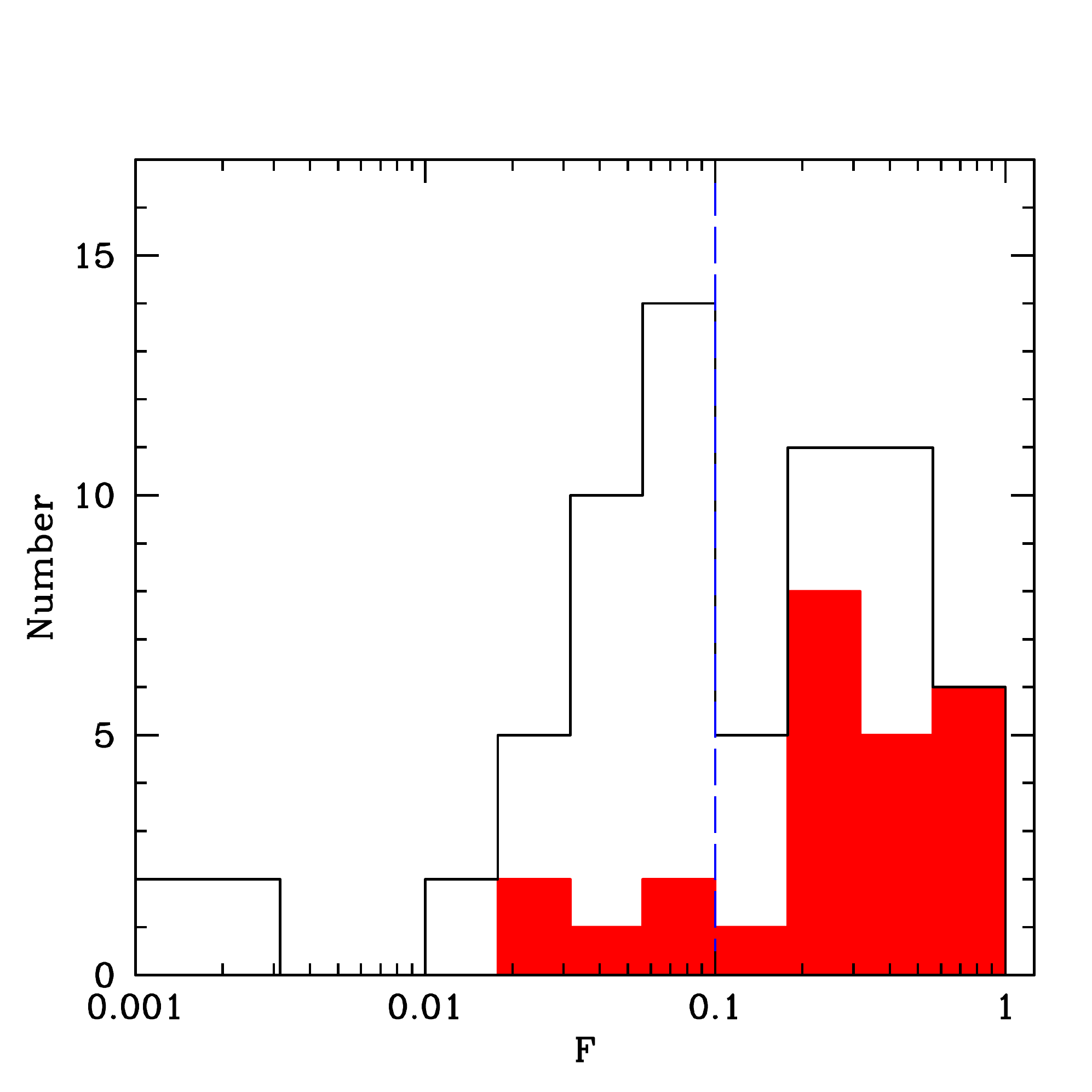}
\caption{Distribution of the parameter $F$ computed for the HSPs observed by VERITAS ({\itshape empty histogram}). The sources that have been actually detected according to \citet[{\itshape red shaded histogram}]{ben19} are highlighted. The threshold adopted in our work to produce the list included in Tab.~\ref{tab:hsp-ma} ({\itshape dashed line}) is also indicated.}
\label{fig:veritas}
\end{figure}

As a test, we have applied the same method to the HSPs observed by VERITAS: 25 HSPs have been detected so far according to \citet{ben19}, while other 43 have only an UL in \citet{arc16}. Using the VERITAS sensitivity curve of 50 h, we computed the values of $F$ for all of these 68 HSPs (see Fig. \ref{fig:veritas}). As expected, the 25 detected sources have values of $F$ significantly larger than the non-detected ones, with 80\% having $F>0.1$ (compared to 30\% of the non-detected). This is a reasonable result, considering the large uncertainties of the method, the variability of the sources and the fact that different exposure times have likely been used for all these targets, while we are using only one sensitivity curve. Based on this result, we have decided to adopt a threshold of $F>0.1$ for the selection of a list of candidates for the ASTRI Mini-Array follow-up. This selection should maximize the detection probability on one hand, and the level of completeness on the other.

For the sources without a redshift estimation but with a lower limit from the literature, we adopt the lower limit as an actual value. If no lower limits are available, we assume a tentative $z$ based on the optical magnitude. By selecting  the sources with at least 10\% of VHE spectral realizations above the sensitivity curve ($F \gtrsim 0.1$) we obtain a list of 11 HSPs with $\delta \gtrsim -20$ deg (i.e. observable from the ASTRI Mini-Array site). The list includes the two brightest and most local HSPs, i.e. Mkn 421 and Mkn 501. As expected, several extreme HSPs like 1ES 1426$+$428 are present in the list, but we also find HSPs with not-so extreme synchrotron peak (between 10$^{16}$ and 10$^{17}$ Hz) like RGB J1117+202, that will be discussed in more details in the next sections.

All but two sources in the selected list are detected by {\itshape Fermi}-LAT, although one (BZB J2347$+$5142) only appears in the preliminary version of the 4FGL catalogue \citep{FermiColl19} and not in the final one. Finally, 8 out of 11 selected HSPs have been already detected also at TeV energies with the current generation of Cherenkov telescopes (see Tab.~\ref{tab:hsp-ma}). As discussed above, we expect that most of the sources listed in Tab.~\ref{tab:hsp-ma} will be actually detected in sufficiently deep ASTRI Mini-Array pointings, although the objects with the lowest values of $F$ ($<$0.2) may require a preventive MWL monitoring to detect them during a high state. In conclusion, while the above studies show that ASTRI can re-detect the brightest TeV emitters and possibly add individual sources to the TeV catalog, it is unlikely that the number of newly detected sources will significantly increase the currently known catalog of ~50 BL Lac objects.

\subsubsection{Extreme blazars}\label{sec:blazres}
The second selection is focused on the ``extreme'' HSP. Recently, the third edition of the HSP/HBL and extreme blaz\-ar catalog \citep[3HSP;][]{cha19} has been released. The catalog contains more than 2000 HSP blazar candidates, with more than 300 classified as EHSPs. Most of the catalog sources also report a redshift estimation (a photometric one whenever the spectroscopic measure is not available) and their $\gamma$-ray counterpart, based on the cross-match with the first release of the Fourth Catalog of 
{\it Fermi}-LAT Sources \citep[4FGL;][]{FermiColl19}, the Second and Third Catalog of Hard {\it Fermi}-LAT Sources \citep[2FHL and 3FHL;][]{Fermi_2FHL, Fermi_3FHL}, and the First Brazil ICRANet $\gamma$-ray blazar catalog \citep[1BIGB;][]{1BIGB_CAT}. A useful quantity provided with the catalog is the figure of merit (FoM) parameter\footnote{The FoM is defined as the ratio of the flux at the synchrotron peak ($\nu_{\rm peak}{\rm f_{\nu_{\rm peak}}}$) of a given source to the peak flux of the faintest 1WHSP blazar detected in the TeV Band by the current IACTs array \citep{cha19}.}, which quantifies the level of potential detectability at TeV energies of each HSP object. In what follows, we have used this catalog to select the best possible EHSP targets to be observed with the ASTRI Mini-Array from the {\itshape Observatorio del Teide} site.

\begin{table*}[width=17cm,align=\centering]
\centering
\caption{List of candidate EHSP targets extracted from the 3HSP catalogue, sorted by decreasing value of the FoM parameter (see text). We note that one source (3HSPJ064326.7+421418) is in common with the list presented in Tab. \ref{tab:hsp-ma}.}
\label{tab:ehsp-ma}
\resizebox{\textwidth}{!}{
\begin{tabular}{ccccccccc}
\hline
\hline
\multicolumn{8}{l}{ }\\
Target & R.A. J2000 &  dec. J2000 & Min. ZA & $z$ & {\itshape Fermi} / TeV & FoM & Min. Det. Time \\
 & (deg) & (deg) & (deg) &  &  &  & [hrs] \\ 
\multicolumn{8}{l}{ }\\
\hline
\multicolumn{8}{l}{ }\\
 3HSPJ064007.2$-$125315 & 100.03   &  $-$12.89  & 41.19 &    0.110 & Y / N  & 5.01 & $<$50 \\
 3HSPJ151148.6$-$051346 & 227.95   &   $-$5.23  & 33.53 &    --    & Y / N  & 2.51 & -- \\
 3HSPJ180408.9$+$004222 & 271.04   &   $+$0.71  & 27.59 &    0.087 & N / N  & 2.51 & -- \\ 
 3HSPJ001827.8$+$294730 &   4.62   &  $+$29.79  &  1.49 &    0.100 & Y / N  & 1.58 & $<$100 \\
 3HSPJ034819.9$+$603508 &  57.08   &  $+$60.59  & 32.29 &    --    & Y / N  & 1.58 & -- \\ 
 3HSPJ050021.5$+$523801 &  75.09   &  $+$52.63  & 24.33 &    0.150 & Y / N  & 1.58 & $<$200 \\
 3HSPJ204206.0$+$242652 & 310.53   &  $+$24.45  &  3.85 &    0.104 & Y / N  & 1.58 & $<$200 \\ 
 3HSPJ044127.5$+$150455 &  70.36   &  $+$15.08  & 13.22 &    0.109 & Y / N  & 1.26 & $<$200 \\
 3HSPJ151041.1$+$333504 & 227.67   &  $+$33.58  &  5.28 &    0.114 & N / N  & 1.26 & -- \\
 3HSPJ151845.7$+$061356 & 229.69   &   $+$6.23  & 22.06 &    0.102 & Y / N  & 1.26 & $\gg$200 \\
 3HSPJ064326.7$+$421418 & 100.86   &  $+$42.24  & 18.44 &    0.089 & N / N  & 1.00 & -- \\   
 3HSPJ005916.9$-$015017 &  14.82   &   $-$1.84  & 30.13 &    0.114 & Y / N  & 0.79 & $<$200 \\
 3HSPJ102212.6$+$512400 & 155.55   &  $+$51.40  & 23.10 &    0.142 & Y / N  & 0.79 & $\gg$200$^*$ \\
 3HSPJ090802.2$-$095937 & 137.01   &   $-$9.99  & 38.29 &    0.054 & N / N  & 0.63 & -- \\
 3HSPJ122514.2$+$721447 & 186.31   &  $+$72.25  & 43.95 &    0.114 & N / N  & 0.63 & -- \\
 3HSPJ190411.8$+$362658 & 286.05   &  $+$36.45  &  8.15 &    0.130 & Y / N  & 0.63 & $<$100 \\ 
\multicolumn{8}{l}{ }\\
\hline
\end{tabular}
}
\begin{flushleft}
{\scriptsize $^*$The 4FGL counterpart of this candidate target (4FGL J1021.9+5123) shows a very poor spectrum determination, 
and it is flagged as 2048 (highly curved spectrum) in the DR3 Catalogue \citep{Fermi4FGLDR3}.}
\end{flushleft}
\end{table*}

We start to select only objects with $\delta \gtrsim -20^\circ$ and redshift $z < 0.15$, since above $\sim$1 TeV (i.e. the nominal energy threshold of the ASTRI Mini-Array) the $\gamma$-ray absorption due to the EBL of sources above a redshift of 0.1 is already severe \citep[see e.g. figures 11 and 12 in][]{fra17}. With this first filtering, we end with a sample of 258 3HSP objects accessible from the ASTRI Mini-Array at Teide. Applying a further selection on FoM $\gtrsim 0.5$, we eventually end with a sample of 146 3HSP sources, which shares most of the well-known TeV-emitting HBLs, like e.g. Mkn 421, Mkn 501, 1ES 1426$+$428, Mkn 180, 
1ES 1959$+$650, 1ES 2344$+$514 (already included in Tab. \ref{tab:hsp-ma}) and 1ES 0229$+$200 (with a $z=0.139$ and FoM $= 3.98$ from the 3HSP Catalog); in particular, the latter is considered as the prototype of such 
extreme TeV sources\footnote{1ES 0229$+$200 also appears as one of the core-science target discussed in {\bfseries Paper II}.}.

The scientific aim of targeted observations of EHSP blaz\-ars with the ASTRI Mini-Array is two-fold: first, spectral 
characterization above several TeV of a few selected EHSP blaz\-ars already detected at TeV energies; 
secondly, detection of new TeV sources belonging to this class of objects. In particular, we want to concentrate the 
observations on already known TeV-extreme blazars like 1ES 0229$+$200 plus the observations of a selected 
sample of EHSP candidates extracted by the 3HSP catalog yet undetected at TeV energies. Detailed simulations of 
the expected spectrum to be observed with the ASTRI Mini-Array in the case of 1ES 0229$+$200, along with the estimate 
of the observing time needed to reach our first scientific objective (extended spectral measurements of EHSP target), 
will be given in the next section. In what follow, we will further process our sample of 3HSP visible for the 
Mini-Array site in order to select highly potential detectable objects not yet detected at TeV energies.

Using the selection criteria described above on source declination, redshift and FoM, and further filtering the remaining sample by selecting only the extreme blazars with an estimated $\nu_{\rm peak}$ above 10$^{17}$~Hz, we 
finally end with a sample of 16 EHSP sources visible from the ASTRI Mini-Array site not yet detected at TeV energies. The complete list is reported in Tab. \ref{tab:ehsp-ma}. Then, taking advantage of the very recent Data Release 3 of the Fourth {\itshape Fermi}-LAT Catalog \citep[4FGL-DR3;][]{Fermi4FGLDR3}, covering 12 yrs of observations, we have reviewed and updated the $\gamma$-ray counterparts of the selected 3HSP sample. We ended with 11 out of 16 targets with a counterpart in the 4FGL-DR3 Catalog (see last column in Tab. \ref{tab:ehsp-ma}).

Still exploiting the 4FGL-DR3 catalog, we have perform\-ed a detectability study of the selected sample for all the sources which both have a Fermi-LAT counterpart and a redshift estimation (9 out of 16 targets). Most of the 4FGL counterparts are significantly detected up to 1 TeV; hence, we have made an estimation of the intrinsic 3HSP multi-TeV source spectrum extrapolating the power-law spectrum measured by {\itshape Fermi}-LAT, and then correcting it by the EBL absorption. The expectation for the EBL-absorbed target spectra above 1 TeV are then compared with the ASTRI sensitivity curves for different exposure times (50, 100, 200, and 500 hours). In the last column of Tab. \ref{tab:ehsp-ma}, we report the minimum number of hours required for a detection at 5$\sigma$ level estimated in this way. The overall result of our study, although limited to the sources with {\itshape Fermi}-LAT counterpart and known redshift, is that at least 40\% of the candidate EHSP targets displayed in Tab. \ref{tab:ehsp-ma} are detectable within 200 hours of observation. 

Since the full ASTRI Mini-Array will be likely operated in $2 \div 3$ years from now, some of the EHSP targets listed above might be already detected at TeV energies by the current Cherenkov detectors operating in the Northern hemisphere (MAGIC, VERITAS, HAWC). If this would be the case, the EHSP targets proposed to be observed might change accordingly. Observation priority should be given in any case to the targets with the highest FoM and the hardest $\gamma$-ray spectra (when available). Finally, the large FoV of the ASTRI Mini-Array can be exploited to perform joint observations of 3HSP blazars with different classes of possible TeV emitters see Sect. \ref{sec:simobs}. In this way, the very long integration times needed to achieve significant results in the other fields of interest (see Sect. \ref{sec:ngc1068} and \ref{sec:dmctools}) might be used to observe more than a target at the same time.

\subsubsection{Simulations of representative cases of observable blazars}
In order to show the actual capability of the ASTRI Mini-Array to detect HSPs/EHSPs, we present here the detailed simulations of two representative cases, namely an extreme HSP (1ES 0229$+$200) and a non-extreme one (RGB J1117$+$\\202). The observing simulation of RGB J1117$+$202 presented in this paper is of particular interest, since it allows us to assess the performance of the ASTRI Mini-Array for the study of similar objects in their low state. The results obtained show the improved sensitivity in the multi-TeV range of the ASTRI Mini-Array compared to the current IACTs. With this new facility, we will be able to extend to higher energies the spectral study of known TeV emitting HSPs, and also to detect in the VHE range some new HSP candidates such as those recently selected for VHE observations \citep[see e.g.][]{cos18,bal19,cha19}. In particular, the improved performances at VHE will be a significant advantage for the study of extreme HSP and, among them, of the so called hard-TeV BL Lacs.\\

\begin{center}
    {\itshape 1ES 0229+200}
\end{center}
The VHE $\gamma$-ray emission of 1ES 0229$+$200 was discovered by H.E.S.S. in 2007 \citep{2007A&A...475L...9A}. Later, the source was observed also by VERITAS \citep{aliu14} and MAGIC \citep{mag19}. By correcting the observed spectrum taking into account the EBL absorption effect, a very hard intrinsic spectrum with a photon index $\Gamma_{\rm intr} = 1.5$ is obtained. Thus, the $\gamma$-ray emission peak results to be located in the multi-TeV range. Due to its spectral characteristics, this source is classified as a hard-TeV BL Lac.

\begin{figure}
\centering
\includegraphics[scale=0.45]{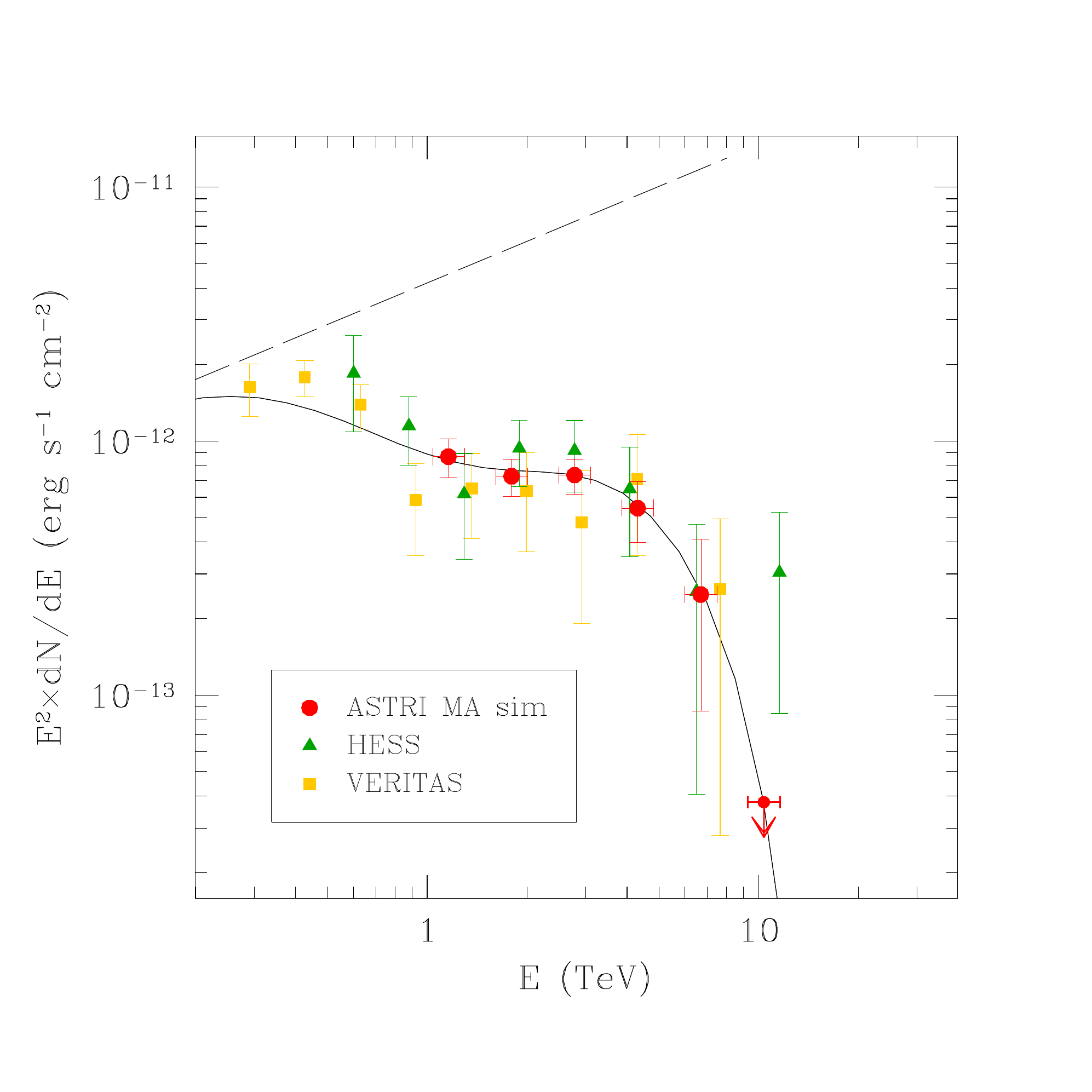}
\caption{Simulated spectrum of 1ES 0229$+$200 as it would be detected in 200 h with the ASTRI Mini-Array ({\itshape red dots}). The assumed spectrum of the source, both intrinsic ({\itshape dashed line}) and modified by EBL absorption ({\itshape solid line}) is also shown, along with H.E.S.S. ({\itshape green triangles}) and VERITAS data ({\itshape yellow squares}) extracted through the SSDC SED builder tool ({\ttfamily  https://tools.ssdc.asi.it/SED/}).}
\label{fig:1es0229_simul}
\end{figure}

To perform the simulation, we provide in input, as spectral model, an EBL-absorbed power law obtained from the results of the VERITAS observations. The significance of the detection is estimated through the test statistics (TS). The TS values derived ensure a solid detection ($>$5$\sigma$) in each of the simulated energy bins. From this study, we expect to detect this source with observations of less than 100-h duration; observations of about 200 h will allow a good characterization of the spectrum. Fig. \ref{fig:1es0229_simul} shows the results of our simulation for 200-h observations with the ASTRI Mini-Array compared to existing datasets. Given the assumed input model and being the VHE emission suppressed for the EBL absorption, the improved  ASTRI Mini-Array performance at VHE is not fully exploited in this case. However, the possibility to  extend to higher energies the spectral study  of this class of objects is important. Since this source is also included in the ASTRI Mini-Array core science programme (see {\bfseries Paper II}), it will be therefore possible to collect data for even longer exposure time, thus enabling accurate studies to investigate in detail its fundamental properties. 

The simulation results for this source are presented mainly as a reference for the study of possible new candidates of this class. Only few objects with a similar VHE spectral shape have been observed so far, and thus the detection of similar sourc\-es would be very useful for several purposes. An extended sample of hard-TeV sources would allow a significant progress for the investigation on fundamental physics and for testing of EBL models, discussed in {\bfseries Paper II}, and would be crucial for the study of emission processes at VHE and, in general, for the knowledge of the blazar population in more detail.

\begin{center}
    {\itshape RGB J1117+202}
\end{center}
As described in Sect. \ref{sec:bzcat}, we expect that the ASTRI Mini-Array will be able to detect not only EHSPs but also less extreme sources. Among the objects included in Tab. \ref{tab:hsp-ma}, we decided to focus on RGB J1117$+$202 ($z=0.138$), a non-extreme HSP (the synchrotron peak falls between 10$^{16}$ and 10$^{17}$ Hz) that has been observed at TeV energies with H.E.S.S. \citep{aha05b} but not detected yet. Our procedure has selected it as potential target for the ASTRI Mini-Array thanks to its bright radio flux combined to a relatively flat $\alpha_{\rm RX}$ value (i.e. relatively high X-ray-to-radio flux ratio). Indeed, it turned out to be a strong $\gamma$-ray source, clearly detected by {\itshape Fermi}-LAT. We decided to simulate this source to show the actual capability of the ASTRI Mini-Array to detect HSPs not necessarily extreme and not yet observed with the current generation of Cherenkov telescopes. This will demonstrate the actual potentialities of the Mini-Array in improving our knowledge of the VHE properties of the blazar population.

\begin{figure}
\centering
\includegraphics[scale=0.4]{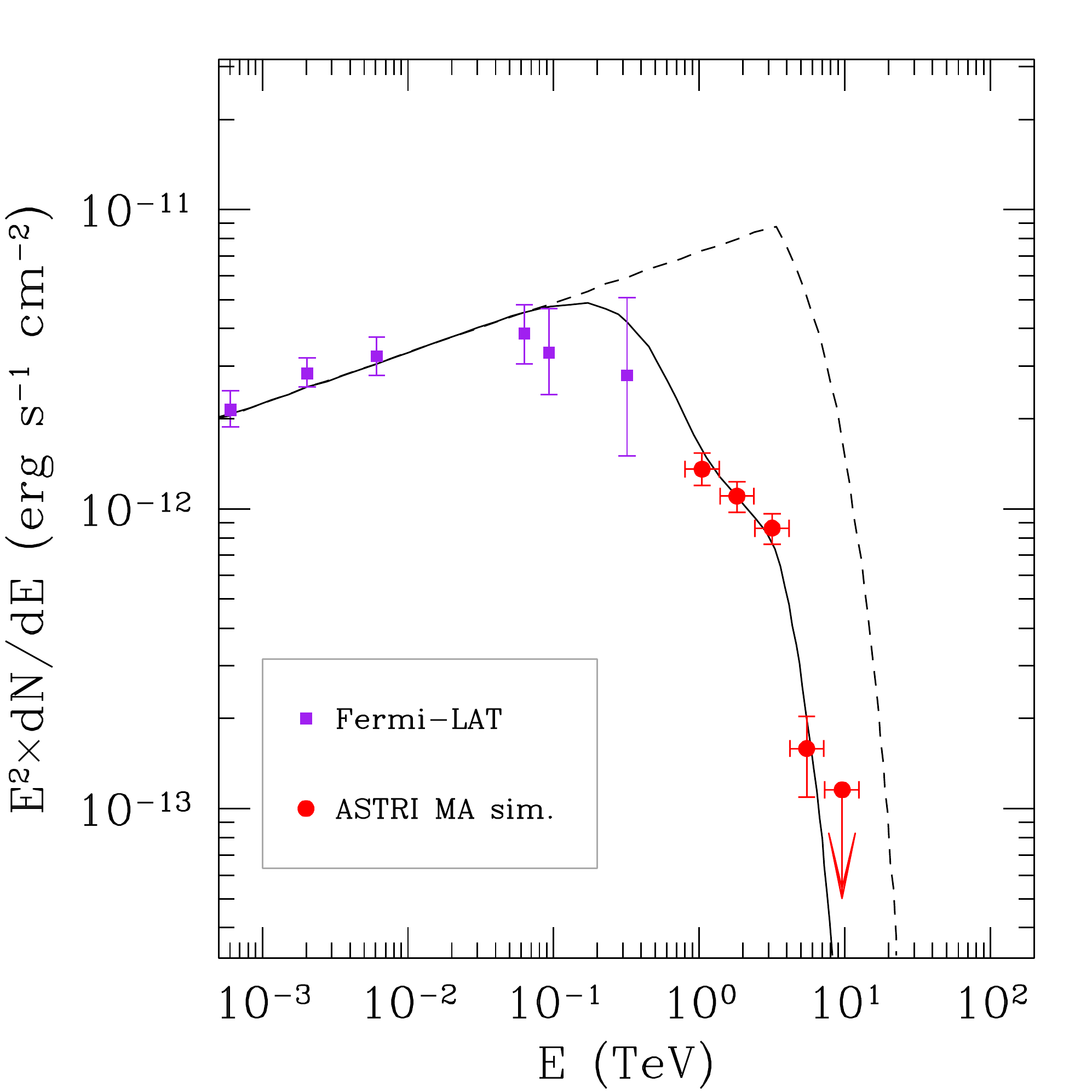}
\caption{SED of RGB J1117+202 in the $\gamma$-rays. The {\itshape Fermi}-LAT data from 3FGL and 2FHL catalogues ({\itshape purple squares}) as drawn from by the SSDC SED builder tool ({\ttfamily https://tools.ssdc.asi.it/SED/}) and the simulated points as observed by the ASTRI Mini-Array in 200 h ({\itshape red dots}) are plotted over the adopted emission models before ({\itshape black dashed line}) and after the EBL absorption ({\itshape black solid line}), respectively (see text for details). }
\label{fig:J1117}
\end{figure}

For the simulation, we extrapolate the observed {\itshape Fermi}-LAT spectrum at higher energies using a power law with an exponential cut-off:
\begin{equation}
\Phi_\gamma(E_\gamma)=k_0\left(\frac{E_\gamma}{E_0}\right)^\Gamma e^{-E_\gamma/E_{\rm cut}}
\end{equation}
where $k_0$ is the normalization, $\Gamma$ is the power-law index, $E_0$ the pivot energy and $E_{\rm cut}$ is the energy of the exponential cut-off. We set the cut-off at 3 TeV on the basis of the statistical relations described in \citet[][see their Eqs. 2 and 3]{bal20}, taking into account the correction for the effect of EBL absorption (see Sect. \ref{sec:intro}). The resulting spectrum given as input for our simulation is reported in Fig. \ref{fig:J1117}. Using this model, we simulate the observation of the source with the ASTRI Mini-Array considering 50 h and 200 h of exposure time producing a single event list for photon energies $E_\gamma \gtrsim 0.8$ TeV. Then, we estimate the detection significance of the source through the TS value: we obtain a significance $\gtrsim$12$\sigma$ already for a 50-h total exposure, confirming that RGB J1117$+$202 can be clearly detected with such an amount of observing time. The simulated spectrum for a 200-h exposure is presented in Fig. \ref{fig:J1117}, showing how it allows a good characterization of the observed $\gamma$-ray emission.

We note that the H.E.S.S. observation of RGB J1117$+$\\202 provided an UL on the integrated flux above 0.61~TeV of $1.44 \times 10^{-12}$ cm$^{-2}$ s$^{-1}$ \citep{aha05b}. This value corresponds to a flux density at 0.61~TeV which is $\sim$40\% lower than our model. Interestingly, also the GeV flux density reported in the 10-yr version of the {\itshape Fermi}-LAT catalogue \citep[4FGL-DR2;][]{abd20} is lower by $\sim$30\% than the 3FGL flux density that we used to constrain the model. This difference, considering the errors on the fluxes reported by these catalogues (4\%-7\%), seems to be significant, suggesting that the source has varied. This is further confirmed by the high value of the {\itshape variability index} given in the 4FGL (138.7) which is well above the threshold (18.48) used to claim that a source is variable at 99\% confidence level \citep[see][for details]{aha05b}. However, if we renormalize the model by this 30\%-40\%, we are still able to predict a detection at $\gtrsim$8$\sigma$ significance in a 50-h observation with the ASTRI Mini-Array. This confirms that the HSPs listed in Tab.~\ref{tab:hsp-ma} with low values of $F$ (between 0.1 and $\sim$0.2) are difficult targets, but whose detection is nevertheless feasible (especially if the observation is carried out during a high state, see Sect. \ref{sec:bzcat}). As a conclusion, our simulations demonstrate that the detection with the ASTRI Mini-Array of sources belonging to the HSP class revealed by {\itshape Fermi}-LAT, but not yet at higher energies by current generation of Cherenkov telescopes, is possible even with relatively short observations (50 h), at least up to redshifts of $\sim$0.15.

\subsection{NGC 1068}\label{sec:ngc1068}
The prototypical Seyfert 2 galaxy NGC 1068 is a nearby galaxy ($d_\odot = 14.4$ Mpc, corresponding to $z = 0.0034$) hosting a luminous AGN \citep[$L_{\rm bol} \simeq 10^{45}$ erg s$^{-1}$;][]{rig09}. Its hard X-ray-to-[O {\scriptsize IV}] luminosity ratio is about 500 times lower than the value expected for unobscured AGN of this luminosity, implying that an extremely high column density $N_{\rm H} \gtrsim 10^{25}$ cm$^2$ \citep[the so-called Compton-thick regime; e.g.,][]{com04} is blocking the line of sight to the nucleus. This source exhibits also starburst activity in its central region. Interferometric observations in the millimetric band identified a $\sim$2 kpc-wide starburst ring that surrounds a circumnuclear disk (CND) of $\sim$100 pc diameter. A sizeable fraction the molecular gas in the CND is observed to be involved in a massive AGN-driven wind \citep{kri11,Garcia14} which causes shocks at the interface with the quiescent gas. In the radio band, structures similar to collimated outflows but weaker and slower than the jets observed in blazars, have been detected \citep{gal96}.

In the $\gamma$-ray domain, NGC 1068 was observed in the HE  band by {\itshape Fermi}-LAT, and in the VHE band by the MAGIC telescopes and the HAWC $\gamma$-ray Observatory. NGC 1068 is the brightest of the Seyfert/starburst galaxies detected by {\itshape Fermi}-LAT. The  spectral analysis based on 10 yr of {\itshape Fermi}-LAT data  yields a power-law index of $-2.3$ and a energy  flux integrated between 100 MeV and 100 GeV of $6.5\times 10^{-12}$ erg cm$^{-2}$ s$^{-1}$ \citep{FermiColl19,Ballet20}. The MAGIC telescopes observed  NGC 1068 for 125 hours. No significant $\gamma$-ray emission was detected, and an UL at the 95\% confidence level to the $\gamma$-ray flux above 200 GeV of $<$5.1$\times 10^{-13}$ cm$^{-2}$ s$^{-1}$ was derived \citep{MAGIC19}.

\begin{figure}
\centering
\includegraphics[scale=0.55]{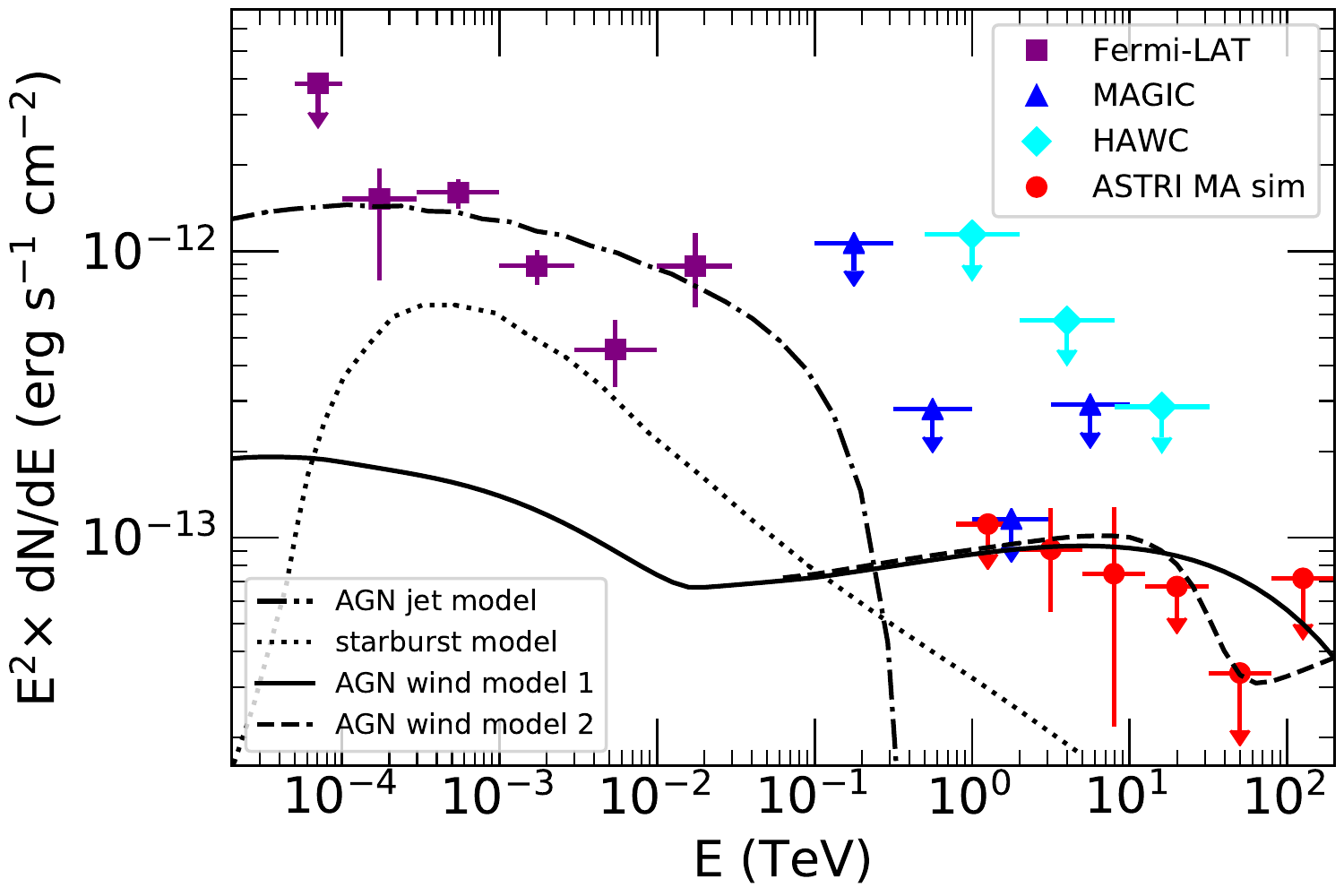}
\caption{$\gamma$-ray spectrum of NGC 1068 in the HE and VHE band. The {\it Fermi}-LAT data ({\itshape purple squares}) are from \citet{Ballet20}. The upper limits are from MAGIC  \citep[\textbf{125 hours}, {\itshape blue triangles};][]{MAGIC19} and  HAWC \citep[\textbf{1523 days}, {\itshape cyan diamonds};][]{2021ApJ...907...67A}. Red points represent the signal that can be observed by ASTRI Mini-Array after 200 hours of observation, and assuming the spectrum predicted by the revised AGN wind model \citep[{\itshape dashed line};][see text for details]{lam16,lam19}. The spectrum predicted by the AGN jet model \citep[{\itshape dot-dashed line};][]{len10}, and the starburst model \citep[{\itshape dotted line};][]{eic16}  are also shown.}
\label{fig:ngc1068}
\end{figure}

The origin of the $\gamma$-ray emission in NGC 1068 is still undetermined, owing to the simultaneous presence of different particle acceleration sites (starburst ring, CND, jets). Figure \ref{fig:ngc1068} shows the  $\gamma$-ray spectrum of NGC 1068 in the HE and VHE band, as well as the spectra predicted by the starburst \citep{eic16}, AGN jet \citep{len10}, and AGN wind \citep{lam16,lam19} models that have been proposed in the literature to explain the $\gamma$-ray emission. The predictions of the theoretical models  differ significantly in the VHE band. The leptonic AGN jet model is characterized by a sharp cutoff at energies $\sim$100 GeV, while the hadronic starburst and AGN wind models extend to the VHE band, but with different spectral slopes.

MAGIC observations of NGC 1068 put stringent const\-raints on the AGN wind model parameters such as the proton spectral index $p$, cut-off energy $E_{\rm cut}$, calorimetric fraction $F_{\rm cal}$, and proton acceleration efficiency $\eta$. In this paper we simulated the VHE spectrum predicted by the AGN wind model that takes into account the MAGIC constraints (see Figure \ref{fig:ngc1068}). Following  \cite{MAGIC19} we adopted $p=2$, $E_{\rm cut}=2\times 10^6$ GeV, $\eta=0.1$, and $F_{\rm cal}=0.5$. This model predicts a $\gamma$-ray flux lower by about an order of magnitude than that measured by {\it Fermi}-LAT, requiring another mechanism(s) to explain $\gamma$-rays in the HE band as AGN jet and star formation activity, but provides a quite flat spectrum that extends in the energy band covered by the ASTRI Mini-Array. 

As discussed in \cite{lam19}, although NGC 1068 is a relatively nearby source, above few tens of TeV the effects of the interaction of $\gamma$-ray photons with the EBL start to be important and determine the absorption of a substantial fraction of the flux.  The electron-positron pairs produced in this way, however, scatter off the photons of the CMB, triggering an electromagnetic cascade that reprocesses the absorbed flux. These pairs can be deflected by intergalactic magnetic fields (IGMFs), which are very uncertain~\citep[see][for a review]{alvesbatista2021a}. Convers\-ely, it is possible to use $\gamma$-ray observations to constrain IGMF intensities \citep{neronov2010a,tavecchio2010a,hess2014a,veritas2017a,fermi2018a, alvesbatista2020a,cta2021b}, as discussed in {\bfseries Paper II}. It has been suggested that electromagnetic cascades could also be qu\-enched by plasma instabilities \citep{broderick2012a,schlickeiser2012a}, which would ultimately cool down the pairs, hardening the spectrum at lower energies. However, the importance of this effect is disputed \citep{miniati2013a,alvesbatista2019g,alawashra2022a}. Here, given the uncertainties associated to both IGMFs and plasma instabilities, we neglect these effects and consider only EBL absorption. The $\gamma$-ray spectrum obtained in this way, which is shown with the dashed line in Figure \ref{fig:ngc1068}, is used to carry out the simulation following the procedure already used for starburst galaxies and other scientific cases presented in {\bfseries Paper II}. We simulated  NGC 1068 as a point-like source, located at the known coordinates, and without considering the energy dispersion. 

\begin{table*}[width=17cm,align=\centering]
\centering
\caption{Relevant quantities (see text) for the three optimal dSphs selected in the Northern hemisphere: Ursa Minor (UMi), Coma Berenices (CBe), and Ursa Major II (UMa~II). The astrophysical factors for DM annihilation $J$ and decay $D$ are reported for both integration angles of 0.1 deg and optimal angles $\alpha_J$ and $\alpha_D$ as defined in \citet{bon15c}.}
\label{tab:dsph}
\resizebox{\textwidth}{!}{
\begin{tabular}{lccccccc}
\hline
\hline
\multicolumn{8}{l}{ }\\
Name & Type & $r_h$ & $\sigma_v$ & $M_V$ & $M/L$ & $\epsilon$ & Ref.\\
 & & (pc) & (km s$^{-1}$) & (mag) & (M$_\odot$/L$_\odot$) & & \\
\multicolumn{8}{l}{ }\\
\hline
\multicolumn{8}{l}{ }\\
UMi & cls & $181 \pm 27$ & $9.5 \pm 1.2$ & $-8.8 \pm 0.5$ & 34 & $0.56 \pm 0.05$ & 1\\
CBe & uft & $77 \pm 10$ & $4.6 \pm 0.8$ & $-4.1 \pm 0.5$ & 252 & $0.38 \pm 0.14$ & 1\\
UMa II & uft & $149 \pm 21$ & $6.7 \pm 1.4$ & $-4.2 \pm 0.6$ & 953 & $0.63 \pm 0.05$ & 1\\
\multicolumn{8}{l}{ }\\
\hline
\hline
\multicolumn{8}{l}{ }\\
Name & $\alpha_J$ & $\log{J(0^\circ .1)}$ & $\log{J(\alpha_J)}$ & $\alpha_D$ & $\log{D(0^\circ .1)}$ & $\log{D(\alpha_D)}$ & Ref.\\
 & (deg) & (GeV$^2$ cm$^{-5}$) & (GeV$^2$ cm$^{-5}$) & (deg) & (GeV cm$^{-2}$) & (GeV cm$^{-2}$) & \\
\multicolumn{8}{l}{ }\\
\hline
\multicolumn{8}{l}{ }\\
UMi & $0.49$ & $18.6^{+0.3}_{-0.2}$ & $19.1^{+0.1}_{-0.1}$ & $0.25$ & $17.4^{+0.1}_{-0.1}$ & $18.1^{+0.1}_{-0.1}$ & 2\\
CBe & $0.20$ & $18.7^{+0.5}_{-0.4}$ & $19.2^{+0.6}_{-0.5}$ & $0.10$ & $17.7^{+0.5}_{-0.4}$ & $17.7^{+0.5}_{-0.4}$ & 2\\
UMa II & $0.53$ & $18.9^{+0.5}_{-0.4}$ & $20.1^{+0.7}_{-0.6}$ & $0.27$ & $17.8^{+0.5}_{-0.3}$ & $18.7^{+0.5}_{-0.4}$ & 2\\
\multicolumn{8}{l}{ }\\
\hline
\multicolumn{8}{l}{$^1$\citet{mcc12}.}\\
\multicolumn{8}{l}{$^2$\citet{bon15c}.}\\
\end{tabular}
}
\end{table*}

To reduce the impact of variations between individual simulations, we performed sets of $N = 100$ statistically independent realisations (see section 3.2.1 of {\bfseries Paper II} for a detailed description of the simulation setup). Briefly, the spectrum is calculated in 6 energy bins logarithmically spaced between 0.8 and 200 TeV. In each bin, we first create event lists based on our input model, and then fit a power-law model by using an unbinned maximum-likelihood approach. For each realisation, the power-law best-fit spectral parameters are used to calculate 100 values of flux and TS in the given energy bin. 
When the mean TS value in a given bin is greater than 9, we calculate the flux value and associated uncertainty, respectively, as the mean $\overline{F_{\rm sim}}$ and the standard deviation $\sigma_{\rm sim}$ obtained from the distribution of the 100 simulated fluxes. When the mean TS value is below 9,  an UL at 95\% confidence level on flux is calculated as \citep{Bevington}:
\begin{equation}\label{eqn:bevington}
    F_{\rm UL}=\overline{F_{sim}}+1.96 \times \frac{\sigma_{\rm sim}}{\sqrt{N}}
\end{equation}
Such a procedure is always applied to the entire sample of simulated flux values in each energy bin, regardless of the statistical significance of the single realizations.

It should be noted that, for very weak sources such as NGC 1068, perhaps more than 100 realisations are needed to obtain a reliable average, as indicated by the fact that the simulated points are slightly below the input model. We find in this way that, with an exposure time of 200 h, the ASTRI Mini-Array is able to measure the source spectrum in the energy bins $\sim$2--5 TeV and $\sim$5--13 TeV, though with a low detection significance of ${\rm TS} = 11$ and 12, respectively (corresponding to a detection significance of $\sim$4.8$\sigma$). We therefore conclude that, with $\sim$10\% more exposure time (i.e. a total of $\sim$220 h), we are able to detect the source with a detection significance of 5$\sigma$.

Confirmed observation of VHE emission from NGC 1068 would represent an observational evidence of particle acceleration, and interaction of accelerated protons with the interstellar matter, in an AGN-driven outflow, that requires a dedicated long-term observing campaign. This  observational effort can contribute significantly to improve  our understanding of AGN feedback mechanisms and the extragalactic $\gamma$-ray and neutrino backgrounds. The  $\gamma$-ray emission may represent the on-set of the interaction between AGN winds and gas in their host galaxy, which is often identified as the main mechanism responsible for suppressing the star formation in AGN host. The resulting hadronuclear $\gamma$-ray and neutrino emission is expected to contribute to the corresponding diffuse fluxes \citep{tamborra14,wang16,wang_gamma,lam17,liu18}.

\section{Dark matter in dwarf spheroidal galaxies}\label{sec:dmctools}
In order to assess the capabilities of the ASTRI Mini-Array to search for DM in dSphs, in the context of the widely studied WIMP scenario (see Sect. \ref{indirectDM}), we consider three optimal targets observable from the Northern hemisphere: Ursa Minor (UMi), Coma Berenices (CBe), and Ursa Major II (UMa II). The targets were selected among the dSphs with the highest values of astrophysical factor, as reported in \citet{bon15b}. The relevant kinematic (velocity dispersion $\sigma_v$, mass-to-light ratio $M/L$ and ellipticity $\epsilon$) and brightness properties (half-light radius $r_h$ and $V$-band absolute magnitude $M_V$) of the three selected targets are reported in Tab. \ref{tab:dsph}.

In our analysis, we take advantage of the full-likelihood method presented in \citet{ale12} and implemented into the {\ttfamily ctools} analysis chain. This method is derived from the likelihood maximization procedure commonly adopted in the analysis of $\gamma$-ray emission from astrophysical sources, and relies on the evaluation of a model-dependent Poisson likelihood function:
\begin{equation}\label{eqn:fulllike}
    \mathcal{L}\left[
    N_{\rm e}, M(\mathbf{\theta}) | N_{\rm o}, E_{1 \rightarrow N_{\rm o}}
    \right] = \frac{N_{\rm e}^{N_{\rm o}}}{N_{\rm o}!} e^{-N_{\rm e}} \prod_{i=1}^{N_{\rm o}} \mathcal{P}_i 
\end{equation}
Here, $N_{\rm e}$ and $N_{\rm o}$ are the total number of estimated and observed events in the regions of interest (source and background) respectively, and $\mathcal{P}_i = \mathcal{P}[E_i, M(\mathbf{\theta})]$ is the value of the probability density function (PDF) associated to the $i$-th event with measured energy $E_i$ according to the DM emission model $M(\mathbf{\theta})$. Such a method is particularly well suited for DM studies, since it fully profits from the potential presence of DM spectral features in VHE data, since it is able to quantitatively compare expected and measured energy distributions in place of number of events.
 
\begin{figure}
\centering
\includegraphics[scale=0.4]{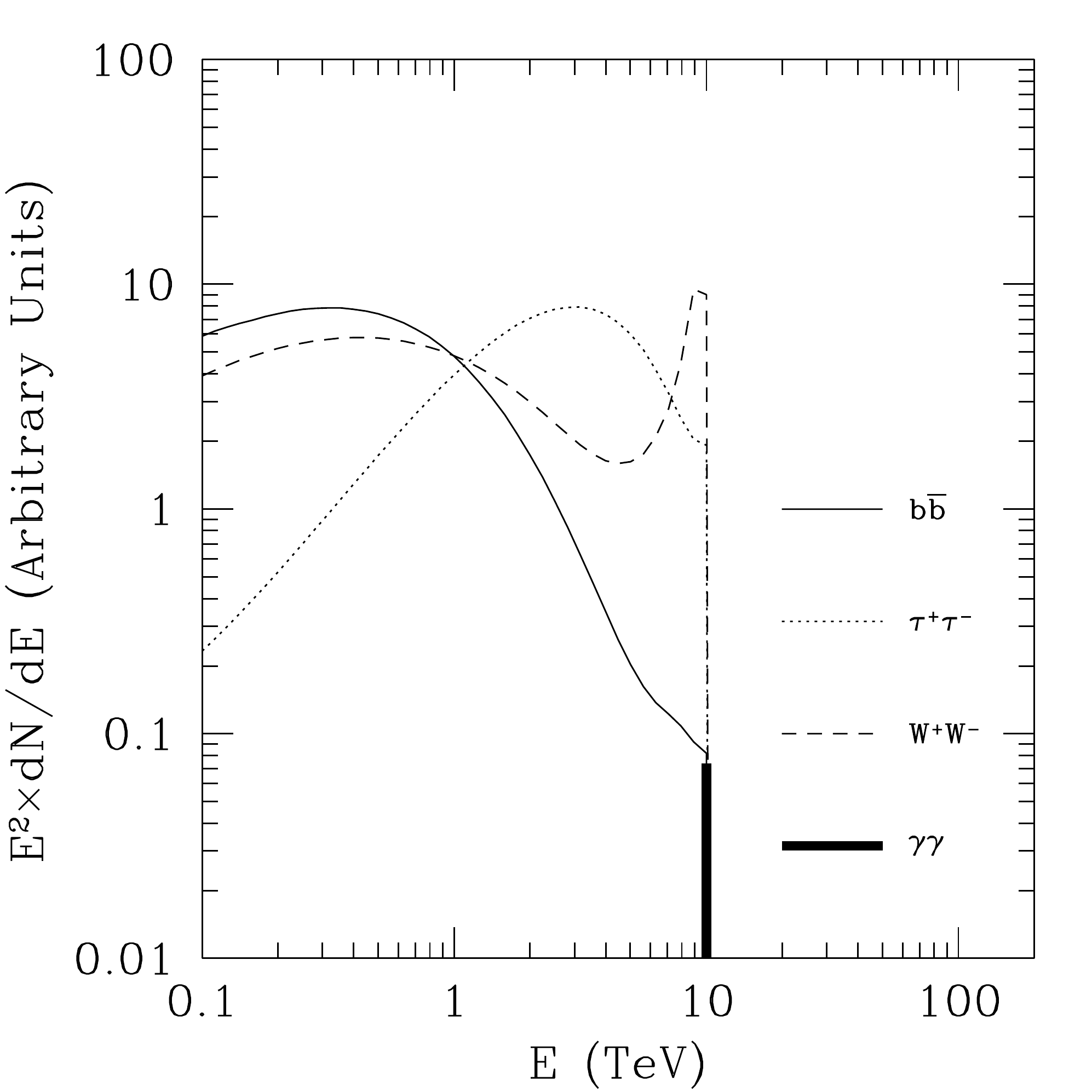}
\caption{Example spectra produced by DM annihilation for $m_\chi = 10$ TeV for the four SM channels considered: $b\bar{b}$ {\itshape thin solid line}), $\tau^+\tau^-$ ({\itshape dotted line}), $W^+W^-$ ({\itshape dashed line}) and $\gamma\gamma$ ({\itshape thick solid line}).}
\label{fig:dmspec}
\end{figure}

We include the DM spectral models for the case of annihilation of particles with masses in the range $0.55 - 100$ TeV. Each model is numerically computed for different DM interaction channels \citep{cem11,cir11,cir12}, and is characterized by the particle mass $m_\chi$ and the velocity-averaged cross section $\langle \sigma_{\rm ann} v \rangle$. We take the single-interaction photon counts for the $b\bar{b}$, $\tau^+\tau^-$, $W^+W^-$ and $\gamma\gamma$ interaction channels from \citet[see Fig. \ref{fig:dmspec}]{cia11}\footnote{Available at {\ttfamily http://www.marcocirelli.net/PPPC4DMID.html}.}, and convert them to fluxes thro\-ugh Eqs. \ref{eqn:dmfluxann} and \ref{eqn:dmfluxdec} taking into account the mass-dependent ``thermal relic'' $\langle \sigma_{\rm ann} v\rangle$ by \citet[$\langle \sigma_{\rm ann} v\rangle \simeq 2.2 \times 10^{-26}$ cm$^3$ s$^{-1}$ for $m_\chi \gtrsim 0.1$ TeV]{ste12}. We then simulate event lists for 100 h observations of UMi and CBe, and for both 100 h and 300 h in the case of UMa II; for this task, we assume the dSph halos to be point-like with respect to the ASTRI Mini-Array PSF of $\sim$0.1 deg. Although this assumption is not strictly valid for the dSphs in object, given their typical projected angular halo extension $\alpha_{\rm opt} \gtrsim 0.1$ deg (see Tab. \ref{tab:dsph}), it provides consistent results as long as the majority of the expected DM signal is enclosed in a region of integration $0.1 \times 0.1$ deg$^2$ wide. We demand the analysis of dSphs as extended targets to future publications.

The simulated observations of the considered dSphs provide no evidence of detection for signals from DM annihilation or decay coming from the regions of interest; therefore, we use the signal ULs to derive constraints on the interaction parameters of the DM particles -- cross section $\langle \sigma_{\rm ann} v \rangle$ and particle lifetime $\tau_{\rm dec}$ -- as a function of the DM particle mass $m_\chi$. We thus produce the expected ASTRI Mini-Array sensitivities to the DM parameters for each interaction channel, using the maximum-likelihood evaluation model to solve the equation:
\begin{equation}\label{eqn:loglikeulim}
-2\ln{\mathcal{L}} = 2.71    
\end{equation}
looking for the largest solution at each DM particle mass. This procedure yields flux ULs to the DM signal integrated over the ASTRI Mini-Array energy range (0.65 TeV $\div$ 200 TeV) that are then converted to a minimal cross section or maximal lifetime of the DM particle at a given mass:
\begin{equation}\label{eqn:xsec}
    \langle\sigma_{\rm ann} v\rangle_{\rm lim} = \langle\sigma_{\rm ann} v\rangle_{\rm thr} \cdot \frac{{\rm UL}\left(
    m_\chi
    \right)}{\int_{E_{\rm min}}^{E_{\rm max}} \frac{d\Phi_{\rm ann}\left(
    m_\chi
    \right)}{dE_\gamma} dE_\gamma}
\end{equation}
\begin{equation}\label{eqn:tdec}
    \tau_{\rm lim} = \frac{D\left(
    \Delta\Omega
    \right)}{4 \pi m_\chi {\rm UL}\left(
    m_\chi
    \right)} \cdot \int_{E_{\rm min}}^{E_{\rm max}} \frac{dN_\gamma\left(
    m_\chi
    \right)}{dE_\gamma} dE_\gamma
\end{equation}
with $\langle \sigma_{\rm ann} v \rangle_{\rm thr} \simeq 2.2 \times 10^{-26}$ cm$^3$ s$^{-1}$ for continuous spectra \citep{ste12} and $\sim$1.2$\times 10^{-30}$ cm$^3$ s$^{-1}$ for monochromatic emission lines \citep[see e.g. section 4.1.5 by][]{cta19}.

\begin{figure*}[width=17cm,align=\centering]
    \begin{minipage}{0.49\textwidth}
        \centering
        \includegraphics[width=\textwidth]{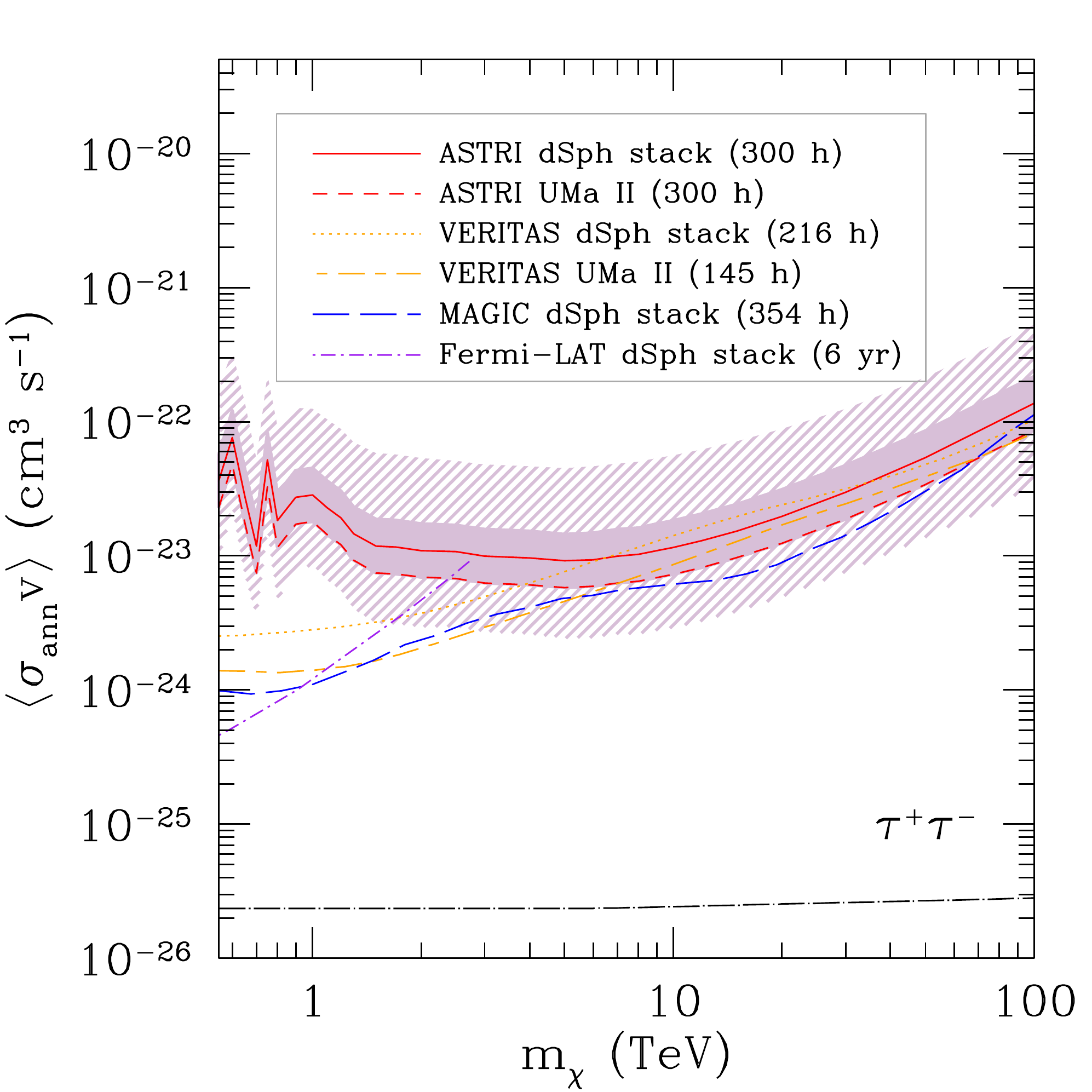}
    \end{minipage}
    \begin{minipage}{0.49\textwidth}
        \centering
        \includegraphics[width=\textwidth]{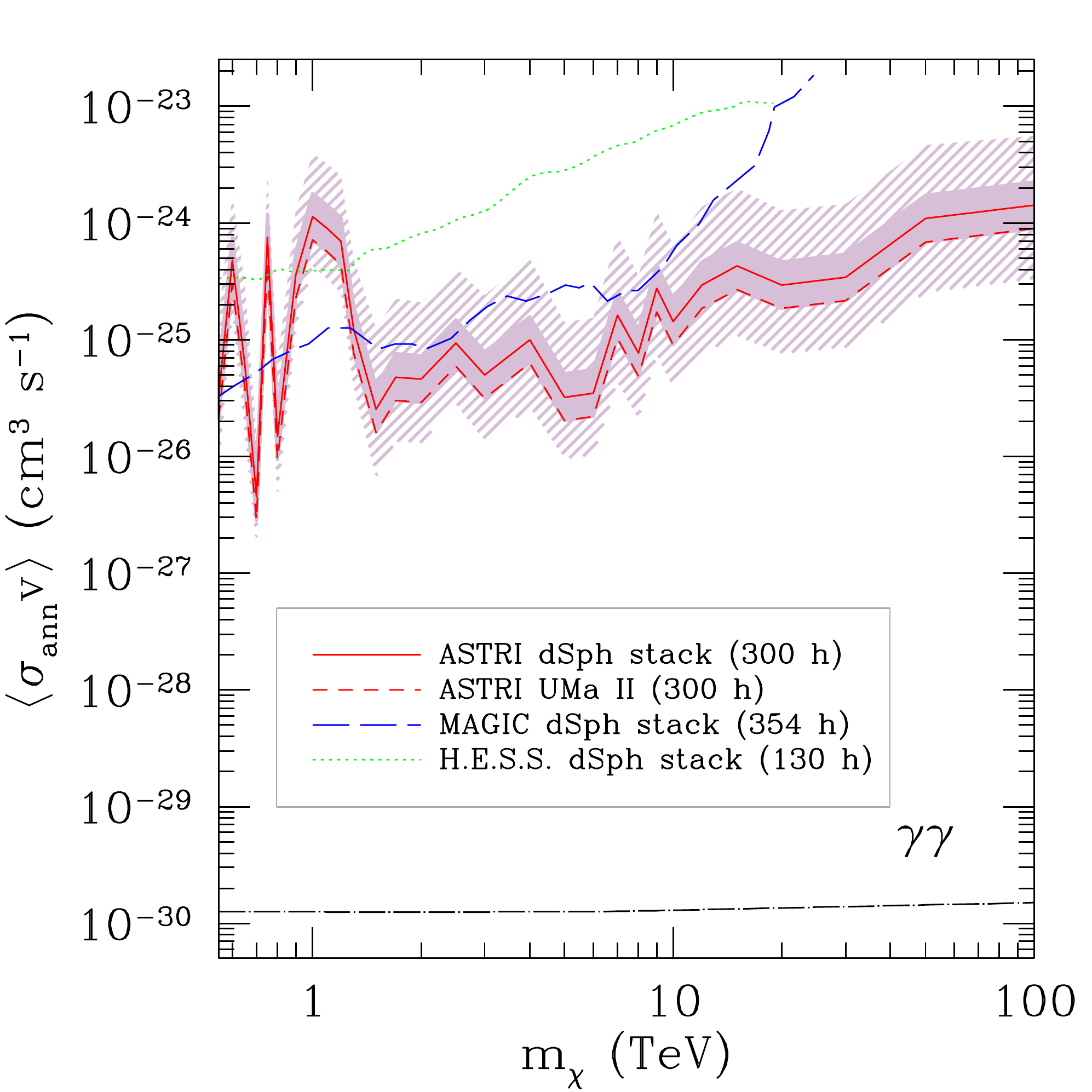}
    \end{minipage}
    \begin{minipage}{0.49\textwidth}
        \centering
        \includegraphics[width=\textwidth]{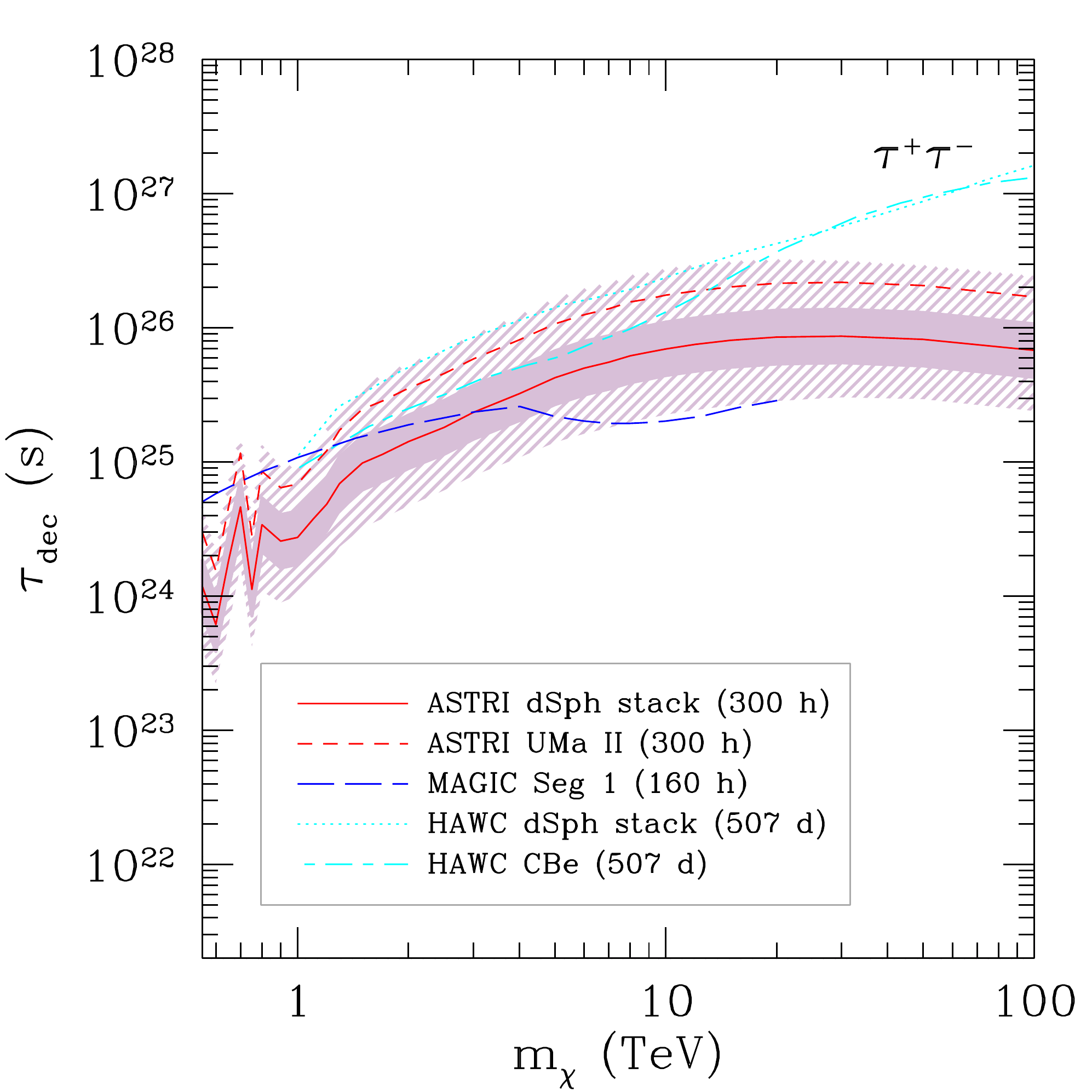}
    \end{minipage}
    \begin{minipage}{0.49\textwidth}
        \centering
        \includegraphics[width=\textwidth]{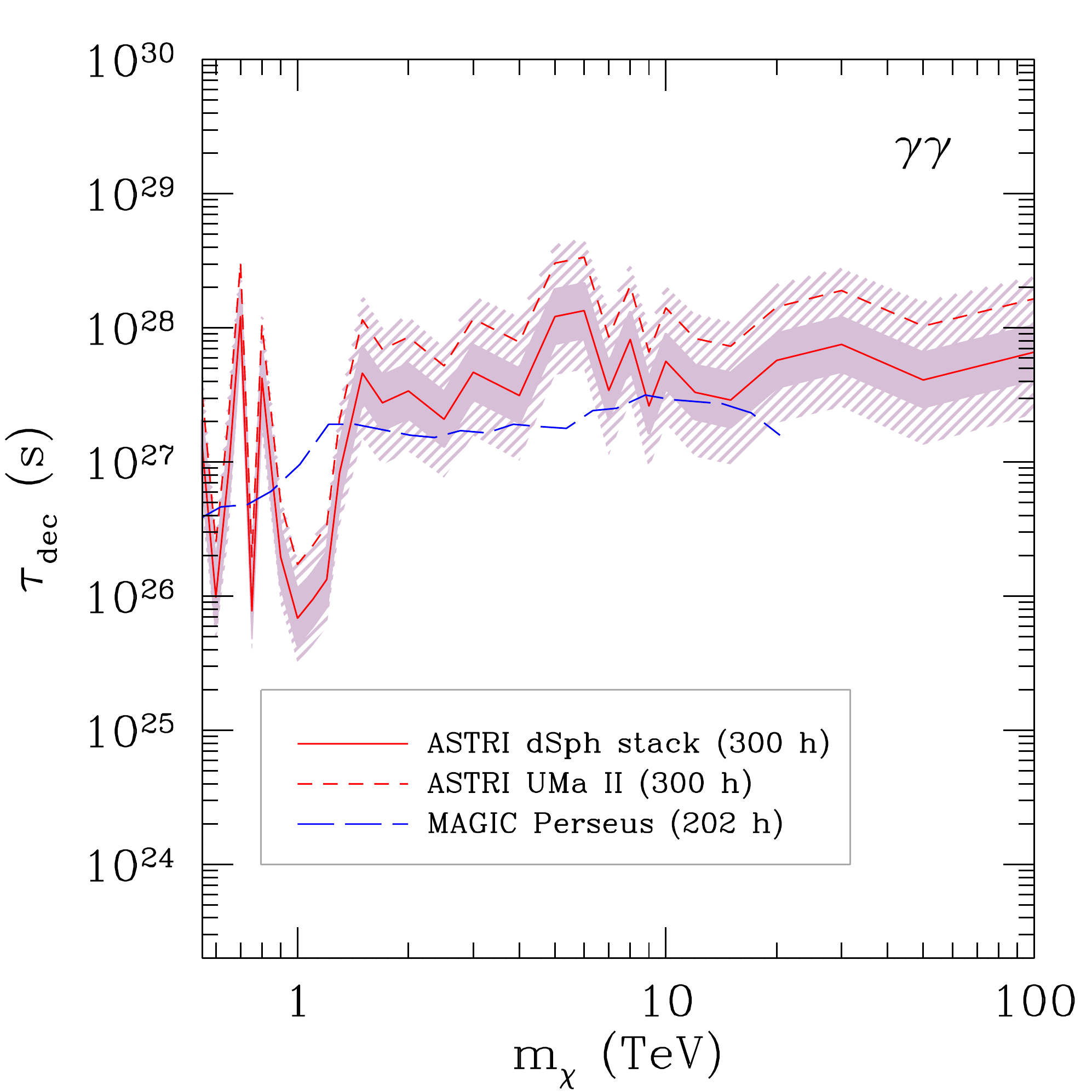}
    \end{minipage}
    \caption{Comparison between the 300-h ASTRI Mini-Array sensitivity limits to DM particle cross section and lifetime for interactions going 100\% into a single SM channel and the corresponding limits obtained by current space- and ground-based facilities for $\gamma$-ray observations \citep{ale14,fer17b,zit17,abd18,alb18,acc18,mag22}. {\itshape Upper panels:} sensitivity plots for annihilating DM searches in the representative SM channels $\tau^+\tau^-$ and $\gamma\gamma$, in which the thermal-relic cross section ({\itshape black dot-dashed line}) is indicated. {\itshape Lower panels:} the same for DM decay. In all panels, the limits for the stacked exposures ({\itshape red solid lines}) are reported, with their uncertainties at 68\% CL from IRF photon statistics ({\itshape pink solid-shaded area}) and astrophysical factor calculations ({\itshape pink dash-shaded area}), along with the limits from deep observations of UMa II ({\itshape red dashed lines}).}
    \label{fig:dmcomp}
\end{figure*}

For the dSph stacking analysis, we adopt as average astrophysical factors an arithmetic mean of the logarithmic values of each target measured at $0^\circ.1$, and weighted by their average logarithmic errors. The associated uncertainties are then derived as direct sum of the logarithmic intrinsic dispersion of such factors with the average logarithmic errors, yielding $\langle\log{J(0^\circ.1)}\rangle = 18.7^{+0.5}_{-0.4}$ and $\langle\log{D(0^\circ.1)}\rangle = 17.4^{+0.4}_{-0.3}$ respectively. This allows us to estimate uncertainties at 68\% confidence level (CL) on the DM parameters taking into account both the IRF photon statistics ($\sim$0.2 dex uncertainty) and the error on the modeling of the DM distribution. We present in Fig. \ref{fig:dmcomp} the final ASTRI Mini-Array sensitivity curves at 300 h to DM annihilation cross sections and decay lifetimes for the $\tau^+\tau^-$ and $\gamma\gamma$ channels, both in the case of single-target (UMa~II) and stacked observations of 3 dSph halos (UMi, CBe and UMa~II).

We then outline how the foreseen prospects on DM se\-arches with the ASTRI Mini-Array compare to current results. As a visual guidance, in Fig. \ref{fig:dmcomp} we show such a comparison in a graphical way. Concerning the scenario of DM particles annihilating into SM pairs, the most stringent limits on the cross section at TeV energies have been currently established by 145-h VERITAS observations of the dSph UMa II \citep[$\langle \sigma v \rangle \lesssim 5 \times 10^{-24}$ cm$^3$ s$^{-1}$ in the $\tau^+\tau^-$ channel for $m_\chi \lesssim 10$ TeV;][]{zit17} for single-target observations, and by 354-h MAGIC combined exposures \citep[$\langle \sigma v \rangle \lesssim 2 \times10^{-25}$ cm$^3$ s$^{-1}$ in the $\gamma\gamma$ channel for $m_\chi \lesssim 10$ TeV;][]{mag22} for the case of stacking analysis. This latter result in particular may be significantly improved by the prospects of 300-h ASTRI Mini-Array observat\-ions presented here ($\langle \sigma v \rangle \lesssim 5 \times 10^{-25}$ cm$^3$ s$^{-1}$), especially at the highest masses ($\langle \sigma v \rangle \lesssim 10^{-24}$ cm$^3$ s$^{-1}$ for $m_\chi \lesssim 100$ TeV).

Regarding the scenario of DM particle decay, for the case of continuous spectra the highest limits on the DM particle lifetime set with both single-target observations of the CBe dSph and dSph stacking analysis come from a 507-day integration with the HAWC $\gamma$-Ray Observatory \citep[$\tau \gtrsim 10^{27}$ s for $m_\chi \sim 100$ TeV in the $\tau^+\tau^-$ channel;][]{alb18}; such a value is $\gtrsim$10 times larger than the corresponding limit derived from ASTRI Mini-Array observations, albeit obtained in a much larg\-er time span of $\gtrsim$12,000 h. No comparable lifetime limits on DM decay into monochromatic lines have been obtained through single-targ\-et observations or stacking analyses of dSph halos, only existing in the literature for extended and cont\-aminated sources like the Perseus galaxy cluster \citep{acc18}.

Coupled with observations of the Galactic center and halo (see {\bfseries Paper III} for a discussion), the search of $\gamma$-ray signals at multi-TeV energies from DM-dominated extragalactic sourc\-es is a science topic in which the ASTRI Mini-Array can give interesting contributions before the CTA era, provided that long-term duration exposures are carried out. Such searches will also take advantage from the large ASTRI Mini-Array FoV of $\sim$10$^\circ$, which will allow to simultaneously observe multiple targets falling in the same sky region (see Sect. \ref{sec:simobs}). In addition, the ASTRI Mini-Array angular resolution of $\lesssim$0.1 deg above 1 TeV and energy resolution of $\sim$10\% are particularly suited for an efficient search of monochromatic $\gamma$-ray emission lines, whose expected fluxes at $m_\chi \gtrsim 10$ TeV can be enhanced by fundament\-al-physics mechanisms \citep[see e.g. section 4.1.5 of][and refs. therein]{cta19}.

\section{Summary and conclusions}\label{sec:conc}
In this paper, we have explored in detail the scientific prospects of extragalactic astrophysics at multi-TeV energies that are within the reach of the next-generation IACT ASTRI Mini-Array, to be deployed at the {\itshape Observatorio del Teide}. In particular, we have focused on the observing feasibility of scientifically interesting targets that will be taken into account in the observatory phase of the array, subsequent to the experiment phase in which the observation of core-science targets (see {\bfseries Paper II}) will be prioritized. The $\gamma$-ray emission properties of such observatory targets can be characterized with exposure times from $\sim$1 h to $\sim$200 h, covering a variety of sky objects (blazars, Seyfert galaxies, DM-dominated halos) and science topics (from the study of the VHE spectral emission of AGN to the indirect searches of self-interacting DM).

Albeit some of the sources potentially detectable with the ASTRI Mini-Array are already being observed since several years with previous and existing Cherenkov facilities (e.g., HEGRA, MAGIC, H.E.S.S., VERITAS and HAWC), the enhanced capabilities of this new instrument will be exploitable during the observatory phase for:
\begin{itemize}
    \item better characterizing the spectral shape and features (such as bumps) of the multi-TeV $\gamma$-ray emission from extreme blazars (HSPs and EHSPs) with respect to current measurements, and extending the search for VHE signals up to $\sim$10$\div$20 TeV for the closest targets;
    \item studying the $\gamma$-ray emission from AGN- or starburst-pow\-ered outflows in Seyfert galaxies;
    \item obtaining new independent observations aimed at improving the constraints on the parameters of particle DM annihilating or decaying into SM products, especially in the case of two VHE photons (monochromatic emission lines), thro\-ugh long-term observation of dwarf sph\-eroidal galaxies.
\end{itemize}

Such observations will greatly benefit of the large ASTRI FoV ($\sim$10$^\circ$ diameter) and almost uniform instrument response up to $\sim$5$^\circ$ off-axis (see {\bfseries Paper II}): in fact, these characteristics will allow us to obtain a relevant fraction of ``free'' observing time for those extragalactic sources located close to core-science targets already during the ASTRI Mini-Array experiment phase. In addition, the same exposures may provide useful insight on the properties of weaker ancillary sour\-ces falling in the same FoV for which interesting flux ULs can be derived. Finally, an observing strategy optimized to take full advantage of the ASTRI Mini-Array capabilities may be foreseen to point at several extragalactic targets at once in the observatory phase, in order to increase the number of observed sources without requesting large amounts of dedicated exposure time.

\section*{Acknowledgements}
\noindent
{\small The ASTRI project is becoming a reality thanks to Giovanni ``Nanni'' Bignami and Nicol{\`o} ``Nichi'' D'Amico, two outstanding scientists who, in their capability of INAF Presidents, provided continuous support and invaluable guidance. While Nan\-ni was instrumental to start the ASTRI telescope, Nichi transformed it into the Mini-Array in Tenerife. Now the project is being built owing to the unfaltering support of Marco Tavani, the current INAF President. Paolo Vettolani and Filippo Zerbi, the past and current INAF Science Directors, as well as Massimo Cappi, the Coordinator of the High Energy branch of INAF, have been also very supportive to our work. We are very grateful to all of them. Nanni and Nichi, unfortunately, passed away but their vision is still guiding us. This work was conducted in the context of the ASTRI Project, and is supported by the Italian Ministry of Education, University, and Research (MIUR) with funds specifically assigned to the Italian National Institute of Astrophysics (INAF). We acknowledge support from the Brazilian Funding Agency FAPESP (Grant 2013/10559-5) and from the South African Department of Science and Technology through Funding Ag\-reement 0227/2014 for the South African Gamma-Ray Astronomy Programme. This work has been supported by H2020-ASTER\-ICS, a project funded by the European Commission Framework Programme Horizon 2020 Research and Innovation action under grant agreement no. 653477. IAC is supported by the Spanish Ministry of Science and Innovation (MCIU). JBG acknowledges the support of the ``Viera y Clavijo'' program, funded by ACIISI and ULL. RAB acknowledges funding by the ``la Caixa'' Foundation (ID 100010434) and the European Union's Horizon~2020 Research and Innovation Programme under the Marie Sk{\l}odowska-Curie grant agreement No. 847648, fellowship code LCF/BQ/PI21/\\11830030. This research made use of the {\ttfamily ctools} \citep{kno16} and {\ttfamily GammaPy} software \citep{gammapy:2017,gammapy:2019}, community-develop\-ed analysis packages for IACT data. The {\ttfamily ctools} software is based on {\ttfamily Gamma\-Lib}, a community-develop\-ed toolbox for the scientific analysis of astronomical $\gamma$-ray data \citep{kno11,kno16b}. \textcopyright\ 2022. This manuscript version is made available under the CC-BY-NC-ND 4.0 license ({\ttfamily https://cre\-ativecommons.org/licenses/by-nc-nd/4.0/}).}

\bibliographystyle{cas-model2-names}
\bibliography{ASTRI_extragal_biblio}

\begin{thebibliography}{182}
\expandafter\ifx\csname natexlab\endcsname\relax\def\natexlab#1{#1}\fi
\providecommand{\url}[1]{\texttt{#1}}
\providecommand{\href}[2]{#2}
\providecommand{\path}[1]{#1}
\providecommand{\DOIprefix}{doi:}
\providecommand{\ArXivprefix}{arXiv:}
\providecommand{\URLprefix}{URL: }
\providecommand{\Pubmedprefix}{pmid:}
\providecommand{\doi}[1]{\href{http://dx.doi.org/#1}{\path{#1}}}
\providecommand{\Pubmed}[1]{\href{pmid:#1}{\path{#1}}}
\providecommand{\bibinfo}[2]{#2}
\ifx\xfnm\relax \def\xfnm[#1]{\unskip,\space#1}\fi
\bibitem[{{Aartsen} et~al.(2020){Aartsen}, {Ackermann}, {Adams} and {et
  al.}}]{ice10}
\bibinfo{author}{{Aartsen}, M.G.}, \bibinfo{author}{{Ackermann}, M.},
  \bibinfo{author}{{Adams}, J.}, \bibinfo{author}{{et al.}},
  \bibinfo{year}{2020}.
\newblock \bibinfo{journal}{Phys. Rev. Lett.} \bibinfo{volume}{124},
  \bibinfo{pages}{051103}.
\bibitem[{{Abdalla} et~al.(2018){Abdalla}, {Aharonian}, {Ait Benkhali} and {et
  al.}}]{abd18}
\bibinfo{author}{{Abdalla}, H.}, \bibinfo{author}{{Aharonian}, F.},
  \bibinfo{author}{{Ait Benkhali}, F.}, \bibinfo{author}{{et al.}},
  \bibinfo{year}{2018}.
\newblock \bibinfo{journal}{JCAP} \bibinfo{volume}{2018}, \bibinfo{pages}{037}.
\bibitem[{{Abdollahi} et~al.(2020a){Abdollahi}, {Acero}, {Ackermann} and {et
  al.}}]{FermiColl19}
\bibinfo{author}{{Abdollahi}, S.}, \bibinfo{author}{{Acero}, F.},
  \bibinfo{author}{{Ackermann}, M.}, \bibinfo{author}{{et al.}},
  \bibinfo{year}{2020}a.
\newblock \bibinfo{journal}{ApJS} \bibinfo{volume}{247}, \bibinfo{pages}{33}.
\bibitem[{{Abdollahi} et~al.(2020b){Abdollahi}, {Acero}, {Ackermann} and {et
  al.}}]{abd20}
\bibinfo{author}{{Abdollahi}, S.}, \bibinfo{author}{{Acero}, F.},
  \bibinfo{author}{{Ackermann}, M.}, \bibinfo{author}{{et al.}},
  \bibinfo{year}{2020}b.
\newblock \bibinfo{journal}{ApJS} \bibinfo{volume}{247}, \bibinfo{pages}{33}.
\bibitem[{{Acero} et~al.(2009){Acero}, {Aharonian}, {Akhperjanian} and {et
  al.}}]{2009Sci...326.1080A}
\bibinfo{author}{{Acero}, F.}, \bibinfo{author}{{Aharonian}, F.},
  \bibinfo{author}{{Akhperjanian}, A.G.}, \bibinfo{author}{{et al.}},
  \bibinfo{year}{2009}.
\newblock \bibinfo{journal}{Science} \bibinfo{volume}{326},
  \bibinfo{pages}{1080}.
\bibitem[{{Ackermann} et~al.(2012){Ackermann}, {Ajello}, {Allafort} and {et
  al.}}]{ack12}
\bibinfo{author}{{Ackermann}, M.}, \bibinfo{author}{{Ajello}, M.},
  \bibinfo{author}{{Allafort}, A.}, \bibinfo{author}{{et al.}},
  \bibinfo{year}{2012}.
\newblock \bibinfo{journal}{ApJ} \bibinfo{volume}{755}, \bibinfo{pages}{164}.
\bibitem[{{Ackermann} et~al.(2016){Ackermann}, {Ajello}, {Atwood} and {et
  al.}}]{Fermi_2FHL}
\bibinfo{author}{{Ackermann}, M.}, \bibinfo{author}{{Ajello}, M.},
  \bibinfo{author}{{Atwood}, W.B.}, \bibinfo{author}{{et al.}},
  \bibinfo{year}{2016}.
\newblock \bibinfo{journal}{ApJS} \bibinfo{volume}{222}, \bibinfo{pages}{5}.
\bibitem[{{Adam} et~al.(2021){Adam}, {Goksu}, {Brown} and {et al.}}]{ada21}
\bibinfo{author}{{Adam}, R.}, \bibinfo{author}{{Goksu}, H.},
  \bibinfo{author}{{Brown}, S.}, \bibinfo{author}{{et al.}},
  \bibinfo{year}{2021}.
\newblock \bibinfo{journal}{A\&A} \bibinfo{volume}{648}, \bibinfo{pages}{A60}.
\bibitem[{{Aharonian} et~al.(2007a){Aharonian}, {Akhperjanian}, {Barres de
  Almeida} and {et al.}}]{2007A&A...475L...9A}
\bibinfo{author}{{Aharonian}, F.}, \bibinfo{author}{{Akhperjanian}, A.G.},
  \bibinfo{author}{{Barres de Almeida}, U.}, \bibinfo{author}{{et al.}},
  \bibinfo{year}{2007}a.
\newblock \bibinfo{journal}{A\&A} \bibinfo{volume}{475},
  \bibinfo{pages}{L9--L13}.
\bibitem[{{Aharonian} et~al.(2005){Aharonian}, {Akhperjanian}, {Bazer-Bachi}
  and {et al.}}]{aha05b}
\bibinfo{author}{{Aharonian}, F.}, \bibinfo{author}{{Akhperjanian}, A.G.},
  \bibinfo{author}{{Bazer-Bachi}, A.R.}, \bibinfo{author}{{et al.}},
  \bibinfo{year}{2005}.
\newblock \bibinfo{journal}{A\&A} \bibinfo{volume}{441},
  \bibinfo{pages}{465--472}.
\bibitem[{{Aharonian} et~al.(2007b){Aharonian}, {Akhperjanian}, {Bazer-Bachi}
  and {et al.}}]{aha07}
\bibinfo{author}{{Aharonian}, F.}, \bibinfo{author}{{Akhperjanian}, A.G.},
  \bibinfo{author}{{Bazer-Bachi}, A.R.}, \bibinfo{author}{{et al.}},
  \bibinfo{year}{2007}b.
\newblock \bibinfo{journal}{ApJL} \bibinfo{volume}{664}, \bibinfo{pages}{L71}.
\bibitem[{{Aharonian} et~al.(2007c){Aharonian}, {Akhperjanian}, {Bazer-Bachi}
  and {et al.}}]{aharonian_etal_07}
\bibinfo{author}{{Aharonian}, F.}, \bibinfo{author}{{Akhperjanian}, A.G.},
  \bibinfo{author}{{Bazer-Bachi}, A.R.}, \bibinfo{author}{{et al.}},
  \bibinfo{year}{2007}c.
\newblock \bibinfo{journal}{ApJL} \bibinfo{volume}{664},
  \bibinfo{pages}{L71--L74}.
\bibitem[{{Aharonian}(2000)}]{aha00}
\bibinfo{author}{{Aharonian}, F.A.}, \bibinfo{year}{2000}.
\newblock \bibinfo{journal}{New Astronomy} \bibinfo{volume}{5},
  \bibinfo{pages}{377--395}.
\bibitem[{{Aharonian} et~al.(1999){Aharonian}, {Akhperjanian}, {Barrio} and {et
  al.}}]{aha99}
\bibinfo{author}{{Aharonian}, F.A.}, \bibinfo{author}{{Akhperjanian}, A.G.},
  \bibinfo{author}{{Barrio}, J.A.}, \bibinfo{author}{{et al.}},
  \bibinfo{year}{1999}.
\newblock \bibinfo{journal}{A\&A} \bibinfo{volume}{349},
  \bibinfo{pages}{11--28}.
\bibitem[{{Aharonian} et~al.(1992){Aharonian}, {Chilingarian}, {Mirzoyan} and
  {et al.}}]{aha92}
\bibinfo{author}{{Aharonian}, F.A.}, \bibinfo{author}{{Chilingarian}, A.A.},
  \bibinfo{author}{{Mirzoyan}, R.G.}, \bibinfo{author}{{et al.}},
  \bibinfo{year}{1992}.
\newblock \bibinfo{journal}{Experimental Astronomy} \bibinfo{volume}{2},
  \bibinfo{pages}{331--344}.
\bibitem[{{Ajello} et~al.(2020a){Ajello}, {Angioni}, {Axelsson} and {et
  al.}}]{fer20}
\bibinfo{author}{{Ajello}, M.}, \bibinfo{author}{{Angioni}, R.},
  \bibinfo{author}{{Axelsson}, M.}, \bibinfo{author}{{et al.}},
  \bibinfo{year}{2020}a.
\newblock \bibinfo{journal}{ApJ} \bibinfo{volume}{892}, \bibinfo{pages}{105}.
\bibitem[{{Ajello} et~al.(2017){Ajello}, {Atwood}, {Baldini} and {et
  al.}}]{Fermi_3FHL}
\bibinfo{author}{{Ajello}, M.}, \bibinfo{author}{{Atwood}, W.B.},
  \bibinfo{author}{{Baldini}, L.}, \bibinfo{author}{{et al.}},
  \bibinfo{year}{2017}.
\newblock \bibinfo{journal}{ApJS} \bibinfo{volume}{232}, \bibinfo{pages}{18}.
\bibitem[{{Ajello} et~al.(2020b){Ajello}, {Di Mauro}, {Paliya} and {et
  al.}}]{aje20}
\bibinfo{author}{{Ajello}, M.}, \bibinfo{author}{{Di Mauro}, M.},
  \bibinfo{author}{{Paliya}, V.S.}, \bibinfo{author}{{et al.}},
  \bibinfo{year}{2020}b.
\newblock \bibinfo{journal}{ApJ} \bibinfo{volume}{894}, \bibinfo{pages}{88}.
\bibitem[{{Alawashra} and {Pohl}(2022)}]{alawashra2022a}
\bibinfo{author}{{Alawashra}, M.}, \bibinfo{author}{{Pohl}, M.},
  \bibinfo{year}{2022}.
\newblock \bibinfo{journal}{arXiv e-prints} ,
  \bibinfo{pages}{arXiv:2203.01022}.
\bibitem[{{Albert} et~al.(2018){Albert}, {Alfaro}, {Alvarez} and {et
  al.}}]{alb18}
\bibinfo{author}{{Albert}, A.}, \bibinfo{author}{{Alfaro}, R.},
  \bibinfo{author}{{Alvarez}, C.}, \bibinfo{author}{{et al.}},
  \bibinfo{year}{2018}.
\newblock \bibinfo{journal}{ApJ} \bibinfo{volume}{853}, \bibinfo{pages}{154}.
\bibitem[{{Albert} et~al.(2007){Albert}, {Aliu}, {Anderhub} and {et
  al.}}]{Albert2007}
\bibinfo{author}{{Albert}, J.}, \bibinfo{author}{{Aliu}, E.},
  \bibinfo{author}{{Anderhub}, H.}, \bibinfo{author}{{et al.}},
  \bibinfo{year}{2007}.
\newblock \bibinfo{journal}{ApJ} \bibinfo{volume}{669},
  \bibinfo{pages}{862--883}.
\bibitem[{{Aleksi{\'c}} et~al.(2012a){Aleksi{\'c}}, {Alvarez}, {Antonelli} and
  {et al.}}]{Aleksic2012}
\bibinfo{author}{{Aleksi{\'c}}, J.}, \bibinfo{author}{{Alvarez}, E.A.},
  \bibinfo{author}{{Antonelli}, L.A.}, \bibinfo{author}{{et al.}},
  \bibinfo{year}{2012}a.
\newblock \bibinfo{journal}{A\&A} \bibinfo{volume}{542}, \bibinfo{pages}{A100}.
\bibitem[{{Aleksi{\'c}} et~al.(2010){Aleksi{\'c}}, {Anderhub}, {Antonelli} and
  {et al.}}]{Aleksic2010}
\bibinfo{author}{{Aleksi{\'c}}, J.}, \bibinfo{author}{{Anderhub}, H.},
  \bibinfo{author}{{Antonelli}, L.A.}, \bibinfo{author}{{et al.}},
  \bibinfo{year}{2010}.
\newblock \bibinfo{journal}{A\&A} \bibinfo{volume}{519}, \bibinfo{pages}{A32}.
\bibitem[{{Aleksi{\'c}} et~al.(2014a){Aleksi{\'c}}, {Ansoldi}, {Antonelli} and
  {et al.}}]{2014A&A...567A.135A}
\bibinfo{author}{{Aleksi{\'c}}, J.}, \bibinfo{author}{{Ansoldi}, S.},
  \bibinfo{author}{{Antonelli}, L.A.}, \bibinfo{author}{{et al.}},
  \bibinfo{year}{2014}a.
\newblock \bibinfo{journal}{A\&A} \bibinfo{volume}{567}, \bibinfo{pages}{A135}.
\bibitem[{{Aleksi{\'c}} et~al.(2014b){Aleksi{\'c}}, {Ansoldi}, {Antonelli} and
  {et al.}}]{ale14}
\bibinfo{author}{{Aleksi{\'c}}, J.}, \bibinfo{author}{{Ansoldi}, S.},
  \bibinfo{author}{{Antonelli}, L.A.}, \bibinfo{author}{{et al.}},
  \bibinfo{year}{2014}b.
\newblock \bibinfo{journal}{JCAP} \bibinfo{volume}{2014}, \bibinfo{pages}{008}.
\bibitem[{{Aleksi{\'c}} et~al.(2011){Aleksi{\'c}}, {Antonelli}, {Antoranz} and
  {et al.}}]{2011ApJ...730L...8A}
\bibinfo{author}{{Aleksi{\'c}}, J.}, \bibinfo{author}{{Antonelli}, L.A.},
  \bibinfo{author}{{Antoranz}, P.}, \bibinfo{author}{{et al.}},
  \bibinfo{year}{2011}.
\newblock \bibinfo{journal}{ApJL} \bibinfo{volume}{730}, \bibinfo{pages}{L8}.
\bibitem[{{Aleksi{\'c}} et~al.(2012b){Aleksi{\'c}}, {Rico} and
  {Martinez}}]{ale12}
\bibinfo{author}{{Aleksi{\'c}}, J.}, \bibinfo{author}{{Rico}, J.},
  \bibinfo{author}{{Martinez}, M.}, \bibinfo{year}{2012}b.
\newblock \bibinfo{journal}{JCAP} \bibinfo{volume}{2012}, \bibinfo{pages}{032}.
\bibitem[{{Alexander} and {Hickox}(2012)}]{Alexander12}
\bibinfo{author}{{Alexander}, D.M.}, \bibinfo{author}{{Hickox}, R.C.},
  \bibinfo{year}{2012}.
\newblock \bibinfo{journal}{New Astron. Rev.} \bibinfo{volume}{56},
  \bibinfo{pages}{93--121}.
\bibitem[{{Aliu} et~al.(2014){Aliu}, {Archambault}, {Arlen} and {et
  al.}}]{aliu14}
\bibinfo{author}{{Aliu}, E.}, \bibinfo{author}{{Archambault}, S.},
  \bibinfo{author}{{Arlen}, T.}, \bibinfo{author}{{et al.}},
  \bibinfo{year}{2014}.
\newblock \bibinfo{journal}{ApJ} \bibinfo{volume}{782}, \bibinfo{pages}{13}.
\bibitem[{{Alves Batista} and {Saveliev}(2020)}]{alvesbatista2020a}
\bibinfo{author}{{Alves Batista}, R.}, \bibinfo{author}{{Saveliev}, A.},
  \bibinfo{year}{2020}.
\newblock \bibinfo{journal}{ApJL} \bibinfo{volume}{902}, \bibinfo{pages}{L11}.
\bibitem[{{Alves Batista} and {Saveliev}(2021)}]{alvesbatista2021a}
\bibinfo{author}{{Alves Batista}, R.}, \bibinfo{author}{{Saveliev}, A.},
  \bibinfo{year}{2021}.
\newblock \bibinfo{journal}{Universe} \bibinfo{volume}{7},
  \bibinfo{pages}{223}.
\bibitem[{{Alves Batista} et~al.(2019){Alves Batista}, {Saveliev} and {de
  Gouveia Dal Pino}}]{alvesbatista2019g}
\bibinfo{author}{{Alves Batista}, R.}, \bibinfo{author}{{Saveliev}, A.},
  \bibinfo{author}{{de Gouveia Dal Pino}, E.M.}, \bibinfo{year}{2019}.
\newblock \bibinfo{journal}{MNRAS} \bibinfo{volume}{489},
  \bibinfo{pages}{3836}.
\bibitem[{{Arsioli, B.} and {Chang, Y.-L.}(2017)}]{1BIGB_CAT}
\bibinfo{author}{{Arsioli, B.}}, \bibinfo{author}{{Chang, Y.-L.}},
  \bibinfo{year}{2017}.
\newblock \bibinfo{title}{Searching for signature in whsp blazars - fermi-lat
  detection of 150 excess signal in the 0.3-500 gev band}.
\newblock \bibinfo{journal}{A\&A} \bibinfo{volume}{598}, \bibinfo{pages}{A134}.
\newblock \URLprefix \url{https://doi.org/10.1051/0004-6361/201628691},
  \DOIprefix\doi{10.1051/0004-6361/201628691}.
\bibitem[{{Baghmanyan} et~al.(2021){Baghmanyan}, {Zargaryan}, {Aharonian} and
  {et al.}}]{bag21}
\bibinfo{author}{{Baghmanyan}, V.}, \bibinfo{author}{{Zargaryan}, D.},
  \bibinfo{author}{{Aharonian}, F.}, \bibinfo{author}{{et al.}},
  \bibinfo{year}{2021}.
\newblock \bibinfo{journal}{arXiv e-prints} ,
  \bibinfo{pages}{arXiv:2110.00309}.
\bibitem[{{Balmaverde} et~al.(2020){Balmaverde}, {Caccianiga}, {Della Ceca} and
  {et al.}}]{bal20}
\bibinfo{author}{{Balmaverde}, B.}, \bibinfo{author}{{Caccianiga}, A.},
  \bibinfo{author}{{Della Ceca}, R.}, \bibinfo{author}{{et al.}},
  \bibinfo{year}{2020}.
\newblock \bibinfo{journal}{MNRAS} \bibinfo{volume}{492},
  \bibinfo{pages}{3728}.
\bibitem[{{Balmaverde} et~al.(2019){Balmaverde}, {Capetti}, {Marconi} and {et
  al.}}]{bal19}
\bibinfo{author}{{Balmaverde}, B.}, \bibinfo{author}{{Capetti}, A.},
  \bibinfo{author}{{Marconi}, A.}, \bibinfo{author}{{et al.}},
  \bibinfo{year}{2019}.
\newblock \bibinfo{journal}{A\&A} \bibinfo{volume}{632}, \bibinfo{pages}{A124}.
\bibitem[{{Benbow}(2019)}]{ben19}
\bibinfo{author}{{Benbow}, W.}, \bibinfo{year}{2019}.
\newblock in: \bibinfo{booktitle}{36th International Cosmic Ray Conference
  (ICRC2019)}, p. \bibinfo{pages}{632}.
\bibitem[{{Bergstr{\"o}m} et~al.(1998){Bergstr{\"o}m}, {Ullio} and
  {Buckley}}]{ber97}
\bibinfo{author}{{Bergstr{\"o}m}, L.}, \bibinfo{author}{{Ullio}, P.},
  \bibinfo{author}{{Buckley}, J.H.}, \bibinfo{year}{1998}.
\newblock \bibinfo{journal}{Astroparticle Physics} \bibinfo{volume}{9},
  \bibinfo{pages}{137}.
\bibitem[{{Bertone} et~al.(2005){Bertone}, {Hooper} and {Silk}}]{ber05}
\bibinfo{author}{{Bertone}, G.}, \bibinfo{author}{{Hooper}, D.},
  \bibinfo{author}{{Silk}, J.}, \bibinfo{year}{2005}.
\newblock \bibinfo{journal}{Phys. Rep.} \bibinfo{volume}{405},
  \bibinfo{pages}{279--390}.
\bibitem[{{Bevington}(1969)}]{Bevington}
\bibinfo{author}{{Bevington}, P.R.}, \bibinfo{year}{1969}.
\newblock \bibinfo{title}{{Data reduction and error analysis for the physical
  sciences}}.
\bibitem[{{Biteau} et~al.(2020){Biteau}, {Prandini}, {Costamante} and {et
  al.}}]{Biteau2020}
\bibinfo{author}{{Biteau}, J.}, \bibinfo{author}{{Prandini}, E.},
  \bibinfo{author}{{Costamante}, L.}, \bibinfo{author}{{et al.}},
  \bibinfo{year}{2020}.
\newblock \bibinfo{journal}{Nature Astronomy} \bibinfo{volume}{4},
  \bibinfo{pages}{124--131}.
\bibitem[{{Bonnivard} et~al.(2015a){Bonnivard}, {Combet}, {Daniel} and {et
  al.}}]{bon15a}
\bibinfo{author}{{Bonnivard}, V.}, \bibinfo{author}{{Combet}, C.},
  \bibinfo{author}{{Daniel}, M.}, \bibinfo{author}{{et al.}},
  \bibinfo{year}{2015}a.
\newblock \bibinfo{journal}{MNRAS} \bibinfo{volume}{453}, \bibinfo{pages}{849}.
\bibitem[{{Bonnivard} et~al.(2015b){Bonnivard}, {Combet}, {Maurin} and {et
  al.}}]{bon15c}
\bibinfo{author}{{Bonnivard}, V.}, \bibinfo{author}{{Combet}, C.},
  \bibinfo{author}{{Maurin}, D.}, \bibinfo{author}{{et al.}},
  \bibinfo{year}{2015}b.
\newblock \bibinfo{journal}{MNRAS} \bibinfo{volume}{446},
  \bibinfo{pages}{3002}.
\bibitem[{{Bonnivard} et~al.(2015c){Bonnivard}, {Combet}, {Maurin} and {et
  al.}}]{bon15b}
\bibinfo{author}{{Bonnivard}, V.}, \bibinfo{author}{{Combet}, C.},
  \bibinfo{author}{{Maurin}, D.}, \bibinfo{author}{{et al.}},
  \bibinfo{year}{2015}c.
\newblock \bibinfo{journal}{ApJL} \bibinfo{volume}{808}, \bibinfo{pages}{L36}.
\bibitem[{{B{\"o}ttcher} et~al.(2013){B{\"o}ttcher}, {Reimer}, {Sweeney} and
  {et al.}}]{bot13}
\bibinfo{author}{{B{\"o}ttcher}, M.}, \bibinfo{author}{{Reimer}, A.},
  \bibinfo{author}{{Sweeney}, K.}, \bibinfo{author}{{et al.}},
  \bibinfo{year}{2013}.
\newblock \bibinfo{journal}{ApJ} \bibinfo{volume}{768}, \bibinfo{pages}{54}.
\bibitem[{{Broderick} et~al.(2012){Broderick}, {Chang} and
  {Pfrommer}}]{broderick2012a}
\bibinfo{author}{{Broderick}, A.E.}, \bibinfo{author}{{Chang}, P.},
  \bibinfo{author}{{Pfrommer}, C.}, \bibinfo{year}{2012}.
\newblock \bibinfo{journal}{ApJ} \bibinfo{volume}{752}, \bibinfo{pages}{22}.
\bibitem[{{Brunetti} and {Blasi}(2005)}]{bru05}
\bibinfo{author}{{Brunetti}, G.}, \bibinfo{author}{{Blasi}, P.},
  \bibinfo{year}{2005}.
\newblock \bibinfo{journal}{MNRAS} \bibinfo{volume}{363},
  \bibinfo{pages}{1173--1187}.
\bibitem[{{Brunetti} et~al.(2012){Brunetti}, {Blasi}, {Reimer} and {et
  al.}}]{bru12}
\bibinfo{author}{{Brunetti}, G.}, \bibinfo{author}{{Blasi}, P.},
  \bibinfo{author}{{Reimer}, O.}, \bibinfo{author}{{et al.}},
  \bibinfo{year}{2012}.
\newblock \bibinfo{journal}{MNRAS} \bibinfo{volume}{426}, \bibinfo{pages}{956}.
\bibitem[{{Brunetti} and {Jones}(2014)}]{bru14}
\bibinfo{author}{{Brunetti}, G.}, \bibinfo{author}{{Jones}, T.W.},
  \bibinfo{year}{2014}.
\newblock \bibinfo{journal}{International Journal of Modern Physics D}
  \bibinfo{volume}{23}, \bibinfo{pages}{1430007--98}.
\bibitem[{{Brunetti} and {Lazarian}(2011)}]{bru11}
\bibinfo{author}{{Brunetti}, G.}, \bibinfo{author}{{Lazarian}, A.},
  \bibinfo{year}{2011}.
\newblock \bibinfo{journal}{MNRAS} \bibinfo{volume}{412}, \bibinfo{pages}{817}.
\bibitem[{{Brunetti} et~al.(2017){Brunetti}, {Zimmer} and {Zandanel}}]{bru17}
\bibinfo{author}{{Brunetti}, G.}, \bibinfo{author}{{Zimmer}, S.},
  \bibinfo{author}{{Zandanel}, F.}, \bibinfo{year}{2017}.
\newblock \bibinfo{journal}{MNRAS} \bibinfo{volume}{472},
  \bibinfo{pages}{1506}.
\bibitem[{{Celotti} and {Ghisellini}(2008)}]{cel08}
\bibinfo{author}{{Celotti}, A.}, \bibinfo{author}{{Ghisellini}, G.},
  \bibinfo{year}{2008}.
\newblock \bibinfo{journal}{MNRAS} \bibinfo{volume}{385}, \bibinfo{pages}{283}.
\bibitem[{{Cembranos} et~al.(2011){Cembranos}, {de La Cruz-Dombriz}, {Dobado}
  and {et al.}}]{cem11}
\bibinfo{author}{{Cembranos}, J.A.R.}, \bibinfo{author}{{de La Cruz-Dombriz},
  A.}, \bibinfo{author}{{Dobado}, A.}, \bibinfo{author}{{et al.}},
  \bibinfo{year}{2011}.
\newblock \bibinfo{journal}{Phys. Rev. D} \bibinfo{volume}{83},
  \bibinfo{pages}{083507}.
\bibitem[{{Chang} et~al.(2019){Chang}, {Arsioli}, {Giommi} and {et
  al.}}]{cha19}
\bibinfo{author}{{Chang}, Y.L.}, \bibinfo{author}{{Arsioli}, B.},
  \bibinfo{author}{{Giommi}, P.}, \bibinfo{author}{{et al.}},
  \bibinfo{year}{2019}.
\newblock \bibinfo{journal}{A\&A} \bibinfo{volume}{632}, \bibinfo{pages}{A77}.
\bibitem[{Chang et~al.(2020)Chang, Brandt and Giommi}]{Chang_2020}
\bibinfo{author}{Chang, Y.L.}, \bibinfo{author}{Brandt, C.},
  \bibinfo{author}{Giommi, P.}, \bibinfo{year}{2020}.
\newblock \bibinfo{journal}{Astronomy and Computing} \bibinfo{volume}{30},
  \bibinfo{pages}{100350}.
\bibitem[{{Ciafaloni} et~al.(2011){Ciafaloni}, {Comelli}, {Riotto} and {et
  al.}}]{cia11}
\bibinfo{author}{{Ciafaloni}, P.}, \bibinfo{author}{{Comelli}, D.},
  \bibinfo{author}{{Riotto}, A.}, \bibinfo{author}{{et al.}},
  \bibinfo{year}{2011}.
\newblock \bibinfo{journal}{JCAP} \bibinfo{volume}{2011}, \bibinfo{pages}{019}.
\bibitem[{{Cirelli} et~al.(2011){Cirelli}, {Corcella}, {Hektor} and {et
  al.}}]{cir11}
\bibinfo{author}{{Cirelli}, M.}, \bibinfo{author}{{Corcella}, G.},
  \bibinfo{author}{{Hektor}, A.}, \bibinfo{author}{{et al.}},
  \bibinfo{year}{2011}.
\newblock \bibinfo{journal}{JCAP} \bibinfo{volume}{2011}, \bibinfo{pages}{051}.
\bibitem[{{Cirelli} et~al.(2012){Cirelli}, {Corcella}, {Hektor} and {et
  al.}}]{cir12}
\bibinfo{author}{{Cirelli}, M.}, \bibinfo{author}{{Corcella}, G.},
  \bibinfo{author}{{Hektor}, A.}, \bibinfo{author}{{et al.}},
  \bibinfo{year}{2012}.
\newblock \bibinfo{journal}{JCAP} \bibinfo{volume}{2012}, \bibinfo{pages}{E01}.
\bibitem[{{Clowe} et~al.(2004){Clowe}, {Gonzalez} and {Markevitch}}]{clo04}
\bibinfo{author}{{Clowe}, D.}, \bibinfo{author}{{Gonzalez}, A.},
  \bibinfo{author}{{Markevitch}, M.}, \bibinfo{year}{2004}.
\newblock \bibinfo{journal}{ApJ} \bibinfo{volume}{604},
  \bibinfo{pages}{596--603}.
\bibitem[{{Comastri}(2004)}]{com04}
\bibinfo{author}{{Comastri}, A.}, \bibinfo{year}{2004}.
\newblock in: \bibinfo{editor}{{Barger}, A.J.} (Ed.),
  \bibinfo{booktitle}{Supermassive Black Holes in the Distant Universe}, p.
  \bibinfo{pages}{245}.
\bibitem[{{Costamante} et~al.(2018){Costamante}, {Bonnoli}, {Tavecchio} and {et
  al.}}]{cos18}
\bibinfo{author}{{Costamante}, L.}, \bibinfo{author}{{Bonnoli}, G.},
  \bibinfo{author}{{Tavecchio}, F.}, \bibinfo{author}{{et al.}},
  \bibinfo{year}{2018}.
\newblock \bibinfo{journal}{MNRAS} \bibinfo{volume}{477},
  \bibinfo{pages}{4257}.
\bibitem[{{Costamante} et~al.(2001){Costamante}, {Ghisellini}, {Giommi} and {et
  al.}}]{Costamante2001}
\bibinfo{author}{{Costamante}, L.}, \bibinfo{author}{{Ghisellini}, G.},
  \bibinfo{author}{{Giommi}, P.}, \bibinfo{author}{{et al.}},
  \bibinfo{year}{2001}.
\newblock \bibinfo{journal}{A\&A} \bibinfo{volume}{371},
  \bibinfo{pages}{512--526}.
\bibitem[{{Crocker} et~al.(2022){Crocker}, {Macias}, {Mackey} and {et
  al.}}]{cro22}
\bibinfo{author}{{Crocker}, R.M.}, \bibinfo{author}{{Macias}, O.},
  \bibinfo{author}{{Mackey}, D.}, \bibinfo{author}{{et al.}},
  \bibinfo{year}{2022}.
\newblock \bibinfo{journal}{arXiv e-prints} ,
  \bibinfo{pages}{arXiv:2204.12054}.
\bibitem[{{CTA Consortium}(2019)}]{cta19}
\bibinfo{author}{{CTA Consortium}}, \bibinfo{year}{2019}.
\newblock \bibinfo{title}{{Science with the Cherenkov Telescope Array}}.
\bibitem[{{CTA Consortium}(2021)}]{cta2021b}
\bibinfo{author}{{CTA Consortium}}, \bibinfo{year}{2021}.
\newblock \bibinfo{journal}{JCAP} \bibinfo{volume}{02}, \bibinfo{pages}{048}.
\bibitem[{{de Cea del Pozo} et~al.(2009){de Cea del Pozo}, {Torres} and
  {Rodriguez Marrero}}]{cea09}
\bibinfo{author}{{de Cea del Pozo}, E.}, \bibinfo{author}{{Torres}, D.F.},
  \bibinfo{author}{{Rodriguez Marrero}, A.Y.}, \bibinfo{year}{2009}.
\newblock \bibinfo{journal}{ApJ} \bibinfo{volume}{698},
  \bibinfo{pages}{1054--1060}.
\bibitem[{{de Gouveia Dal Pino} et~al.(2019){de Gouveia Dal Pino}, {Alves
  Batista}, {Kowal} and {et al.}}]{2019arXiv190308982D}
\bibinfo{author}{{de Gouveia Dal Pino}, E.M.}, \bibinfo{author}{{Alves
  Batista}, R.}, \bibinfo{author}{{Kowal}, G.}, \bibinfo{author}{{et al.}},
  \bibinfo{year}{2019}.
\newblock \bibinfo{journal}{arXiv e-prints} ,
  \bibinfo{pages}{arXiv:1903.08982}.
\bibitem[{{de Gouveia Dal Pino} and {Lazarian}(2005)}]{dgdp_lazarian_05}
\bibinfo{author}{{de Gouveia Dal Pino}, E.M.}, \bibinfo{author}{{Lazarian},
  A.}, \bibinfo{year}{2005}.
\newblock \bibinfo{journal}{A\&A} \bibinfo{volume}{441},
  \bibinfo{pages}{845--853}.
\bibitem[{{Deil} et~al.(2017){Deil}, {Zanin}, {Lefaucheur} and {et
  al.}}]{gammapy:2017}
\bibinfo{author}{{Deil}, C.}, \bibinfo{author}{{Zanin}, R.},
  \bibinfo{author}{{Lefaucheur}, J.}, \bibinfo{author}{{et al.}},
  \bibinfo{year}{2017}.
\newblock in: \bibinfo{booktitle}{35th International Cosmic Ray Conference
  (ICRC2017)}, p. \bibinfo{pages}{766}.
\bibitem[{{Dom{\'i}nguez} et~al.(2011){Dom{\'i}nguez}, {Primack}, {Rosario} and
  {et al.}}]{2011MNRAS.410.2556D}
\bibinfo{author}{{Dom{\'i}nguez}, A.}, \bibinfo{author}{{Primack}, J.R.},
  \bibinfo{author}{{Rosario}, D.J.}, \bibinfo{author}{{et al.}},
  \bibinfo{year}{2011}.
\newblock \bibinfo{journal}{MNRAS} \bibinfo{volume}{410},
  \bibinfo{pages}{2556}.
\bibitem[{{Eichmann} and {Becker Tjus}(2016)}]{eic16}
\bibinfo{author}{{Eichmann}, B.}, \bibinfo{author}{{Becker Tjus}, J.},
  \bibinfo{year}{2016}.
\newblock \bibinfo{journal}{ApJ} \bibinfo{volume}{821}, \bibinfo{pages}{87}.
\bibitem[{{Elmouttie} et~al.(1998){Elmouttie}, {Koribalski}, {Gordon} and {et
  al.}}]{elm98}
\bibinfo{author}{{Elmouttie}, M.}, \bibinfo{author}{{Koribalski}, B.},
  \bibinfo{author}{{Gordon}, S.}, \bibinfo{author}{{et al.}},
  \bibinfo{year}{1998}.
\newblock \bibinfo{journal}{MNRAS} \bibinfo{volume}{297}, \bibinfo{pages}{49}.
\bibitem[{{Evans} et~al.(2004){Evans}, {Ferrer} and {Sarkar}}]{eva04}
\bibinfo{author}{{Evans}, N.W.}, \bibinfo{author}{{Ferrer}, F.},
  \bibinfo{author}{{Sarkar}, S.}, \bibinfo{year}{2004}.
\newblock \bibinfo{journal}{Phys. Rev. D} \bibinfo{volume}{69},
  \bibinfo{pages}{123501}.
\bibitem[{{Faucher-Gigu{\`e}re} and {Quataert}(2012)}]{fau12}
\bibinfo{author}{{Faucher-Gigu{\`e}re}, C.A.}, \bibinfo{author}{{Quataert},
  E.}, \bibinfo{year}{2012}.
\newblock \bibinfo{journal}{MNRAS} \bibinfo{volume}{425}, \bibinfo{pages}{605}.
\bibitem[{{Fermi-LAT collaboration} et~al.(2022){Fermi-LAT collaboration}, {:},
  {Abdollahi}, {Acero}, {Baldini} and {et al.}}]{Fermi4FGLDR3}
\bibinfo{author}{{Fermi-LAT collaboration}}, \bibinfo{author}{{:}},
  \bibinfo{author}{{Abdollahi}, S.}, \bibinfo{author}{{Acero}, F.},
  \bibinfo{author}{{Baldini}, L.}, \bibinfo{author}{{et al.}},
  \bibinfo{year}{2022}.
\newblock \bibinfo{title}{{Incremental Fermi Large Area Telescope Fourth Source
  Catalog}}.
\newblock \bibinfo{journal}{arXiv e-prints} ,
  \bibinfo{pages}{arXiv:2201.11184}\href{http://arxiv.org/abs/2201.11184}{\tt
  arXiv:2201.11184}.
\bibitem[{{Fermi-LAT Collaboration}(2010)}]{abd10}
\bibinfo{author}{{Fermi-LAT Collaboration}}, \bibinfo{year}{2010}.
\newblock \bibinfo{journal}{ApJL} \bibinfo{volume}{709},
  \bibinfo{pages}{L152--L157}.
\bibitem[{{Fermi-LAT Collaboration}(2020)}]{Ballet20}
\bibinfo{author}{{Fermi-LAT Collaboration}}, \bibinfo{year}{2020}.
\newblock \bibinfo{journal}{ApJS} \bibinfo{volume}{247}, \bibinfo{pages}{33}.
\bibitem[{{Fermi-LAT Collaboration} and {DES Collaboration}(2017)}]{fer17b}
\bibinfo{author}{{Fermi-LAT Collaboration}}, \bibinfo{author}{{DES
  Collaboration}}, \bibinfo{year}{2017}.
\newblock \bibinfo{journal}{ApJ} \bibinfo{volume}{834}, \bibinfo{pages}{110}.
\bibitem[{{Fermi-LAT Collaboration} et~al.(2014){Fermi-LAT Collaboration},
  {Pinzke} and {Pfrommer}}]{fer14}
\bibinfo{author}{{Fermi-LAT Collaboration}}, \bibinfo{author}{{Pinzke}, A.},
  \bibinfo{author}{{Pfrommer}, C.}, \bibinfo{year}{2014}.
\newblock \bibinfo{journal}{ApJ} \bibinfo{volume}{787}, \bibinfo{pages}{18}.
\bibitem[{{Fomin} et~al.(1994){Fomin}, {Stepanian}, {Lamb} and {et
  al.}}]{fom94}
\bibinfo{author}{{Fomin}, V.}, \bibinfo{author}{{Stepanian}, A.},
  \bibinfo{author}{{Lamb}, R.}, \bibinfo{author}{{et al.}},
  \bibinfo{year}{1994}.
\newblock \bibinfo{journal}{Astropart. Phys.} \bibinfo{volume}{2},
  \bibinfo{pages}{137}.
\bibitem[{{Franceschini} and {Rodighiero}(2017)}]{fra17}
\bibinfo{author}{{Franceschini}, A.}, \bibinfo{author}{{Rodighiero}, G.},
  \bibinfo{year}{2017}.
\newblock \bibinfo{journal}{A\&A} \bibinfo{volume}{603}, \bibinfo{pages}{A34}.
\bibitem[{{Gallimore} et~al.(1996){Gallimore}, {Baum}, {O'Dea} and {et
  al.}}]{gal96}
\bibinfo{author}{{Gallimore}, J.F.}, \bibinfo{author}{{Baum}, S.A.},
  \bibinfo{author}{{O'Dea}, C.P.}, \bibinfo{author}{{et al.}},
  \bibinfo{year}{1996}.
\newblock \bibinfo{journal}{ApJ} \bibinfo{volume}{458}, \bibinfo{pages}{136}.
\bibitem[{{Gao} and {Solomon}(2004)}]{gao04}
\bibinfo{author}{{Gao}, Y.}, \bibinfo{author}{{Solomon}, P.M.},
  \bibinfo{year}{2004}.
\newblock \bibinfo{journal}{ApJS} \bibinfo{volume}{152},
  \bibinfo{pages}{63--80}.
\bibitem[{{Garc{\'\i}a-Burillo} et~al.(2014){Garc{\'\i}a-Burillo}, {Combes},
  {Usero} and {et al.}}]{Garcia14}
\bibinfo{author}{{Garc{\'\i}a-Burillo}, S.}, \bibinfo{author}{{Combes}, F.},
  \bibinfo{author}{{Usero}, A.}, \bibinfo{author}{{et al.}},
  \bibinfo{year}{2014}.
\newblock \bibinfo{journal}{A\&A} \bibinfo{volume}{567}, \bibinfo{pages}{A125}.
\bibitem[{{Ghisellini} et~al.(2005){Ghisellini}, {Tavecchio} and
  {Chiaberge}}]{GTC05}
\bibinfo{author}{{Ghisellini}, G.}, \bibinfo{author}{{Tavecchio}, F.},
  \bibinfo{author}{{Chiaberge}, M.}, \bibinfo{year}{2005}.
\newblock \bibinfo{journal}{A\&A} \bibinfo{volume}{432},
  \bibinfo{pages}{401--410}.
\bibitem[{{Giannios} et~al.(2009){Giannios}, {Uzdensky} and
  {Begelman}}]{giannios_etal_09}
\bibinfo{author}{{Giannios}, D.}, \bibinfo{author}{{Uzdensky}, D.A.},
  \bibinfo{author}{{Begelman}, M.C.}, \bibinfo{year}{2009}.
\newblock \bibinfo{journal}{MNRAS} \bibinfo{volume}{395}, \bibinfo{pages}{L29}.
\bibitem[{{Giommi} et~al.(2019){Giommi}, {Brandt}, {Barres de Almeida} and {et
  al.}}]{gio19}
\bibinfo{author}{{Giommi}, P.}, \bibinfo{author}{{Brandt}, C.H.},
  \bibinfo{author}{{Barres de Almeida}, U.}, \bibinfo{author}{{et al.}},
  \bibinfo{year}{2019}.
\newblock \bibinfo{journal}{A\&A} \bibinfo{volume}{631}, \bibinfo{pages}{A116}.
\bibitem[{{HAWC Collaboration}(2021)}]{2021ApJ...907...67A}
\bibinfo{author}{{HAWC Collaboration}}, \bibinfo{year}{2021}.
\newblock \bibinfo{journal}{ApJ} \bibinfo{volume}{907}, \bibinfo{pages}{67}.
\bibitem[{{Hayashida} et~al.(2013){Hayashida}, {Stawarz}, {Cheung} and {et
  al.}}]{hay13}
\bibinfo{author}{{Hayashida}, M.}, \bibinfo{author}{{Stawarz}, {\L}.},
  \bibinfo{author}{{Cheung}, C.C.}, \bibinfo{author}{{et al.}},
  \bibinfo{year}{2013}.
\newblock \bibinfo{journal}{ApJ} \bibinfo{volume}{779}, \bibinfo{pages}{131}.
\bibitem[{{Heckmann} et~al.(2022){Heckmann}, {Paneque}, {Gasparyan} and {et
  al.}}]{hec22}
\bibinfo{author}{{Heckmann}, L.}, \bibinfo{author}{{Paneque}, D.},
  \bibinfo{author}{{Gasparyan}, S.}, \bibinfo{author}{{et al.}},
  \bibinfo{year}{2022}.
\newblock in: \bibinfo{booktitle}{37th International Cosmic Ray Conference.
  12-23 July 2021. Berlin}, p. \bibinfo{pages}{844}.
\bibitem[{{Henkel} et~al.(2018){Henkel}, {M{\"u}hle}, {Bendo} and {et
  al.}}]{hen18}
\bibinfo{author}{{Henkel}, C.}, \bibinfo{author}{{M{\"u}hle}, S.},
  \bibinfo{author}{{Bendo}, G.}, \bibinfo{author}{{et al.}},
  \bibinfo{year}{2018}.
\newblock \bibinfo{journal}{A\&A} \bibinfo{volume}{615}, \bibinfo{pages}{A155}.
\bibitem[{{H.E.S.S. Collaboration}(2014)}]{hess2014a}
\bibinfo{author}{{H.E.S.S. Collaboration}}, \bibinfo{year}{2014}.
\newblock \bibinfo{journal}{A\&A} \bibinfo{volume}{562}, \bibinfo{pages}{A145}.
\bibitem[{{Hussain} et~al.(2022){Hussain}, {Alves Batista}, {de Gouveia Dal
  Pino} and {et al.}}]{hus22}
\bibinfo{author}{{Hussain}, S.}, \bibinfo{author}{{Alves Batista}, R.},
  \bibinfo{author}{{de Gouveia Dal Pino}, E.M.}, \bibinfo{author}{{et al.}},
  \bibinfo{year}{2022}.
\newblock \bibinfo{journal}{arXiv e-prints} ,
  \bibinfo{pages}{arXiv:2203.01260}.
\bibitem[{{IceCube Collaboration} et~al.(2018){IceCube Collaboration},
  {Fermi-LAT Collaboration}, {MAGIC Collaboration}, {AGILE Team}, {ASAS-SN
  Team}, {HAWC Collaboration}, {H.~E.~S.~S. Collaboration}, {INTEGRAL Team},
  {VERITAS Collaboration} and {VLA/B Team}}]{ice18}
\bibinfo{author}{{IceCube Collaboration}}, \bibinfo{author}{{Fermi-LAT
  Collaboration}}, \bibinfo{author}{{MAGIC Collaboration}},
  \bibinfo{author}{{AGILE Team}}, \bibinfo{author}{{ASAS-SN Team}},
  \bibinfo{author}{{HAWC Collaboration}}, \bibinfo{author}{{H.~E.~S.~S.
  Collaboration}}, \bibinfo{author}{{INTEGRAL Team}}, \bibinfo{author}{{VERITAS
  Collaboration}}, \bibinfo{author}{{VLA/B Team}}, \bibinfo{year}{2018}.
\newblock \bibinfo{journal}{Science} \bibinfo{volume}{361},
  \bibinfo{pages}{eaat1378}.
\bibitem[{{Inoue} et~al.(2005){Inoue}, {Aharonian} and {Sugiyama}}]{ino05}
\bibinfo{author}{{Inoue}, S.}, \bibinfo{author}{{Aharonian}, F.A.},
  \bibinfo{author}{{Sugiyama}, N.}, \bibinfo{year}{2005}.
\newblock \bibinfo{journal}{ApJL} \bibinfo{volume}{628}, \bibinfo{pages}{L9}.
\bibitem[{{{\itshape Fermi}-LAT Collaboration}(2018)}]{fermi2018a}
\bibinfo{author}{{{\itshape Fermi}-LAT Collaboration}}, \bibinfo{year}{2018}.
\newblock \bibinfo{journal}{ApJS} \bibinfo{volume}{237}, \bibinfo{pages}{32}.
\bibitem[{{Jeltema} et~al.(2009){Jeltema}, {Kehayias} and {Profumo}}]{jel09}
\bibinfo{author}{{Jeltema}, T.E.}, \bibinfo{author}{{Kehayias}, J.},
  \bibinfo{author}{{Profumo}, S.}, \bibinfo{year}{2009}.
\newblock \bibinfo{journal}{Phys. Rev. D} \bibinfo{volume}{80},
  \bibinfo{pages}{023005}.
\bibitem[{{Kadowaki} et~al.(2021){Kadowaki}, {de Gouveia Dal Pino},
  {Medina-Torrej{\'o}n} and {et al.}}]{kad21}
\bibinfo{author}{{Kadowaki}, L.H.S.}, \bibinfo{author}{{de Gouveia Dal Pino},
  E.M.}, \bibinfo{author}{{Medina-Torrej{\'o}n}, T.E.}, \bibinfo{author}{{et
  al.}}, \bibinfo{year}{2021}.
\newblock \bibinfo{journal}{ApJ} \bibinfo{volume}{912}, \bibinfo{pages}{109}.
\bibitem[{{Kadowaki} et~al.(2015){Kadowaki}, {de Gouveia Dal Pino} and
  {Singh}}]{2015ApJ...802..113K}
\bibinfo{author}{{Kadowaki}, L.H.S.}, \bibinfo{author}{{de Gouveia Dal Pino},
  E.M.}, \bibinfo{author}{{Singh}, C.B.}, \bibinfo{year}{2015}.
\newblock \bibinfo{journal}{ApJ} \bibinfo{volume}{802}, \bibinfo{pages}{113}.
\bibitem[{{Kelner} et~al.(2006){Kelner}, {Aharonian} and {Bugayov}}]{kel06}
\bibinfo{author}{{Kelner}, S.R.}, \bibinfo{author}{{Aharonian}, F.A.},
  \bibinfo{author}{{Bugayov}, V.V.}, \bibinfo{year}{2006}.
\newblock \bibinfo{journal}{Phys. Rev. D} \bibinfo{volume}{74},
  \bibinfo{pages}{034018}.
\bibitem[{{Khiali} and {de Gouveia Dal Pino}(2016)}]{2016MNRAS.455..838K}
\bibinfo{author}{{Khiali}, B.}, \bibinfo{author}{{de Gouveia Dal Pino}, E.M.},
  \bibinfo{year}{2016}.
\newblock \bibinfo{journal}{MNRAS} \bibinfo{volume}{455}, \bibinfo{pages}{838}.
\bibitem[{{Kn{\"o}dlseder}(2011)}]{kno11}
\bibinfo{author}{{Kn{\"o}dlseder}, J.}, \bibinfo{year}{2011}.
\newblock \href{http://arxiv.org/abs/1110.007}{\tt arXiv:1110.007}.
\bibitem[{{Kn{\"o}dlseder} et~al.(2016a){Kn{\"o}dlseder}, {Mayer}, {Deil} and
  {et al.}}]{kno16}
\bibinfo{author}{{Kn{\"o}dlseder}, J.}, \bibinfo{author}{{Mayer}, M.},
  \bibinfo{author}{{Deil}, C.}, \bibinfo{author}{{et al.}},
  \bibinfo{year}{2016}a.
\newblock \bibinfo{journal}{A\&A} \bibinfo{volume}{593}, \bibinfo{pages}{A1}.
\bibitem[{{Kn{\"o}dlseder} et~al.(2016b){Kn{\"o}dlseder}, {Mayer}, {Deil} and
  {et al.}}]{kno16b}
\bibinfo{author}{{Kn{\"o}dlseder}, J.}, \bibinfo{author}{{Mayer}, M.},
  \bibinfo{author}{{Deil}, C.}, \bibinfo{author}{{et al.}},
  \bibinfo{year}{2016}b.
\newblock \href{http://arxiv.org/abs/1601.005}{\tt arXiv:1601.005}.
\bibitem[{{Krips} et~al.(2011){Krips}, {Mart{\'{\i}}n}, {Eckart} and {et
  al.}}]{kri11}
\bibinfo{author}{{Krips}, M.}, \bibinfo{author}{{Mart{\'{\i}}n}, S.},
  \bibinfo{author}{{Eckart}, A.}, \bibinfo{author}{{et al.}},
  \bibinfo{year}{2011}.
\newblock \bibinfo{journal}{ApJ} \bibinfo{volume}{736}, \bibinfo{pages}{37}.
\bibitem[{{Lamastra} et~al.(2016){Lamastra}, {Fiore}, {Guetta} and {et
  al.}}]{lam16}
\bibinfo{author}{{Lamastra}, A.}, \bibinfo{author}{{Fiore}, F.},
  \bibinfo{author}{{Guetta}, D.}, \bibinfo{author}{{et al.}},
  \bibinfo{year}{2016}.
\newblock \bibinfo{journal}{A\&A} \bibinfo{volume}{596}, \bibinfo{pages}{A68}.
\bibitem[{{Lamastra} et~al.(2017){Lamastra}, {Menci}, {Fiore} and {et
  al.}}]{lam17}
\bibinfo{author}{{Lamastra}, A.}, \bibinfo{author}{{Menci}, N.},
  \bibinfo{author}{{Fiore}, F.}, \bibinfo{author}{{et al.}},
  \bibinfo{year}{2017}.
\newblock \bibinfo{journal}{A\&A} \bibinfo{volume}{607}, \bibinfo{pages}{A18}.
\bibitem[{{Lamastra} et~al.(2019){Lamastra}, {Tavecchio}, {Romano} and {et
  al.}}]{lam19}
\bibinfo{author}{{Lamastra}, A.}, \bibinfo{author}{{Tavecchio}, F.},
  \bibinfo{author}{{Romano}, P.}, \bibinfo{author}{{et al.}},
  \bibinfo{year}{2019}.
\newblock \bibinfo{journal}{Astroparticle Physics} \bibinfo{volume}{112},
  \bibinfo{pages}{16--23}.
\bibitem[{{Lenain} et~al.(2010){Lenain}, {Ricci}, {T{\"u}rler} and {et
  al.}}]{len10}
\bibinfo{author}{{Lenain}, J.P.}, \bibinfo{author}{{Ricci}, C.},
  \bibinfo{author}{{T{\"u}rler}, M.}, \bibinfo{author}{{et al.}},
  \bibinfo{year}{2010}.
\newblock \bibinfo{journal}{A\&A} \bibinfo{volume}{524}, \bibinfo{pages}{A72}.
\bibitem[{{Lenc} and {Tingay}(2009)}]{len09}
\bibinfo{author}{{Lenc}, E.}, \bibinfo{author}{{Tingay}, S.J.},
  \bibinfo{year}{2009}.
\newblock \bibinfo{journal}{AJ} \bibinfo{volume}{137},
  \bibinfo{pages}{537--553}.
\bibitem[{{Liu} et~al.(2018){Liu}, {Murase}, {Inoue} and {et al.}}]{liu18}
\bibinfo{author}{{Liu}, R.Y.}, \bibinfo{author}{{Murase}, K.},
  \bibinfo{author}{{Inoue}, S.}, \bibinfo{author}{{et al.}},
  \bibinfo{year}{2018}.
\newblock \bibinfo{journal}{ApJ} \bibinfo{volume}{858}, \bibinfo{pages}{9}.
\bibitem[{{Lombardi} et~al.(2020){Lombardi}, {Catalano}, {Scuderi} and {et
  al.}}]{lom20}
\bibinfo{author}{{Lombardi}, S.}, \bibinfo{author}{{Catalano}, O.},
  \bibinfo{author}{{Scuderi}, S.}, \bibinfo{author}{{et al.}},
  \bibinfo{year}{2020}.
\newblock \bibinfo{journal}{A\&A} \bibinfo{volume}{634}, \bibinfo{pages}{A22}.
\bibitem[{{Lunardini} et~al.(2019){Lunardini}, {Vance}, {Emig} and {et
  al.}}]{Lunardini19}
\bibinfo{author}{{Lunardini}, C.}, \bibinfo{author}{{Vance}, G.S.},
  \bibinfo{author}{{Emig}, K.L.}, \bibinfo{author}{{et al.}},
  \bibinfo{year}{2019}.
\newblock \bibinfo{journal}{JCAP} \bibinfo{volume}{2019}, \bibinfo{pages}{073}.
\bibitem[{{Lynden-Bell}(1969)}]{lyn69}
\bibinfo{author}{{Lynden-Bell}, D.}, \bibinfo{year}{1969}.
\newblock \bibinfo{journal}{Nature} \bibinfo{volume}{223},
  \bibinfo{pages}{690}.
\bibitem[{{Madejski} and {Sikora}(2016)}]{mad16}
\bibinfo{author}{{Madejski}, G.G.}, \bibinfo{author}{{Sikora}, M.},
  \bibinfo{year}{2016}.
\newblock \bibinfo{journal}{ARA\&A} \bibinfo{volume}{54},
  \bibinfo{pages}{725--760}.
\bibitem[{{MAGIC Collaboration}(2018)}]{acc18}
\bibinfo{author}{{MAGIC Collaboration}}, \bibinfo{year}{2018}.
\newblock \bibinfo{journal}{Physics of the Dark Universe} \bibinfo{volume}{22},
  \bibinfo{pages}{38--47}.
\bibitem[{{MAGIC Collaboration}(2019a)}]{2019A&A...623A.175M}
\bibinfo{author}{{MAGIC Collaboration}}, \bibinfo{year}{2019}a.
\newblock \bibinfo{journal}{A\&A} \bibinfo{volume}{623}, \bibinfo{pages}{A175}.
\bibitem[{{MAGIC Collaboration}(2019b)}]{mag19}
\bibinfo{author}{{MAGIC Collaboration}}, \bibinfo{year}{2019}b.
\newblock \bibinfo{journal}{MNRAS} \bibinfo{volume}{486},
  \bibinfo{pages}{4233}.
\bibitem[{{MAGIC Collaboration}(2020)}]{2020A&A...637A..86M}
\bibinfo{author}{{MAGIC Collaboration}}, \bibinfo{year}{2020}.
\newblock \bibinfo{journal}{A\&A} \bibinfo{volume}{637}, \bibinfo{pages}{A86}.
\bibitem[{{MAGIC Collaboration}(2022)}]{mag22}
\bibinfo{author}{{MAGIC Collaboration}}, \bibinfo{year}{2022}.
\newblock \bibinfo{journal}{Physics of the Dark Universe} \bibinfo{volume}{35},
  \bibinfo{pages}{100912}.
\bibitem[{{MAGIC Collaboration} et~al.(2020){MAGIC Collaboration}, {Finke},
  {D'Ammando} and {et al.}}]{mag20b}
\bibinfo{author}{{MAGIC Collaboration}}, \bibinfo{author}{{Finke}, J.},
  \bibinfo{author}{{D'Ammando}, F.}, \bibinfo{author}{{et al.}},
  \bibinfo{year}{2020}.
\newblock \bibinfo{journal}{ApJS} \bibinfo{volume}{248}, \bibinfo{pages}{29}.
\bibitem[{{MAGIC Collaboration} et~al.(2019){MAGIC Collaboration}, {Fiore},
  {Feruglio} and {et al.}}]{MAGIC19}
\bibinfo{author}{{MAGIC Collaboration}}, \bibinfo{author}{{Fiore}, F.},
  \bibinfo{author}{{Feruglio}, C.}, \bibinfo{author}{{et al.}},
  \bibinfo{year}{2019}.
\newblock \bibinfo{journal}{ApJ} \bibinfo{volume}{883}, \bibinfo{pages}{135}.
\bibitem[{{Mannheim}(1993)}]{man93}
\bibinfo{author}{{Mannheim}, K.}, \bibinfo{year}{1993}.
\newblock \bibinfo{journal}{A\&A} \bibinfo{volume}{269},
  \bibinfo{pages}{67--76}.
\bibitem[{{Markarian} and {Lipovetskij}(1972)}]{mar72}
\bibinfo{author}{{Markarian}, B.E.}, \bibinfo{author}{{Lipovetskij}, V.A.},
  \bibinfo{year}{1972}.
\newblock \bibinfo{journal}{Astrofizika} \bibinfo{volume}{8},
  \bibinfo{pages}{155--164}.
\bibitem[{{Massaro} et~al.(2009){Massaro}, {Giommi}, {Leto} and {et
  al.}}]{mas09}
\bibinfo{author}{{Massaro}, E.}, \bibinfo{author}{{Giommi}, P.},
  \bibinfo{author}{{Leto}, C.}, \bibinfo{author}{{et al.}},
  \bibinfo{year}{2009}.
\newblock \bibinfo{journal}{A\&A} \bibinfo{volume}{495},
  \bibinfo{pages}{691--696}.
\bibitem[{{Massaro} et~al.(2015){Massaro}, {Maselli}, {Leto} and {et
  al.}}]{mas15}
\bibinfo{author}{{Massaro}, E.}, \bibinfo{author}{{Maselli}, A.},
  \bibinfo{author}{{Leto}, C.}, \bibinfo{author}{{et al.}},
  \bibinfo{year}{2015}.
\newblock \bibinfo{journal}{Ap\&SS} \bibinfo{volume}{357}, \bibinfo{pages}{75}.
\bibitem[{{Mateo}(1998)}]{mat98}
\bibinfo{author}{{Mateo}, M.L.}, \bibinfo{year}{1998}.
\newblock \bibinfo{journal}{ARA\&A} \bibinfo{volume}{36}, \bibinfo{pages}{435}.
\bibitem[{{Matthews} et~al.(2020){Matthews}, {Bell} and
  {Blundell}}]{2020arXiv200306587M}
\bibinfo{author}{{Matthews}, J.H.}, \bibinfo{author}{{Bell}, A.R.},
  \bibinfo{author}{{Blundell}, K.M.}, \bibinfo{year}{2020}.
\newblock \bibinfo{journal}{New Astron. Rev.} \bibinfo{volume}{89},
  \bibinfo{pages}{101543}.
\bibitem[{{McConnachie}(2012)}]{mcc12}
\bibinfo{author}{{McConnachie}, A.W.}, \bibinfo{year}{2012}.
\newblock \bibinfo{journal}{AJ} \bibinfo{volume}{144}, \bibinfo{pages}{4}.
\bibitem[{{Medina-Torrej{\'o}n} et~al.(2021){Medina-Torrej{\'o}n}, {de Gouveia
  Dal Pino}, {Kadowaki} and {et al.}}]{med21}
\bibinfo{author}{{Medina-Torrej{\'o}n}, T.E.}, \bibinfo{author}{{de Gouveia Dal
  Pino}, E.M.}, \bibinfo{author}{{Kadowaki}, L.H.S.}, \bibinfo{author}{{et
  al.}}, \bibinfo{year}{2021}.
\newblock \bibinfo{journal}{ApJ} \bibinfo{volume}{908}, \bibinfo{pages}{193}.
\bibitem[{{Melioli} and {de Gouveia Dal Pino}(2015)}]{mel15}
\bibinfo{author}{{Melioli}, C.}, \bibinfo{author}{{de Gouveia Dal Pino}, E.M.},
  \bibinfo{year}{2015}.
\newblock \bibinfo{journal}{ApJ} \bibinfo{volume}{812}, \bibinfo{pages}{90}.
\bibitem[{{Miniati} and {Elyiv}(2013)}]{miniati2013a}
\bibinfo{author}{{Miniati}, F.}, \bibinfo{author}{{Elyiv}, A.},
  \bibinfo{year}{2013}.
\newblock \bibinfo{journal}{ApJ} \bibinfo{volume}{770}, \bibinfo{pages}{54}.
\bibitem[{{M{\"u}cke} et~al.(2003){M{\"u}cke}, {Protheroe}, {Engel} and {et
  al.}}]{mue03}
\bibinfo{author}{{M{\"u}cke}, A.}, \bibinfo{author}{{Protheroe}, R.J.},
  \bibinfo{author}{{Engel}, R.}, \bibinfo{author}{{et al.}},
  \bibinfo{year}{2003}.
\newblock \bibinfo{journal}{Astropart. Phys.} \bibinfo{volume}{18},
  \bibinfo{pages}{593--613}.
\bibitem[{{Neronov} and {Vovk}(2010)}]{neronov2010a}
\bibinfo{author}{{Neronov}, A.}, \bibinfo{author}{{Vovk}, I.},
  \bibinfo{year}{2010}.
\newblock \bibinfo{journal}{Science} \bibinfo{volume}{328},
  \bibinfo{pages}{73}.
\bibitem[{{Nigro} et~al.(2019){Nigro}, {Deil}, {Zanin} and {et
  al.}}]{gammapy:2019}
\bibinfo{author}{{Nigro}, C.}, \bibinfo{author}{{Deil}, C.},
  \bibinfo{author}{{Zanin}, R.}, \bibinfo{author}{{et al.}},
  \bibinfo{year}{2019}.
\newblock \bibinfo{journal}{A\&A} \bibinfo{volume}{625}, \bibinfo{pages}{A10}.
\bibitem[{{Nims} et~al.(2015){Nims}, {Quataert} and
  {Faucher-Gigu{\`e}re}}]{nim15}
\bibinfo{author}{{Nims}, J.}, \bibinfo{author}{{Quataert}, E.},
  \bibinfo{author}{{Faucher-Gigu{\`e}re}, C.A.}, \bibinfo{year}{2015}.
\newblock \bibinfo{journal}{MNRAS} \bibinfo{volume}{447},
  \bibinfo{pages}{3612}.
\bibitem[{{Pace} and {Strigari}(2019)}]{pac19}
\bibinfo{author}{{Pace}, A.B.}, \bibinfo{author}{{Strigari}, L.E.},
  \bibinfo{year}{2019}.
\newblock \bibinfo{journal}{MNRAS} \bibinfo{volume}{482},
  \bibinfo{pages}{3480}.
\bibitem[{{Padovani} and {Giommi}(1995)}]{pad95}
\bibinfo{author}{{Padovani}, P.}, \bibinfo{author}{{Giommi}, P.},
  \bibinfo{year}{1995}.
\newblock \bibinfo{journal}{MNRAS} \bibinfo{volume}{277},
  \bibinfo{pages}{1477}.
\bibitem[{{Padovani} et~al.(2016){Padovani}, {Resconi}, {Giommi} and {et
  al.}}]{pad16}
\bibinfo{author}{{Padovani}, P.}, \bibinfo{author}{{Resconi}, E.},
  \bibinfo{author}{{Giommi}, P.}, \bibinfo{author}{{et al.}},
  \bibinfo{year}{2016}.
\newblock \bibinfo{journal}{MNRAS} \bibinfo{volume}{457},
  \bibinfo{pages}{3582}.
\bibitem[{{Pareschi} et~al.(2016){Pareschi}, {Bonnoli}, {Vercellone}, {ASTRI
  Project} and {CTA Consortium}}]{par16}
\bibinfo{author}{{Pareschi}, G.}, \bibinfo{author}{{Bonnoli}, G.},
  \bibinfo{author}{{Vercellone}, S.}, \bibinfo{author}{{ASTRI Project}},
  \bibinfo{author}{{CTA Consortium}}, \bibinfo{year}{2016}.
\newblock in: \bibinfo{booktitle}{Journal of Physics Conference Series}, p.
  \bibinfo{pages}{052028}.
\bibitem[{{Persic} et~al.(2008){Persic}, {Rephaeli} and {Arieli}}]{per08}
\bibinfo{author}{{Persic}, M.}, \bibinfo{author}{{Rephaeli}, Y.},
  \bibinfo{author}{{Arieli}, Y.}, \bibinfo{year}{2008}.
\newblock \bibinfo{journal}{A\&A} \bibinfo{volume}{486},
  \bibinfo{pages}{143--149}.
\bibitem[{{Petropoulou} et~al.(2016){Petropoulou}, {Coenders} and
  {Dimitrakoudis}}]{2016APh....80..115P}
\bibinfo{author}{{Petropoulou}, M.}, \bibinfo{author}{{Coenders}, S.},
  \bibinfo{author}{{Dimitrakoudis}, S.}, \bibinfo{year}{2016}.
\newblock \bibinfo{journal}{Astroparticle Physics} \bibinfo{volume}{80},
  \bibinfo{pages}{115--130}.
\bibitem[{{Pierre} et~al.(2014){Pierre}, {Siegal-Gaskins} and {Scott}}]{pie14}
\bibinfo{author}{{Pierre}, M.}, \bibinfo{author}{{Siegal-Gaskins}, J.M.},
  \bibinfo{author}{{Scott}, P.}, \bibinfo{year}{2014}.
\newblock \bibinfo{journal}{JCAP} \bibinfo{volume}{2014}, \bibinfo{pages}{024}.
\bibitem[{{Pierre Auger Collaboration}(2018)}]{augerSBG}
\bibinfo{author}{{Pierre Auger Collaboration}}, \bibinfo{year}{2018}.
\newblock \bibinfo{journal}{ApJL} \bibinfo{volume}{853}, \bibinfo{pages}{L29}.
\bibitem[{{Pintore} et~al.(2020){Pintore}, {Giuliani}, {Belfiore} and {et
  al.}}]{pin20}
\bibinfo{author}{{Pintore}, F.}, \bibinfo{author}{{Giuliani}, A.},
  \bibinfo{author}{{Belfiore}, A.}, \bibinfo{author}{{et al.}},
  \bibinfo{year}{2020}.
\newblock \bibinfo{journal}{Journal of High Energy Astrophysics}
  \bibinfo{volume}{26}, \bibinfo{pages}{83--94}.
\bibitem[{{Pinzke} et~al.(2017){Pinzke}, {Oh} and {Pfrommer}}]{pin17}
\bibinfo{author}{{Pinzke}, A.}, \bibinfo{author}{{Oh}, S.P.},
  \bibinfo{author}{{Pfrommer}, C.}, \bibinfo{year}{2017}.
\newblock \bibinfo{journal}{MNRAS} \bibinfo{volume}{465},
  \bibinfo{pages}{4800}.
\bibitem[{{Pinzke} et~al.(2009){Pinzke}, {Pfrommer} and
  {Bergstr{\"o}m}}]{pin09}
\bibinfo{author}{{Pinzke}, A.}, \bibinfo{author}{{Pfrommer}, C.},
  \bibinfo{author}{{Bergstr{\"o}m}, L.}, \bibinfo{year}{2009}.
\newblock \bibinfo{journal}{Phys. Rev. Lett.} \bibinfo{volume}{103},
  \bibinfo{pages}{181302}.
\bibitem[{{Planck Collaboration} et~al.(2016){Planck Collaboration}, {Ade},
  {Aghanim} and {et al.}}]{pla16}
\bibinfo{author}{{Planck Collaboration}}, \bibinfo{author}{{Ade}, P.A.R.},
  \bibinfo{author}{{Aghanim}, N.}, \bibinfo{author}{{et al.}},
  \bibinfo{year}{2016}.
\newblock \bibinfo{journal}{A\&A} \bibinfo{volume}{594}, \bibinfo{pages}{A13}.
\bibitem[{{Resconi} et~al.(2017){Resconi}, {Coenders}, {Padovani} and {et
  al.}}]{res17}
\bibinfo{author}{{Resconi}, E.}, \bibinfo{author}{{Coenders}, S.},
  \bibinfo{author}{{Padovani}, P.}, \bibinfo{author}{{et al.}},
  \bibinfo{year}{2017}.
\newblock \bibinfo{journal}{MNRAS} \bibinfo{volume}{468}, \bibinfo{pages}{597}.
\bibitem[{{Rigby} et~al.(2009){Rigby}, {Diamond-Stanic} and {Aniano}}]{rig09}
\bibinfo{author}{{Rigby}, J.R.}, \bibinfo{author}{{Diamond-Stanic}, A.M.},
  \bibinfo{author}{{Aniano}, G.}, \bibinfo{year}{2009}.
\newblock \bibinfo{journal}{ApJ} \bibinfo{volume}{700}, \bibinfo{pages}{1878}.
\bibitem[{{Rodr{\'i}guez-Ram{\'i}rez} et~al.(2019){Rodr{\'i}guez-Ram{\'i}rez},
  {de Gouveia Dal Pino} and {Alves Batista}}]{2019ApJ...879....6R}
\bibinfo{author}{{Rodr{\'i}guez-Ram{\'i}rez}, J.C.}, \bibinfo{author}{{de
  Gouveia Dal Pino}, E.M.}, \bibinfo{author}{{Alves Batista}, R.},
  \bibinfo{year}{2019}.
\newblock \bibinfo{journal}{ApJ} \bibinfo{volume}{879}, \bibinfo{pages}{6}.
\bibitem[{{Rubin} et~al.(1980){Rubin}, {Ford} and {Thonnard}}]{rub80}
\bibinfo{author}{{Rubin}, V.C.}, \bibinfo{author}{{Ford}, W.~K., J.},
  \bibinfo{author}{{Thonnard}, N.}, \bibinfo{year}{1980}.
\newblock \bibinfo{journal}{ApJ} \bibinfo{volume}{238},
  \bibinfo{pages}{471--487}.
\bibitem[{{Salpeter}(1964)}]{sal64}
\bibinfo{author}{{Salpeter}, E.E.}, \bibinfo{year}{1964}.
\newblock \bibinfo{journal}{ApJ} \bibinfo{volume}{140},
  \bibinfo{pages}{796--800}.
\bibitem[{{Schlickeiser} et~al.(2012){Schlickeiser}, {Ibscher} and
  {Supsar}}]{schlickeiser2012a}
\bibinfo{author}{{Schlickeiser}, R.}, \bibinfo{author}{{Ibscher}, D.},
  \bibinfo{author}{{Supsar}, M.}, \bibinfo{year}{2012}.
\newblock \bibinfo{journal}{ApJ} \bibinfo{volume}{758}, \bibinfo{pages}{102}.
\bibitem[{{S{\'e}rsic}(1960)}]{ser60}
\bibinfo{author}{{S{\'e}rsic}, J.L.}, \bibinfo{year}{1960}.
\newblock \bibinfo{journal}{Zeitschr. f{\"u}r Astroph.} \bibinfo{volume}{50},
  \bibinfo{pages}{168}.
\bibitem[{{Singh} et~al.(2015){Singh}, {de Gouveia Dal Pino} and
  {Kadowaki}}]{2015ApJ...799L..20S}
\bibinfo{author}{{Singh}, C.B.}, \bibinfo{author}{{de Gouveia Dal Pino}, E.M.},
  \bibinfo{author}{{Kadowaki}, L.H.S.}, \bibinfo{year}{2015}.
\newblock \bibinfo{journal}{ApJL} \bibinfo{volume}{799}, \bibinfo{pages}{L20}.
\bibitem[{{Singh} et~al.(2016){Singh}, {Mizuno} and {de Gouveia Dal
  Pino}}]{2016ApJ...824...48S}
\bibinfo{author}{{Singh}, C.B.}, \bibinfo{author}{{Mizuno}, Y.},
  \bibinfo{author}{{de Gouveia Dal Pino}, E.M.}, \bibinfo{year}{2016}.
\newblock \bibinfo{journal}{ApJ} \bibinfo{volume}{824}, \bibinfo{pages}{48}.
\bibitem[{{Steigman} et~al.(2012){Steigman}, {Dasgupta} and {Beacom}}]{ste12}
\bibinfo{author}{{Steigman}, G.}, \bibinfo{author}{{Dasgupta}, B.},
  \bibinfo{author}{{Beacom}, J.F.}, \bibinfo{year}{2012}.
\newblock \bibinfo{journal}{Phys. Rev. D} \bibinfo{volume}{86},
  \bibinfo{pages}{023506}.
\bibitem[{{Strigari} et~al.(2008){Strigari}, {Bullock}, {Kaplinghat} and {et
  al.}}]{str08}
\bibinfo{author}{{Strigari}, L.E.}, \bibinfo{author}{{Bullock}, J.S.},
  \bibinfo{author}{{Kaplinghat}, M.}, \bibinfo{author}{{et al.}},
  \bibinfo{year}{2008}.
\newblock \bibinfo{journal}{Nature} \bibinfo{volume}{454},
  \bibinfo{pages}{1096--1097}.
\bibitem[{{Tamborra} et~al.(2014){Tamborra}, {Ando} and {Murase}}]{tamborra14}
\bibinfo{author}{{Tamborra}, I.}, \bibinfo{author}{{Ando}, S.},
  \bibinfo{author}{{Murase}, K.}, \bibinfo{year}{2014}.
\newblock \bibinfo{journal}{JCAP} \bibinfo{volume}{2014}, \bibinfo{pages}{043}.
\bibitem[{{Tavecchio} and {Ghisellini}(2015)}]{TG15}
\bibinfo{author}{{Tavecchio}, F.}, \bibinfo{author}{{Ghisellini}, G.},
  \bibinfo{year}{2015}.
\newblock \bibinfo{journal}{MNRAS} \bibinfo{volume}{451},
  \bibinfo{pages}{1502}.
\bibitem[{{Tavecchio} et~al.(2010){Tavecchio}, {Ghisellini}, {Foschini} and {et
  al.}}]{tavecchio2010a}
\bibinfo{author}{{Tavecchio}, F.}, \bibinfo{author}{{Ghisellini}, G.},
  \bibinfo{author}{{Foschini}, L.}, \bibinfo{author}{{et al.}},
  \bibinfo{year}{2010}.
\newblock \bibinfo{journal}{MNRAS} \bibinfo{volume}{406},
  \bibinfo{pages}{L70--L74}.
\bibitem[{{Tavecchio} et~al.(1998){Tavecchio}, {Maraschi} and
  {Ghisellini}}]{1998ApJ...509..608T}
\bibinfo{author}{{Tavecchio}, F.}, \bibinfo{author}{{Maraschi}, L.},
  \bibinfo{author}{{Ghisellini}, G.}, \bibinfo{year}{1998}.
\newblock \bibinfo{journal}{ApJ} \bibinfo{volume}{509},
  \bibinfo{pages}{608--619}.
\bibitem[{{Tavecchio} et~al.(2019){Tavecchio}, {Oikonomou} and
  {Righi}}]{2019MNRAS.488.4023T}
\bibinfo{author}{{Tavecchio}, F.}, \bibinfo{author}{{Oikonomou}, F.},
  \bibinfo{author}{{Righi}, C.}, \bibinfo{year}{2019}.
\newblock \bibinfo{journal}{MNRAS} \bibinfo{volume}{488},
  \bibinfo{pages}{4023}.
\bibitem[{{Tisserand} et~al.(2007){Tisserand}, {Le Guillou}, {Afonso} and {et
  al.}}]{tis07}
\bibinfo{author}{{Tisserand}, P.}, \bibinfo{author}{{Le Guillou}, L.},
  \bibinfo{author}{{Afonso}, C.}, \bibinfo{author}{{et al.}},
  \bibinfo{year}{2007}.
\newblock \bibinfo{journal}{A\&A} \bibinfo{volume}{469},
  \bibinfo{pages}{387--404}.
\bibitem[{{Trumpler}(1935)}]{tru35}
\bibinfo{author}{{Trumpler}, R.J.}, \bibinfo{year}{1935}.
\newblock \bibinfo{journal}{PASP} \bibinfo{volume}{47}, \bibinfo{pages}{219}.
\bibitem[{{Urry} and {Padovani}(1995)}]{urr95}
\bibinfo{author}{{Urry}, C.M.}, \bibinfo{author}{{Padovani}, P.},
  \bibinfo{year}{1995}.
\newblock \bibinfo{journal}{PASP} \bibinfo{volume}{107}, \bibinfo{pages}{803}.
\bibitem[{{van Weeren} et~al.(2019){van Weeren}, {de Gasperin}, {Akamatsu} and
  {et al.}}]{van19}
\bibinfo{author}{{van Weeren}, R.J.}, \bibinfo{author}{{de Gasperin}, F.},
  \bibinfo{author}{{Akamatsu}, H.}, \bibinfo{author}{{et al.}},
  \bibinfo{year}{2019}.
\newblock \bibinfo{journal}{Space Science Reviews} \bibinfo{volume}{215},
  \bibinfo{pages}{16}.
\bibitem[{{Vannoni} et~al.(2011){Vannoni}, {Aharonian}, {Gabici} and {et
  al.}}]{van11}
\bibinfo{author}{{Vannoni}, G.}, \bibinfo{author}{{Aharonian}, F.A.},
  \bibinfo{author}{{Gabici}, S.}, \bibinfo{author}{{et al.}},
  \bibinfo{year}{2011}.
\newblock \bibinfo{journal}{A\&A} \bibinfo{volume}{536}, \bibinfo{pages}{A56}.
\bibitem[{{VERITAS Collaboration}(2009)}]{2009Natur.462..770V}
\bibinfo{author}{{VERITAS Collaboration}}, \bibinfo{year}{2009}.
\newblock \bibinfo{journal}{Nature} \bibinfo{volume}{462},
  \bibinfo{pages}{770--772}.
\bibitem[{{VERITAS Collaboration}(2017)}]{veritas2017a}
\bibinfo{author}{{VERITAS Collaboration}}, \bibinfo{year}{2017}.
\newblock \bibinfo{journal}{ApJ} \bibinfo{volume}{835}, \bibinfo{pages}{288}.
\bibitem[{{VERITAS Collaboration} et~al.(2016){VERITAS Collaboration},
  {Fumagalli} and {Prochaska}}]{arc16}
\bibinfo{author}{{VERITAS Collaboration}}, \bibinfo{author}{{Fumagalli}, M.},
  \bibinfo{author}{{Prochaska}, J.X.}, \bibinfo{year}{2016}.
\newblock \bibinfo{journal}{AJ} \bibinfo{volume}{151}, \bibinfo{pages}{142}.
\bibitem[{{Wang} and {Loeb}(2016a)}]{wang16}
\bibinfo{author}{{Wang}, X.}, \bibinfo{author}{{Loeb}, A.},
  \bibinfo{year}{2016}a.
\newblock \bibinfo{journal}{JCAP} \bibinfo{volume}{2016}, \bibinfo{pages}{012}.
\bibitem[{{Wang} and {Loeb}(2016b)}]{wang_gamma}
\bibinfo{author}{{Wang}, X.}, \bibinfo{author}{{Loeb}, A.},
  \bibinfo{year}{2016}b.
\newblock \bibinfo{journal}{Nat. Phys.} \bibinfo{volume}{12},
  \bibinfo{pages}{1116--1118}.
\bibitem[{{Wilson}(1955)}]{wil55}
\bibinfo{author}{{Wilson}, A.G.}, \bibinfo{year}{1955}.
\newblock \bibinfo{journal}{PASP} \bibinfo{volume}{67},
  \bibinfo{pages}{27--29}.
\bibitem[{{Xi} et~al.(2018){Xi}, {Wang}, {Liang} and {et al.}}]{xi18}
\bibinfo{author}{{Xi}, S.Q.}, \bibinfo{author}{{Wang}, X.Y.},
  \bibinfo{author}{{Liang}, Y.F.}, \bibinfo{author}{{et al.}},
  \bibinfo{year}{2018}.
\newblock \bibinfo{journal}{Phys. Rev. D} \bibinfo{volume}{98},
  \bibinfo{pages}{063006}.
\bibitem[{{Zandanel} and {Ando}(2014)}]{zan14}
\bibinfo{author}{{Zandanel}, F.}, \bibinfo{author}{{Ando}, S.},
  \bibinfo{year}{2014}.
\newblock \bibinfo{journal}{MNRAS} \bibinfo{volume}{440}, \bibinfo{pages}{663}.
\bibitem[{{Zech} et~al.(2017){Zech}, {Cerruti} and
  {Mazin}}]{2017A&A...602A..25Z}
\bibinfo{author}{{Zech}, A.}, \bibinfo{author}{{Cerruti}, M.},
  \bibinfo{author}{{Mazin}, D.}, \bibinfo{year}{2017}.
\newblock \bibinfo{journal}{A\&A} \bibinfo{volume}{602}, \bibinfo{pages}{A25}.
\bibitem[{{Zel'dovich} and {Novikov}(1965)}]{zel65}
\bibinfo{author}{{Zel'dovich}, Y.B.}, \bibinfo{author}{{Novikov}, I.D.},
  \bibinfo{year}{1965}.
\newblock \bibinfo{journal}{Soviet Physics Doklady} \bibinfo{volume}{9},
  \bibinfo{pages}{834}.
\bibitem[{{Zitzer} and {VERITAS Collaboration}(2017)}]{zit17}
\bibinfo{author}{{Zitzer}, B.}, \bibinfo{author}{{VERITAS Collaboration}},
  \bibinfo{year}{2017}.
\newblock in: \bibinfo{booktitle}{35th International Cosmic Ray Conference
  (ICRC2017)}, p. \bibinfo{pages}{904}.
\bibitem[{{Zschaechner} et~al.(2016){Zschaechner}, {Walter}, {Bolatto} and {et
  al.}}]{zsc16}
\bibinfo{author}{{Zschaechner}, L.K.}, \bibinfo{author}{{Walter}, F.},
  \bibinfo{author}{{Bolatto}, A.}, \bibinfo{author}{{et al.}},
  \bibinfo{year}{2016}.
\newblock \bibinfo{journal}{ApJ} \bibinfo{volume}{832}, \bibinfo{pages}{142}.
\bibitem[{{Zwicky}(1933)}]{zwi33}
\bibinfo{author}{{Zwicky}, F.}, \bibinfo{year}{1933}.
\newblock \bibinfo{journal}{Helvetica Physica Acta} \bibinfo{volume}{6},
  \bibinfo{pages}{110}.

\end{thebibliography}

\end{document}